%% file: racahfinal.tex
\documentclass[article,1p,11pt]{elsarticle}
\usepackage{amsfonts,amsthm,amsmath,amssymb,natbib}
\usepackage{color}
\definecolor{MyDarkBlue}{rgb}{0.15,0.15,0.45}
\usepackage[linktocpage=true]{hyperref}
\hypersetup{
colorlinks=true,
citecolor=MyDarkBlue,
linkcolor=MyDarkBlue,
urlcolor=MyDarkBlue
}
\usepackage[enableskew]{youngtab}

\newcommand{\mdot}{\raise1.5pt \hbox{.}}

\makeatletter
\@addtoreset{equation}{section}
\makeatother

\journal{Nuclear Physics B}

\begin{document}

\begin{frontmatter}

\title{$SU(N)$ quantum Racah coefficients \& non-torus links}
\author{Zodinmawia}
\ead{zodin@iitb.ac.in}
\author{P. Ramadevi}
\ead{ramadevi@phy.iitb.ac.in}
\address{Department of Physics, Indian Institute of Technology Bombay,\\
 Mumbai, India, 400076}

\begin{abstract}
It is well-known that the $SU(2)$ quantum Racah coefficients or the Wigner $6j$ symbols
have a closed form expression which enables the evaluation of any knot or link
polynomials in $SU(2)$ Chern-Simons field theory. Using isotopy 
equivalence of $SU(N)$ Chern-Simons functional integrals over three balls with one or more
$S^2$ boundaries with punctures, we obtain identities to be satisfied by the 
$SU(N)$ quantum Racah coefficients. This enables evaluation of the coefficients for 
a class of $SU(N)$ representations. Using these coefficients, 
we can compute the polynomials for some non-torus knots and two-component links. 
These results are useful for verifying conjectures in topological string theory.
\end{abstract}


\end{frontmatter}

\tableofcontents
\section{Introduction}
Following the seminal work of Witten \cite{Witten:1988hf} on Chern-Simons theory as a 
theory of knots and links, generalised invariants \cite{RamaDevi:1992np,Ramadevi:1996:PHD}
for any knot or link can be directly obtained without going through the recursive procedure.
For torus links, which can be wrapped on a two-torus $T^2$,
using the torus link operators \cite{Labastida:1990bt}, explicit
polynomial form of these invariants could be obtained. However for
non-torus links, the generalised invariants \cite{RamaDevi:1992np}
in $SU(N)$ Chern-Simons theory 
involves $SU(N)$ quantum Racah coefficients which are not known in closed form
as known for $SU(2)$ \cite{Kirillov:1989,Kaul:1993hb}. This prevents in writing the
polynomial form for the non-torus links.

\begin{figure}
\centering
	\includegraphics[scale=1]{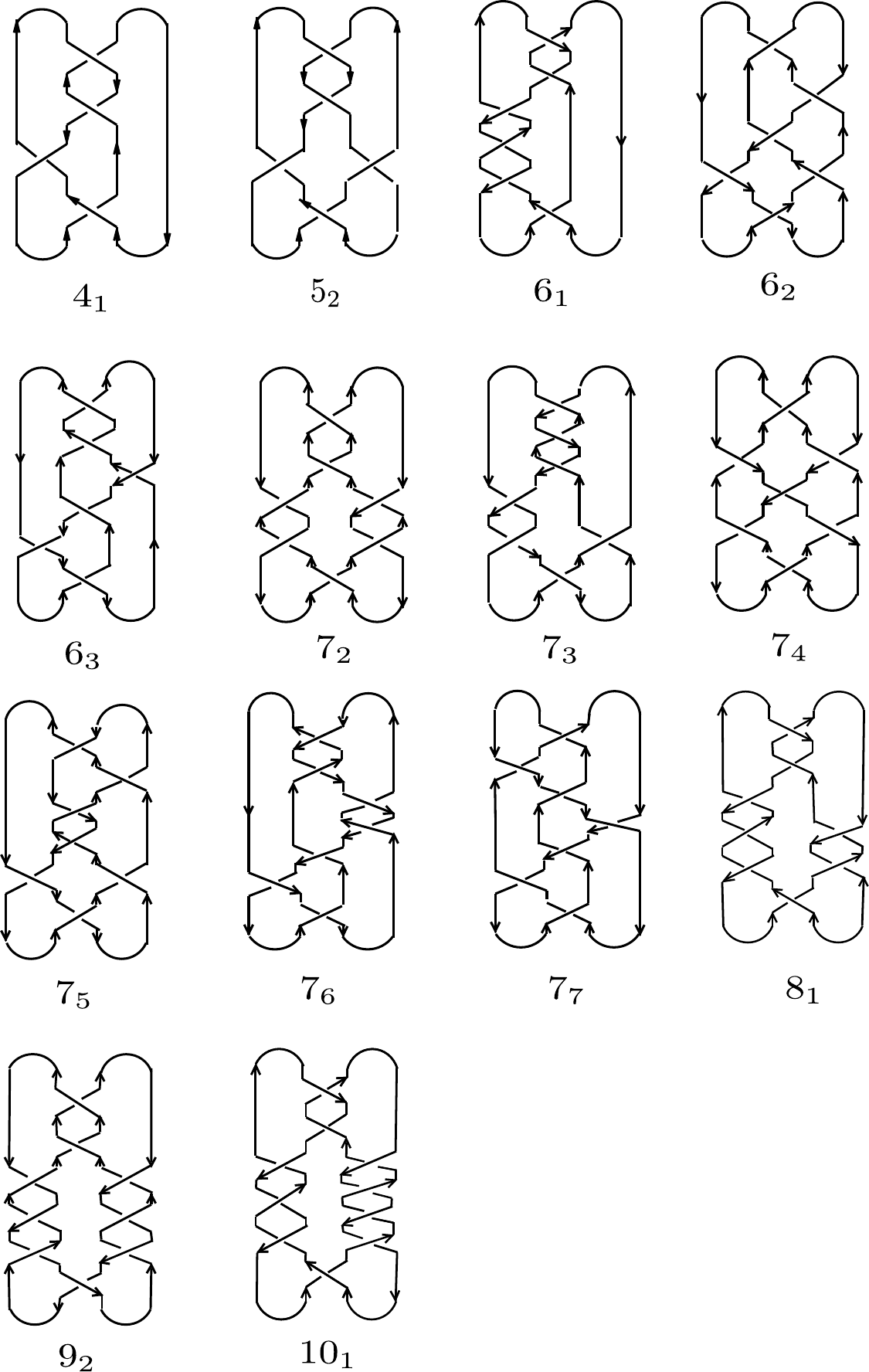} 
\caption{Plat representation for some non-torus knots}
	\label{fig:plat}
\end{figure}

\Yboxdim5pt
For simple $SU(N)$ representations placed on knots, whose Young Tableau are \Yboxdim6pt\yng(1), \Yboxdim6pt\yng(2) and \Yboxdim6pt\yng(1,1), 
we had obtained some Racah coefficients from isotopy equivalence of knots and links 
which was useful to obtain polynomial for few non-torus knots \cite{RamaDevi:1992np,Ramadevi:1996:PHD}.
Going beyond these simple representations and finding the quantum Racah coefficients  
appeared to be a formidable task. In fact, determining these coefficients would help 
in verifying the topological string conjectures for a
general non-torus knot or link
proposed by Ooguri-Vafa \cite{Ooguri:1999bv} and Labastida-Marino-Vafa \cite{Labastida:2000yw}.
Using the few Racah coefficients 
data \cite{RamaDevi:1992np}, Ooguri-Vafa conjecture for $4_1,6_1$ 
non-torus knots as indicated in Figure \ref {fig:plat} were verified \cite{Ramadevi:2000gq}. 
For verifying Labastida-Marinoi-Vafa conjecture for
the non-torus two-component links \cite{Labastida:2000yw},
we need to evaluate the non-torus link whose component knots carry different
representations.

These non-torus links invariants will also
be useful to generalise some 
of the results of recent papers \cite{Witten:2011zz,Brini:2011wi,Aganagic:2011sg,Gaiotto:2011nm}
where torus knots and links are studied.
So, it is very important to  determine the $SU(N)$ quantum Racah coefficients.

Using the correspondence between Chern-Simons functional integral and the
correlator conformal blocks states in the corresponding 
Wess-Zumino conformal field theory \cite{Witten:1988hf}, we can 
derive identities to be obeyed by the $SU(N)$ quantum Racah coefficients. Using the
identities and the the properties of quantum dimensions for any $N$, we could determine
the form of these coefficients for some class of $SU(N)$ representations.
These coefficients are needed to obtain polynomial invariants of some 
non-torus knots and non-torus two-component links.

\begin{figure}
\centering
\includegraphics[scale=1]{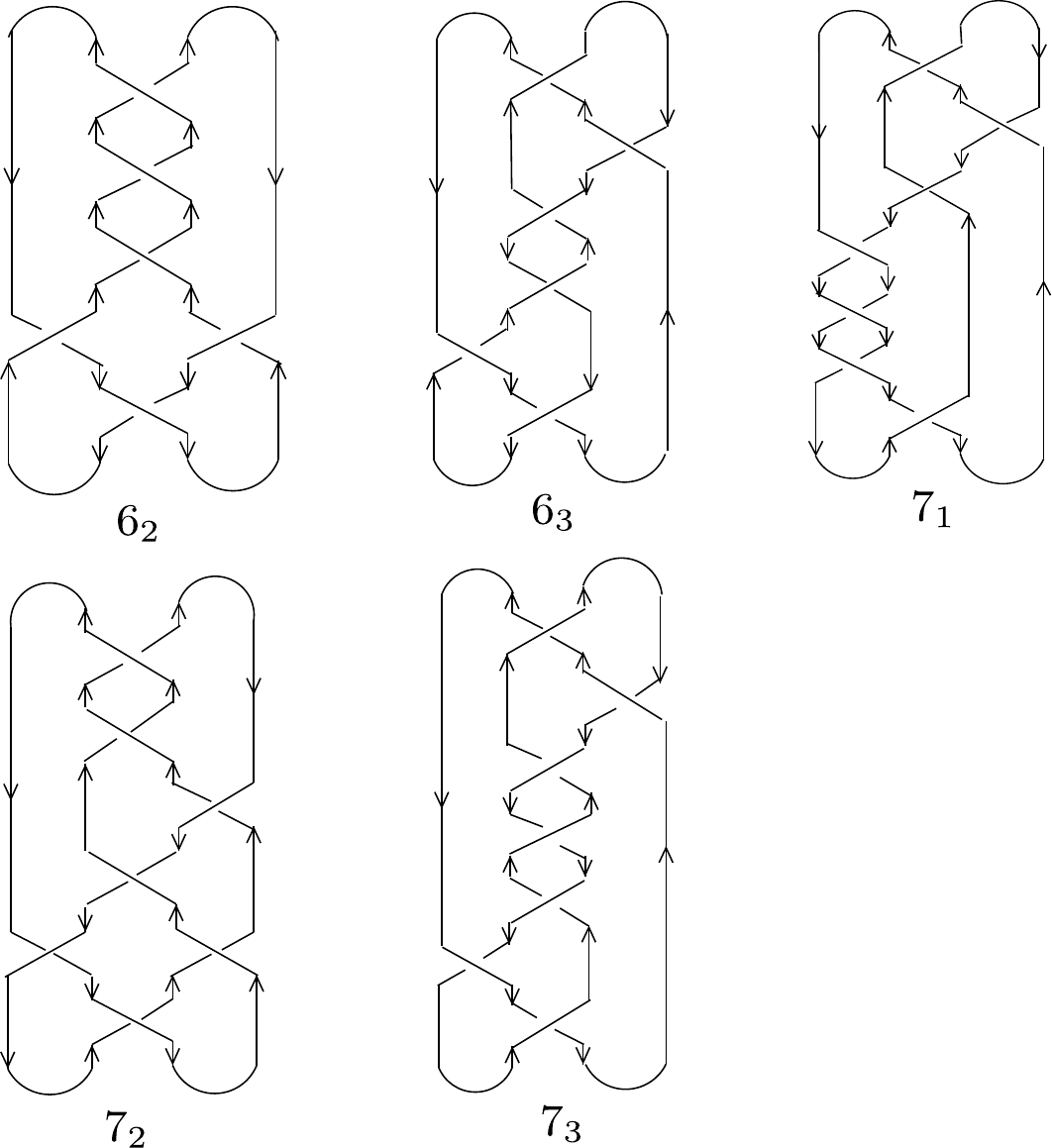} 
\caption{Plat representation for some non-torus links}
	\label{figs:plat}
\end{figure}
The plan of the paper is as follows: In section \ref{sec:cs_theory}, we briefly review
Chern-Simons functional integrals and properties of the Racah coefficients.
In section \ref{sec:duality_matrix}, we systematically study the equivalence of states and
obtain identities to be obeyed by the $SU(N)$ quantum Racah coefficients. 
In section \ref{sec:SU(N)_quantum}, we tabulate the $SU(N)$ quantum Racah coefficients.
In \ref{sec:appenA} we give the
generalised Chern-Simons invariant for the non-torus knots and non-torus links  in Figures 1 \& 2. Then we present
the polynomial form of these invariants for few representations in \ref{sec:appenB} and \ref{sec:appenC}. We also verify Ooguri-Vafa conjecture
for knots and Labastida-Marino-Vafa conjecture for links in \ref{sec:appenD}.  
In the concluding section, we summarize and discuss some of the open problems.

\section{Chern-Simons Field Theory} 
\label{sec:cs_theory}
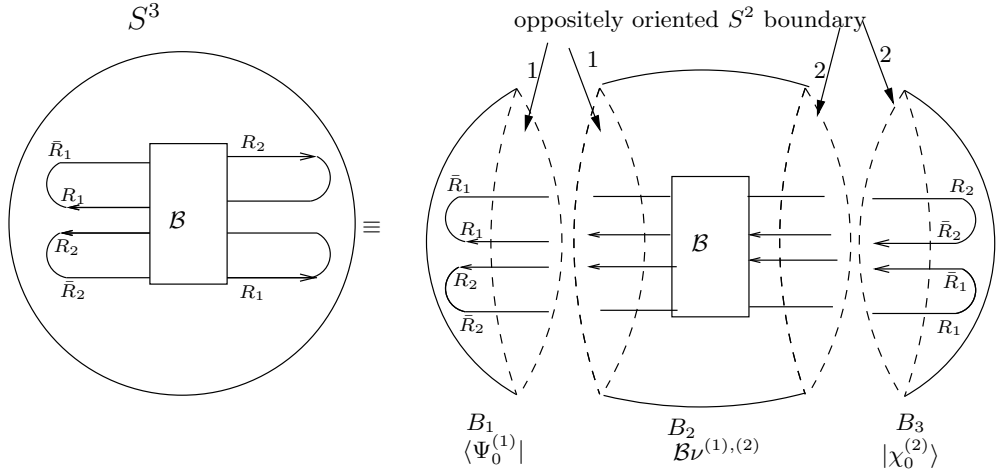
\begin{figure}[htbp]
\centering
	\input{glue.pspdftex}
\caption{Link in $S^3$ viewed as gluing of three-balls with four-punctured $S^2$ boundaries}
	\label{figs:glue}
\end{figure}

\begin{figure}[htbp]
\centering
\includegraphics[scale=1]{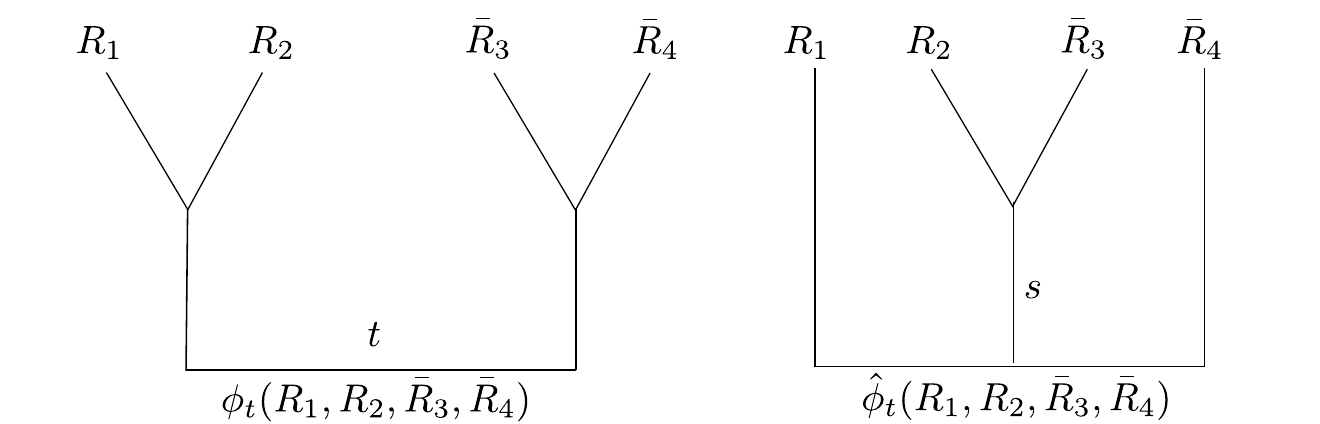} 
\caption{four-point conformal block bases states}
\label{figs:conf}
\end{figure}

Chern-Simons fields on $S^3$ with $U(1)\times SU(N)$ gauge group with levels $k_1,k$
respectively 
is given by
by the following action:
\begin{equation}
 S=\frac{k_1}{4 \pi} \int_{S^3} B \wedge dB+
\frac{k}{4 \pi} \int_{S^3} Tr\left(A \wedge dA+ \frac{2}{3}A \wedge A \wedge A \right)~,
\end{equation}
where $B$ is the $U(1)$ gauge connection and $A$ is the 
$SU(N)$ matrix valued gauge connection. The observables in this theory
are Wilson loop operators:
\begin{equation}
 W_{(R_1,n_1),(R_2,n_2),\ldots
(R_r,n_r)
}[{\cal L}] = \prod_{\beta=1}^{r}Tr_{R_{\beta}}
U^A[{\cal K}_{\beta}]
 Tr_{n_{\beta}}U^B[{\cal K}_{\beta}]~,
\end{equation}
where the  holonomy of the gauge field $A$ around a component knot ${\cal K}_{\beta}$,
carrying a representation $R_{\beta}$, of 
a $r$-component link is denoted by 
$U^A[{\cal K}_{\beta}]=P[\exp \oint_{{\cal K}_{\beta}}A]$
and $n_{\beta}$ is the $U(1)$ charge carried by the component knot ${\cal K}_{\beta}$.
The expectation value of these Wilson loop operators are the link invariants:
\small{
\begin{eqnarray}
V_{(R_1,n_1),\ldots}^{\{U(N)\}}[
{\cal L}](q,\lambda)
&=&
\langle W_{(R_1,n_1),\ldots}[{\cal L}]\rangle=
 \frac{\int[{\cal D}B][{\cal D}A] e^{iS}W_{(R_1,n_1),\ldots (R_r,n_r)}
[{\cal L}]}{\int [{\cal D}B][{\cal D}A] e^{iS}}\nonumber\\
&=&V_{R_1,R_2,\ldots R_r}^{\{SU(N)\}}[{
\cal L}]
V_{n_1,n_2,\ldots ,n_r}^{\{U(1\}}[{\cal L}]~.
\end{eqnarray}
}
We make a specific choice of the U(1) charges and the coupling $k_1$
so that the above invariant is a
polynomial in two variables $q=\exp\left(\frac{2 \pi i}{k+N}\right)$, 
$\lambda=q^N$ \cite{Marino:2001re,Borhade:2003cu} 
\begin{equation}
 n_{\beta}=\frac{l_{\beta}}{\sqrt N} ~;~k_1=k+N~,
\end{equation}
where $l_{\beta}$ is the total number of boxes in the 
Young Tableau representation $R_{\beta}$.
The $U(1)$ link invariant for the link with this substitution gives
{\small
\begin{equation}
 V_{\frac{l_1}{\sqrt N},\ldots ,\frac{l_r}{\sqrt N}}^{\{U(1\}}[
{\cal L}]
=(-1)^{\sum_{\beta} l_{\beta} p_{\beta}}
\exp\left(\frac{i \pi}{k+N}\sum_{\beta=1}^r
\frac{l_{\beta}^2 p_{\beta}}{N}\right) \exp\left( \frac{i \pi}{k+N} \sum_{\alpha \neq \beta}
\frac{l_{\alpha} l_{\beta} {lk}_
{\alpha \beta}}{N} \right)~, \label{u1}
\end{equation}
}
where $p_{\beta}$ is the framing number of the component knot ${\cal K}_{\beta}$
and ${lk}_{\alpha\beta}$ is the linking number between the component knots
${\cal K}_{\alpha}$ and ${\cal K}_{\beta}$.
In order to directly evaluate $SU(N)$ link invariants, we need
to use the following two ingredients:
\begin{enumerate}
\item  The relation between $SU(N)$ Chern-Simons functional integral 
on the three-dimensional ball to the two-dimensional
$SU(N)_k$ Wess-Zumino conformal field theory on the boundary of the three-ball \cite{Witten:1988hf}.
\item  Any knot or link can be drawn as a platting of braids \cite{birman}.
\end{enumerate}
In Figure \ref{fig:plat} and Figure \ref{figs:plat} we have drawn some non-torus knots and 
non-torus links as a plat representation of braids. We have 
labelled them in the Thistlewaithe notation and written their braid words.
We have indicated the orientation and labelled the representation $R_i$
on the component knots in the link.
Note that $b_i^{(-)}$ ($\{b_i^{(-)}\}^{-1}$) in the braid word denotes 
right-handed crossing (left-handed crossing) between $i$-strand and 
$i+1$-th strand which are anti-parallelly oriented. Similarly $b_j^{(+)}$  ($\{b_j^{(+)}\}^{-1}$)  
denotes right-handed crossing (left-handed crossing)  between $j$ and $j+1$-th strand 
which are parallelly oriented. The plat representation of these
 non-torus knots and non-torus links  involves braids with 
four-strands. Hence we can view these knots and links  in 
$S^3$ as gluing of three-balls  with $4$-punctured boundary as 
shown in Figure \ref{figs:glue}.
There are two three-balls $B_1$ and $B_3$
with one $S^2$ boundary. A three-ball denoted as $B_2$
in Figure \ref{figs:glue} with two $S^2$  boundaries with a braid ${\cal B}$ which can represent any
of the braid words corresponding to the non-torus knots and links in Figures \ref{fig:plat}\&\ref{figs:plat}. 
The gluing of the three-balls are along the oppositely oriented $S^2$ boundaries. The functional 
integral over the ball $B_3$ is given by a state $\vert \chi_0^{(2)} \rangle$
where the superscript denotes the label of the
$S^2$ boundary. The representation $R_i$ indicate that the lines are going into the
$S^2$ boundary of the three-ball and the conjugate representation denotes the 
lines going out of the
$S^2$ boundary. The state corresponding to a functional integral on a three-ball
with an oppositely oriented boundary 
is written in a dual space along 
with conjugating all the representations as illustrated for the 
ball $B_1$. The expectation value of the Wilson-loop operator gives the
link invariant for a nontorus link $\mathcal{L}$
\begin{equation}
 V_{R_1,R_2}^{SU(N)}[\mathcal{L}]=\langle \Psi_0^{(1)} \vert 
 {\cal B} \nu^{(1),(2)} \vert \chi_0^{(2)}\rangle~. \label {nontl}
\end{equation}
These invariants multiplied with the $U(1)$ invariant (\ref {u1})
 are polynomials in two variables $q=\exp(2 \pi i/(k+N))$ and 
$\lambda=q^N$. In order to see the polynomial form, we write these states 
on a four-punctured boundary
in a suitable basis of four-point conformal block of the $SU(N)_k$ Wess-Zumino conformal
field theory. There are two different four-point conformal block bases as shown in Figure \ref{figs:conf}
where $t \in (R_1 \otimes R_2) \cap (R_3 \times R_4)$ 
and
$s \in (R_2 \otimes \bar R_3) \cap (\bar R_1 \times R_4)$.
Using these bases, the states corresponding to three balls $B_1,B_2$ and $B_3$ in 
Figure \ref {figs:glue}, can be expanded as \cite{RamaDevi:1992np,Ramadevi:1996:PHD}
\begin{eqnarray}
\langle \Psi_0^{(1)} \vert&=&\sqrt{dim_q R_1 dim_q R_2} \langle 
\phi_0(\bar R_1,R_1, R_2,\bar R_2)^{(1)} \vert \label {st1} \\
~&=&\sum_{s \in R_1 \otimes R_2}\epsilon_s^{(R_1,R_2)}\sqrt {dim_q s} \langle 
\hat{\phi}_s (\bar R_1,R_1,R_2,\bar R_2) ^{(1)}\vert \nonumber\\
{\cal B} \nu^{(1),(2)}&=& {\cal B}\vert \phi_t(R_2,
\bar R_2,\bar R_1,R_1)^{(1)} \rangle \langle
 \phi_t(\bar R_2, R_2, R_1, \bar R_1)^{(2)} \vert\nonumber \\
~&=&{\cal B}
\vert \hat {\phi}_s(R_2,
\bar R_2, \bar R_1,R_1)^{(1)} \rangle \langle
\hat {\phi}_s (\bar R_2, R_2, R_1, \bar R_1)^{(2)} \vert \label{braid}\\
\vert \chi_0^{(2)} \rangle&=&\sqrt{dim_q R_1 dim_q R_2} \vert 
\phi_0(R_2,\bar R_2,\bar R_1,R_1)^{(2)} \rangle \label {st2}\\
&=&\sum_{s \in R_1 \otimes R_2}\epsilon_s^{(R_1,R_2)}\sqrt {dim_q s} \vert \hat{\phi}_s 
(R_2,\bar R_2, \bar R_1,R_1) ^{(2)}\rangle~, 
\end{eqnarray}
where the subscript `0' in the basis state in eqns.(\ref {st1}, \ref {st2}) denotes
the singlet state. 
The states are written such that the  invariant of a simple circle (also called unknot)
carrying representation $R$ is normalised as $dim_q R$, which is the
quantum dimension of the  representation $R$, defined as 
 \begin{equation}
 dim_q R~=~\prod_{\alpha > 0} { [\alpha . (\rho + \Lambda_R)] \over [\alpha . \rho]}~,
\label {qdm}
 \end{equation}
where $\Lambda_R$ denotes the highest weight of the representation $R$,
$\alpha$'s are the positive roots and $\rho$ is equal to the sum of the
fundamental weights of the Lie group $SU(N)$. 
The square bracket refers to quantum number defined as
\begin{equation}
[n] = {q^{n \over 2} - q^{-{n \over 2}} \over 
q^{1 \over 2} - q^{-{1 \over 2}}}~. \label {qun}
\end{equation}
The symbol $\epsilon_s^{(R_1,R_2)}=\pm 1$ which  we will fix from equivalence 
of states in the next section. To operate the braid word
${\cal B}$ in eqn.(\ref {braid}), we need to find the eigenbasis of the braiding generators $b_i^{(\pm)}$'s.

The conformal block $\vert \phi_t(R_1,R_2,\bar R_3, \bar R_4)  \rangle$ is suitable
for the  braiding operator $b_1^{(\pm)}$ and $b_3^{(\pm)}$. Similarly braiding in 
the middle two strands  involving the operator
$b_2^{(\pm)}$ requires the conformal block
$\vert \hat {\phi}_s(R_1,R_2,\bar R_3,\bar R_4) \rangle$. That is,
\begin{eqnarray}
 b_2^{(\pm)} \vert \hat {\phi}_s(R_1,R_2,\bar R_3,\bar R_4) \rangle
&=&
\lambda_s^{(\pm)}(R_2,\bar R_3)
\vert \hat {\phi}_s(R_1,\bar R_3,R_2,\bar R_4) \rangle
~,\nonumber\\
~b_1^
{(\pm)}\vert \phi_t(R_1,R_2,\bar R_3, \bar R_4)  \rangle
&=&
\lambda_t^{(\pm)}(R_1, R_2)
\vert \hat {\phi}_s(R_2,R_1,\bar R_3,\bar R_4) \rangle~,\nonumber\\
b_3^{(\pm)}
\vert \hat {\phi}_s(R_1,R_2,\bar R_3,\bar R_4) \rangle&=&
\lambda_t^{(\pm)}(\bar R_3,\bar R_4)\vert \hat {\phi}_s(R_1,R_2,\bar R_4,\bar R_3) \rangle~,
\end{eqnarray}
where braiding eigenvalues 
$\lambda_t^{(\pm)}(R_1,R_2)$ in vertical framing are 
\begin{equation}
 \lambda_t^{(\pm)}(R_1,R_2)=\epsilon^{(\pm)}_{t;R_1,R_2} \left
(q^{\frac{C_{R_1}+C_{R_2}-C_{R_t}}{2}}\right
)^{\pm 1}~.\label {brev}
\end{equation}
In this framing, framing number $p_{\beta}$ for the component
knot is equal to writhe $w$ of that component knot
which is the difference between total number of left-handed crossings
and total number of right-handed crossing. For example,
torus knots $4_1,5_2$ in Figure \ref{fig:plat}  have writhe
 $w$ equal to 0 and 5 respectively. The symbol
$\epsilon^{(\pm)}_{t;R_1,R_2}$ is a sign which can be fixed by studying
equivalence of states or equivalence of links which we shall elaborate for
a class of representations in the next section and $C_R$ denotes the quadratic casimir 
for the representation $R$ given by
\begin{equation}
C_R= \kappa_R - \frac{l^2}{2N}~,~\kappa_R=\frac{1}{2}[Nl+l+\sum_i (l_i^2-2il_i)]~, \label {casi}
\end{equation}
where $l_i$ is the number of boxes in the i-th row of the Young Tableau representation
$R$ and $l$ is the total number of boxes.
The two bases in Figure \ref{figs:conf} are related by a duality matrix $a$ as follows:
\begin{equation}
\vert \phi_t(R_1,R_2,\bar R_3, \bar R_4)  \rangle= a_{ts}\left[\begin{matrix}R_1&R_2 \cr\bar R_3 &\bar R_4 \end{matrix}\right]  
\vert \hat {\phi}_s(R_1,R_2,\bar R_3,\bar R_4) \rangle~. \label{dual}
\end{equation}
From the definition of $t$,$s$, we can see that the duality matrix $a$ obeys the 
following properties:
\small{
\begin{equation}
 a_{ts}\left[\begin{matrix}R_1&R_2 \cr\bar R_3 &\bar R_4 \end{matrix} \right]=
 a_{st}\left[\begin{matrix}\bar R_3 &R_2 \cr R_1 &\bar R_4 \end{matrix}\right]=
 a_{\bar s \bar t}\left[\begin{matrix}R_1&\bar R_4 \cr \bar R_3 & R_2 \end{matrix}\right]=
 a_{\bar t \bar s}\left[\begin{matrix}\bar R_3&\bar R_4 \cr R_1 & R_2 \end{matrix}\right]=
a_{ t  s}\left[\begin{matrix} R_3& R_4 \cr \bar R_1 & \bar R_2 \end{matrix}\right]~.\label{pro1}
\end{equation}
}
If one of the representation is singlet (denoted by $0$),
we see that the matrix elements are 
\begin{eqnarray}
 a_{ts}\left[\begin{matrix}R_1&0\cr\bar R_3 &\bar R_4 \end{matrix} \right]=
\delta_{t R_1} \delta_{s \bar R_3}~,~
 a_{ts}\left[\begin{matrix}0&R_2 \cr\bar R_3 &\bar R_4 \end{matrix} \right]=
\delta_{t R_2} \delta_{s R_4}~,\nonumber
\\
a_{ts}\left[\begin{matrix}R_1&R_2\cr 0 &\bar R_4 \end{matrix} \right]=
\delta_{t R_4} \delta_{s R_2}~,~
 a_{ts}\left[\begin{matrix}R_1&R_2 \cr\bar R_3 &0 \end{matrix} \right]=
\delta_{t R_3} \delta_{s\bar  R_1}~. \label {sing}
\end{eqnarray}
From the procedure presented in this section, we can 
write the $U(N)$ invariants of  non-torus knots and 
links as a product of $U(1)$ invariant times $SU(N)$ invariant  
(\ref {nontl}). For example, non-torus knot $5_2$ with
framing $p=5$, the invariant will be
\begin{eqnarray}
V_R^{\{U(N)\}}[5_2;p]&=&(-1)^{5l}q^{\frac{5l^2}{2N}}\sum_{s,t,s'} \epsilon_s^{R,R}~ \sqrt{dim_q s}~ 
\epsilon_{s'}^{R,R}~ \sqrt{dim_q s'}
(\lambda_s^{(+)}(R,R))^{-1} \nonumber\\
~&~&a_{t s}\left[\begin{matrix}\bar R& R\cr R&\bar R
               \end{matrix}\right]
(\lambda_t^{(-)}(\bar R,R))^{-2}
a_{t s'}\left[\begin{matrix}R&\bar R\cr \bar R&R
               \end{matrix}\right](\lambda_{s'}^{(+)}(R,R))^{-2}, \label{52k}
\end{eqnarray}
where $l$ is the total number of boxes in the Young Tableau representing $R$
and we have indicated the framing number of the knot.
We could add additional framing $p_1$ by multiplying these invariants by a framing
factor as follows:
\begin{eqnarray}
 V_R^{\{U(N)\}}[K;(p+p_1)]&=&(-1)^{l p_1} q^{p_1 l^2 \over 2N}(\lambda_0^{(-)}(R,\bar R))^{-p_1}
V_R^{\{U(N)\}}[K;p]\nonumber\\
~&=& (-1)^{lp_1}q^{p_1 \kappa_R} V_R^{(U(N)}[K;p]~. \label{52kp}
\end{eqnarray}
So, to obtain $5_2$ invariant with zero framing, we have to take $p_1=-5$.
For all the non-torus knots in Figure \ref{fig:plat}, we have presented zero framed knot invariants in \ref{sec:appenA}.

Similarly the invariant for link $6_2$ with linking number $lk=3$ 
with framing $p_1,p_2$ on the component knots will be
{\small
\begin{eqnarray} 
V_{(R_1,R_2)}^{\{U(N)\}}[6_2;p_1,p_2]&=&\prod_{i=1}^2 \{(-1)^{l_{R_i} p_i} 
q^{p_i \kappa_{R_i}}\}
q^{\frac{3l_{R_1}l_{R_2}}{N}}
\sum_{s,t,s'} \epsilon_s^{R_1,R_2}~ \sqrt{dim_q s}~ \label{62l}\\  
~~~&~&\epsilon_{s'}^{\bar{R}_1,\bar{R}_2}~ 
\sqrt{dim_q s'} (\lambda_s^{(+)}(R_1,R_2))^{-3} 
a_{ts}\left[\begin{array}{cc}R_2&\bar{R}_1\\ \bar{R}_2&R_1\end{array}\right]\nonumber\\
~&~&(\lambda_t^{(-)}(\bar{R}_1,R_2))^{-2}
a_{ts'}\left[\begin{array}{cc}\bar{R}_1&R_2\\R_1&\bar{R}_2\end{array}\right]
(\lambda_{s'}^{(+)}(\bar{R}_1,\bar{R}_2))^{-1}~,~~~~\nonumber
\end{eqnarray}}
where $l_{R_i}$ is the total number of boxes in the Young diagram representation $R_i$
placed on the component knots in link $6_2$. The invariants for the non-torus links in Figure \ref{figs:plat} are
presented in \ref{sec:appenA}.
However to see the polynomial form of  these link invariants, we have to 
determine the coefficients of the
duality matrices. Unlike the
 $SU(2)$ duality matrices \cite{Kirillov:1989,Kaul:1993hb}, we do not 
have a closed form expression for SU(N) duality matrices.

In the following section, we will use equivalence of states
to obtain the sign (\ref {brev}, \ref {st1}) and also 
derive identities satisfied by the coefficients of the duality matrix.
This enables the evaluation of duality matrix elements for some class of
representations. 

Once we have these coefficients, we could evaluate
the framed link invariants and obtain the reformulated invariants \cite{Labastida:2001ts}
\begin{eqnarray}
f_{R_1,R_2, \ldots R_r}(q, \lambda)&=& \sum_{d,m=1}^{\infty} (-1)^{m-1}
{\mu(d) \over dm} \sum_{\{\vec k^{(\alpha j)},R_{\alpha j} \}}\times
\nonumber\\
~&~&\prod_{\alpha=1}^r \chi_{R_{\alpha}} \left( C\left( (\sum_{j=1}^m 
\vec k^{(\alpha j)} )_d \right) \right) \prod_{j=1}^m 
{\vert C(\vec k^{(\alpha j)})\vert \over \ell_{\alpha j}!} \times
\nonumber\\
~&~& \chi_{R_{\alpha j}}(C(\vec k^{(\alpha j)})) ~ V_{R_{1j}, R_{2j},
\ldots R_{rj}}^{(U(N))}[L; \{p_{\alpha} \}](q^d, \lambda^d)~, \label{findd}
\end{eqnarray}
where $\mu(d)$ is the Moebius function defined as follows: 
if $d$ has a prime decomposition ($\{p_i\}$), $d= \prod_{i=1}^a
p_i^{m_i}$, then $\mu(d)=0$ if any of the $m_i$ is greater
than one. If all $m_i=1$, then $\mu(d)=(-1)^a$.
The second sum in the above equation runs over all vectors 
$\vec k^{(\alpha j)}$, with
$\alpha=1 , \ldots r$ and $j=1, \ldots m$, such that 
$\sum_{\alpha=1}^r \vert \vec k^{(\alpha j)} \vert > 0$ for
any $j$ and over representations $R_{\alpha j}$. 
Further $\vec k_d$ is defined as follows: $(\vec k_d)_{di}= k_i$ and has
zero entries for the other components. Therefore, if
$\vec k= (k_1, k_2, \ldots)$, then
\begin{equation}
\vec k_d=(0,\ldots, 0, k_1, 0, \ldots ,0,k_2,0,\ldots),
\end{equation}
where $k_1$ is in the $d$-the entry, $k_2$ in the $2d$-th entry, and so
on. Here  $C(\vec k)$ denotes the conjugacy class determined by the
sequence $(k_1,k_2, \cdots)$ (i.e there are $k_1$ 1-cycles, 
$k_2$ 2-cycles etc) in the permutation group $S_{\ell}$ (${\ell}=\sum_j jk_j$). 
For a Young Tabeleau representation $R$ with
$\ell$ number of boxes,  $\chi_R( C(\vec k))$
gives the character of the conjugacy class $C(\vec k)$ in the representation $R$.
The explicit relation of the  above expression 
in terms of Chern-Simons invariants are presented in appendix D for few
representations.
The reformulated invariants for $r$-component links  are expected to obey 
Labastida-Marino-Vafa conjecture \cite{Labastida:2000yw}
\begin{equation}
f_{(R_1, R_2, \ldots R_r)} (q, \lambda)= (q^{1/2}- q^{-1/2})^{r-2}
\sum_{Q,s} N_{(R_1, \ldots R_r),Q,s} q^s \lambda^Q~, \label{fexp}
\end{equation}
where $N_{(R_1, \ldots R_r),Q,s}$ are integers. After determining the 
identities and some of the matrix elements of the duality matrices in
the following two sections, we will obtain the polynomial invariants of the non-torus knots and links in \ref{sec:appenB} and \ref{sec:appenC}.

\section{Duality Matrix identities}
\label{sec:duality_matrix}
We had elaborated in the previous section on writing states (\ref {st1},\ref{braid},\ref{st2})
corresponding to Chern-Simons functional integral on three balls. We can
determine the following coefficients of the duality matrix by
comparing eqn.(\ref {dual}) and eqns.(\ref {st1},\ref{st2}):
\begin{equation}
a_{0 s}\left[\begin{matrix}\bar R_1& R_1\cr R_2 & \bar R_2
               \end{matrix}\right]=\epsilon_s^{(R_1,R_2)} \frac{\sqrt{dim_q s}}
{\sqrt{dim_q R_1 dim_q R_2}}~. \label {sign}
\end{equation}
This relation along with the property (\ref {sing}) suggests 
that 
\begin{equation}
a_{ts}\left[\begin{matrix}R_1&R_2 \cr\bar R_3 &\bar R_4 \end{matrix} \right]
=\epsilon_{R_1}\epsilon_{R_2}\epsilon_{\bar R_3}\epsilon_{\bar R_4} \sqrt{dim_q s} \sqrt{dim_q t}
 \left\{\begin{matrix}R_1&R_2&t\cr \bar R_3& 
 \bar R_4&s\end{matrix}\right\}~,
\end{equation}
where $\epsilon_{R_i}=\pm 1=\epsilon_{\bar R_i}$ and $\epsilon_0=1$. The term in 
parenthesis is similar to the $SU(2)$ quantum Wigner $6j$ symbol but requires 
appropriate conjugation of representations under interchange of columns in the 
following way:
\begin{equation} 
\left\{\begin{matrix}R_1&R_2&t\cr \bar R_3& 
 \bar R_4&s\end{matrix}\right\}=
 \left\{\begin{matrix}t&\bar R_2& R_1
\cr s&\bar R_4&\bar R_3\end{matrix}\right\}=
 \left\{\begin{matrix}\bar R_1&t&R_2\cr 
\bar R_3&\bar s&R_4 \end{matrix}\right\}~.\label{wigner}
\end{equation}
Using (\ref {sing}) and the relation to quantum Wigner symbol with
the above properties, the $SU(N)$ duality matrix can
be called as SU(N) quantum Racah coefficients and hence we propose 
that the coefficients obey the following property: 
\begin{equation}
a_{ts}\left[\begin{matrix}R_1&R_2 \cr\bar R_3 &\bar R_4 \end{matrix} \right]
=\frac{\sqrt{dim_q t dim_q s}}{\sqrt{dim_q R_1 dim_q R_3}}
\epsilon_{R_1}\epsilon_{\bar R_3}(\epsilon_s \epsilon_t)^{-1} 
a_{R_1 \bar R_3}
\left[\begin{matrix}t&\bar R_2 \cr s &\bar R_4 \end{matrix} \right]~.
\end{equation}
Using this property, we can relate the sign in eqn.(\ref {sign}) as
\begin{equation}
\epsilon_s^{(R_1,R_2)}= \epsilon_{R_1}\epsilon_{R_2}(\epsilon_s)^{-1}~.
\end{equation}
Starting from the state $\nu^{(1),(2)}$ in eqn.(\ref {braid}) and the
duality relation (\ref {dual}), we observe that the 
Racah coefficients must obey the following identities:
\begin{eqnarray}
\sum_s a_{t s}\left[\begin{matrix}R_1&\bar R_2\cr \bar R_3 & \bar R_4
               \end{matrix}\right]
a_{t' s}\left[\begin{matrix}R_1&\bar R_2\cr \bar R_3 & \bar R_4
               \end{matrix}\right]= \delta_{t t'} ~,\label {ortho1}\\
\sum_t
a_{t s}\left[\begin{matrix}R_1&\bar R_2\cr \bar R_3 & \bar R_4
               \end{matrix}\right]
a_{t s'}\left[\begin{matrix}R_1&\bar R_2\cr \bar R_3 & \bar R_4
               \end{matrix}\right]= \delta_{s s'}~.\label {ortho2}
\end{eqnarray}
\subsection{Fixing signs of the braiding eigenvalues}
\begin{figure}[htbp]
\centering
	\input{unknot.pspdftex}
\caption{Unknot drawn in two equivalent ways}
	\label{figs:unk}
\end{figure}
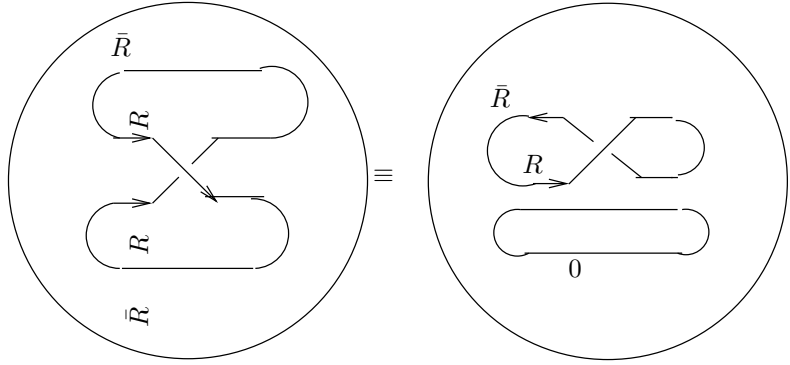
From Figure \ref{figs:unk}, we can write the invariant for the  unknot in two
equivalent ways giving the following constraint equation:
\begin{equation}
 \sum_s dim_q s \lambda_s^{(+)}(R,R)=\lambda_0^{(-)}(R,\bar R) dim_q R~.
\end{equation}
Taking $\epsilon_{0;R,\bar R}^{(-)}=1$, we can determine the 
signs $\epsilon_{s;R,R}^{(+)}$ which satisfies the above equation.
We can write a general form for the sign for a class of representations  
\setlength{\unitlength}{.7cm}
$R_n =
\begin{picture}(2,.7)
\put(0,0){\line(1,0){1.8}}
\put(0,.3){\line(1,0){1.8}}
\multiput(0,0)(.3,0){7}{\line(0,1){.3}}
\put(.7,.5){\vector(-1,0){.7}}
\put(1.1,.5){\vector(1,0){.7}}
\put(.8,.5){$n$}
\end{picture},$
the irreducible representations $\rho_{\ell} \in R_n \otimes R_n$ in 
$SU(N)_k$ Wess-Zumino conformal field theory in the large $k$ limit 
will be 
\setlength{\unitlength}{1cm}
$$\begin{picture}(10,0)
\put(0,0){\line(1,0){1.8}}
\put(0,.3){\line(1,0){1.8}}
\multiput(0,0)(.3,0){7}{\line(0,1){.3}}
\put(.7,.5){\vector(-1,0){.7}}
\put(1.1,.5){\vector(1,0){.7}}
\put(.8,.5){$n$}
\put(2,0){$\otimes$}
\put(.8,-.4){$R_n$}
\put(2.6,0){\line(1,0){1.8}}
\put(2.6,.3){\line(1,0){1.8}}
\multiput(2.6,0)(.3,0){7}{\line(0,1){.3}}
\put(3.3,.5){\vector(-1,0){.7}}
\put(3.7,.5){\vector(1,0){.7}}
\put(3.4,.5){$n$}
\put(3.4,-.4){$R_n$}
\put(4.8,0){$=$}
\put(5.4,0){$\oplus_{\ell=0}^n$}
\put(6.5,0){\line(1,0){3}}
\put(6.5,.3){\line(1,0){3}}
\put(6.5,-.3){\line(1,0){1.8}}
\multiput(6.5,0)(.3,0){11}{\line(0,1){.3}}
\multiput(6.5,0)(.3,0){7}{\line(0,-1){.3}}
\put(8,-.6){$\rho_{\ell}$}
\put(6.8,.5){\vector(-1,0){.3}}
\put(8,.5){\vector(1,0){.3}}
\put(6.9,.5){$n-\ell$}
\put(8.6,.5){\vector(-1,0){.3}}
\put(9.2,.5){\vector(1,0){.3}}
\put(8.8,.5){$2\ell$}
\end{picture}~.$$
\\
The sign $\epsilon_{\rho_{\ell};R_n,R_n}^{(+)}=(-1)^{(n-{\ell})}$. 
Similarly, for antiparallely oriented strands, 
the irreducible representations in  $\tilde {\rho}_\ell \in
R_n \otimes \bar{R}_n$ are
$$
\begin{picture}(10,1.5)
\put(0,0){\line(1,0){1.8}}
\put(0,.3){\line(1,0){1.8}}
\multiput(0,0)(.3,0){7}{\line(0,1){.3}}
\multiput(.15,.15)(.3,0){6}{$.$}
\put(.7,.5){\vector(-1,0){.7}}
\put(1.1,.5){\vector(1,0){.7}}
\put(.8,.5){$n$}
\put(2,0){$\otimes$}
\put(.8,-.4){$\bar R_n$}
\put(2.6,0){\line(1,0){1.8}}
\put(2.6,.3){\line(1,0){1.8}}
\multiput(2.6,0)(.3,0){7}{\line(0,1){.3}}
\put(3.3,.5){\vector(-1,0){.7}}
\put(3.7,.5){\vector(1,0){.7}}
\put(3.4,.5){$n$}
\put(3.4,-.4){$R_n$}
\put(4.8,0){$=$}
\put(5.4,0){$\oplus_{\ell=0}^n$}
\put(6.5,0){\line(1,0){2.4}}
\put(6.5,.3){\line(1,0){2.4}}
\multiput(6.5,0)(.3,0){9}{\line(0,1){.3}}
\multiput(6.65,.15)(.3,0){4}{$.$}
\put(7.5,-.4){${\tilde \rho}_{\ell}$}
\put(6.8,.5){\vector(-1,0){.3}}
\put(7.4,.5){\vector(1,0){.3}}
\put(7,.5){$\ell$}
\put(8,.5){\vector(-1,0){.3}}
\put(8.6,.5){\vector(1,0){.3}}
\put(8.2,.5){$\ell$}
\end{picture}
$$
\noindent Here  boxes with dot represents a column of length $N-1$.
We take the sign $\epsilon_{\tilde \rho_{\ell};R_n,\bar R_n}^{(-)}=(-1)^{\ell}$
which is +1 for the singlet ${\ell}=0$. 
We can generalise these results for tensor product of two different symmetric
representations $\rho_{\ell} \in R_n \otimes R_m$ and $\tilde {\rho}_{\ell}
\in R_n \otimes \bar R_m$  as
\begin{equation}
 \epsilon_{\rho_{\ell}}^{(+)}(R_n,R_m)=(-1)^{\frac{n+m}{2}-{\ell}}~;~
\epsilon_{\tilde{\rho}_{\ell}}^{(-)}(R_n,\bar R_m)=(-1)^{\frac{n-m}{2}-{\ell}}~,~
\end{equation}
where we assume $n\geq m$ and $\ell=(n-m)/2, (n-m)/2 +1, \ldots (n+m)/2$.
Similarly  for
antisymmetric representations $\hat R_n$ placed on antiparallely oriented strands, the
irreducible representations $\tilde {\hat \rho}_{\ell}\in \hat R_n \otimes \bar {\hat R}_n$ are
\vskip.5cm
$$\begin{picture}(10,1.5)
\put(0,0){\line(0,1){1.8}}
\put(.3,0){\line(0,1){1.8}}
\multiput(0,0)(0,.3){7}{\line(1,0){.3}}
\put(.5,.7){\vector(0,-1){.7}}
\put(.5,1.1){\vector(0,1){.7}}
\put(.5,.8){$n$}
\put(1,.8){$\otimes$}
\put(-.6,.8){${\hat R}_n$}
\put(2.3,0){\line(0,1){2.1}}
\put(2.6,0){\line(0,1){2.1}}
\multiput(2.3,0)(0,0.3){8}{\line(1,0){.3}}
\put(3.1,.7){\vector(0,-1){.7}}
\put(3.1,1.1){\vector(0,1){.7}}
\put(3.1,.8){$N-n$}
\put(1.7,.8){${\bar {\hat R}}_n$}
\put(4.4,.6){$=$}
\put(5,.6){$\oplus_{\ell=0}^n$}
\put(6.8,2){\line(0,-1){3}}
\put(7.1,2){\line(0,-1){1.8}}
\put(6.5,-1){\line(0,1){3}}
\put(6.5,2){\line(2,0){.6}}
\put(6.5,1.7){\line(2,0){.6}}
\put(6.5,1.4){\line(2,0){.6}}
\put(6.5,1.1){\line(2,0){.6}}
\put(6.5,.8){\line(2,0){.6}}
\put(6.5,.5){\line(2,0){.6}}
\put(6.5,.2){\line(2,0){.6}}
\put(6.5,-.1){\line(1,0){.3}}
\put(6.5,-.4){\line(1,0){.3}}
\put(6.5,-.7){\line(1,0){.3}}
\put(6.5,-.7){\line(1,0){.3}}
\put(6.5,-1){\line(1,0){.3}}
\put(6.7,-1.5){${\tilde {\hat \rho}}_{\ell}$}
\put(7.2,1.2){\vector(0,1){.9}}
\put(7.2,1){$\ell$}
\put(7.2,.8){\vector(0,-1){.65}}
\put(7.2,-.6){\vector(0,-1){.5}}
\put(7.2,-.2){\vector(0,1){.5}}
\put(7.2,-.4){$N-2\ell$}
\end{picture}~.$$
\vskip1.2cm
For this case, the sign $\epsilon_{\tilde {\hat \rho}_{\ell};\hat R_n,\bar {\hat R}_n}^{(-)}
=(-1)^{\ell}$.
If we replace $N-n=n$ in the above tensor product, we get irreducible representations 
$\hat {\rho}_{\ell} \in \hat R_n \otimes \hat R_n$ for parallelly
oriented strands carrying antisymmetric representation whose sign will be
 $\epsilon_{\hat {\rho}_{\ell};\hat R_n,\hat R_n}^{(+)}=(-1)^{2n-{\ell}}.$
The signs for antisymmetric representations can be similarly generalised
for $\hat {\rho}_{\ell} \in {\hat R}_n \otimes {\hat R}_m$
and $\tilde {\hat \rho}_{\ell} \in \hat R_n \otimes \bar {\hat R}_m$
as
\begin{equation}
\epsilon_{\hat {\rho}_{\ell};\hat R_n,\hat R_m}^{(+)}=(-1)^{n+m-{\ell}}~;~
\epsilon_{\tilde{\hat \rho}_{\ell};\hat R_n,\bar {\hat R}_m}^{(-)}=
(-1)^{n-m-\ell}~, 
\end{equation}
where $n \geq m \& n+m \leq N$ and $\ell=0,1,\ldots m$ for parallel strands.
Similarly for antiparallel strands with $N-m \geq n$, $\ell=n-m, n-m+1,\ldots n$.

It is possible to fix the signs for the
mixed representation but cannot be written in the most general form as done 
for the symmetric and the antisymmetric representations.
Some of the mixed representation signs are given 
in the earlier papers \cite{Ramadevi:2000gq,Paul:2010wr}.
For simplicity, we will confine to the symmetric or the antisymmetric representations
placed on the component knots with the defined signs which will be
useful for writing the Racah coefficients. In the following
subsection, we will study equivalence of states which is needed
to obtain the Racah coefficients.

\subsection{More identities of Racah coefficients from equivalence of states}
We can view the two three balls in Figure \ref {figs:equv}
as two equivalent states:
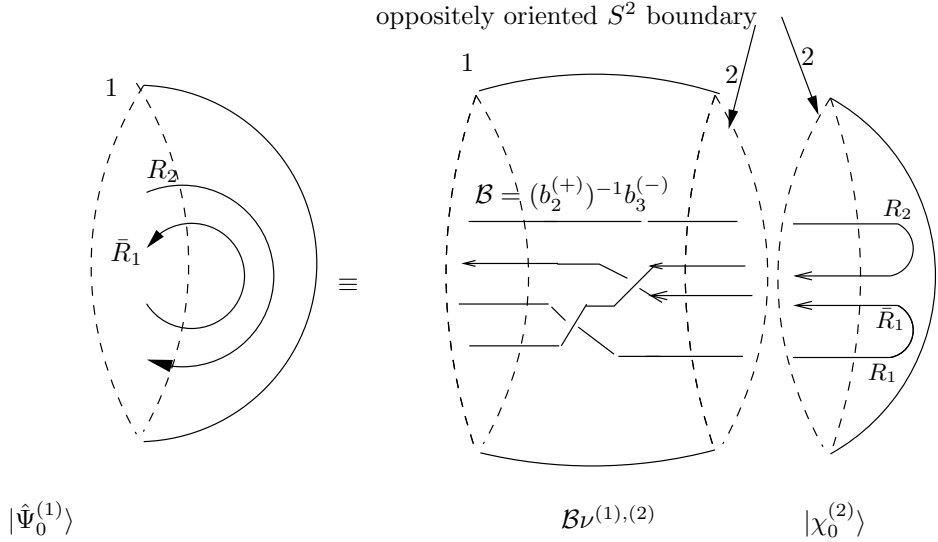
\begin{figure}[htbp]
\centering
	\input{equiva.pspdftex}
\caption{Two equivalent three-balls}
	\label{figs:equv}
\end{figure}
The three-ball corresponding to the 
state $\vert {\hat {\Psi_0}}^{(1)}\rangle$ can be glued with a similar three-ball with
oppositely oriented $S^2$ boundary to give two unknots. So, this
state can be represented as
{\small
\begin{eqnarray}
 \vert {\hat {\Psi_0}^{(1)}}\rangle&=&\epsilon(R_1,R_2)\sqrt{dim_q R_1 dim_q R_2}
\vert \hat {\phi}_0(R_2 \bar R_1 R_1 
\bar R_2) ^{(1)}\rangle \label {equa}\\
&=&(b_2^{(+)})^{-1} b_3^{(-)} \nu^{(1),(2)} \vert \chi_0 ^{(2)}\rangle\nonumber\\
&=&\sum_s
a_{0 s}\left[\begin{matrix}R_2&\bar R_2\cr \bar R_1 & R_1 \end{matrix} \right]
(\lambda_s^{(+)}(R_1,R_2))^{-1} 
a_{ts}
\left[\begin{matrix}R_2& \bar R_1\cr \bar R_2&R_1 \end{matrix} \right]\nonumber\\
~&~&\times \lambda_t^{(-)}(R_2,\bar R_1)
\vert \hat {\phi}_t(R_2 \bar R_1 R_1 
\bar R_2) ^{(1)}\rangle~. \label {eqni}
\end{eqnarray}
From eqns.(\ref {equa},\ref{eqni}) and similar relations for braid word ${\cal B}=
b_2^{(+)}(b_3^{(-)})^{-1}$, we can deduce the following identity
{\small
\begin{eqnarray}
\sum_s
a_{0 s}\left[\begin{matrix}R_2&\bar R_2\cr \bar R_1 & R_1 \end{matrix} \right]
(\lambda_s^{(+)}(R_1,R_2))^{\mp 1}
a_{ts}
\left[\begin{matrix}R_2& \bar R_1\cr \bar R_2&R_1 \end{matrix} \right]~~~~~~~~~~~~~~~\nonumber\\
~~~~~~~~~~~~~~~~~~~~~~~~~~~=\epsilon(R_1,R_2)(\lambda_t^{(-)}(R_2,\bar R_1))^{\mp 1}
a_{t0}\left[\begin{matrix}R_2&\bar R_1\cr R_1 &\bar R_2 \end{matrix} \right]~.
\label {idy1}
\end{eqnarray}
}
Using the data from $SU(2)$ \cite{Kaul:1993hb}, we can fix the signs $\epsilon(R_1,R_2)$ and
the signs in the duality matrix for the class of symmetric or antisymmetric representations.
Suppose we take symmetric representations for $R_1,R_2$ then
the sign $\epsilon(R_n,R_m)=(-1)^{min(n,m)}$ and the signs in the duality matrix is $\epsilon_{R_n}=(-1)^{n/2}$.
Similarly, for antisymmetric representations for $R_1,R_2$, 
$\epsilon(\hat R_n, \hat R_m)=(-1)^{min(2n,2m)}$ and $\epsilon_{\hat R_n}=(-1)^n$.
With this sign convention, the above identity enables fixing some of the 
coefficients of the duality matrix which we tabulate in the next section.

The well-known braiding identity relates the two three-balls with two $S^2$ boundaries
as pictorially shown in Figure \ref {figs:yb}.
\begin{figure}[htbp]
\centering
	\input{ybid.pspdftex}
\caption{Braiding Identity}
\label{figs:yb}
\end{figure}
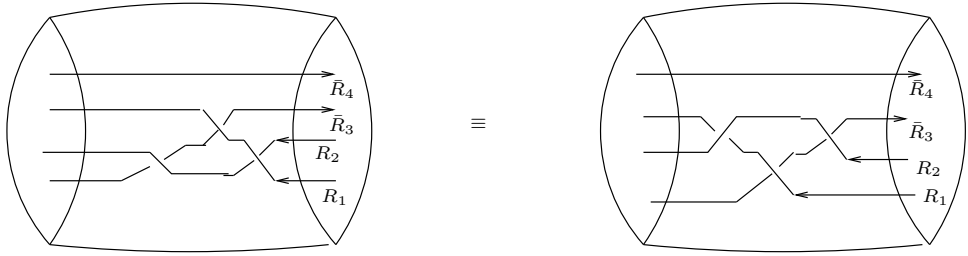
Operating these braiding operators on the two $S^2$ boundary states,
we can obtain the following identity for the duality matrix:
\begin{eqnarray}
\sum_{s,t} \lambda_s^{(+)}(R_1,R_2) a_{st}
\left[\begin{matrix}R_2&R_1\cr \bar R_3&\bar R_4 \end{matrix} \right]
\{\lambda_t^{(-)}(R_1,\bar R_3)\}^{-1} a_{\tilde s t}
\left[\begin{matrix}R_2 &\bar R_3 \cr R_1 &\bar R_4\end{matrix}\right]
\{\lambda_{\tilde s}^{(-)}(R_2,\bar R_3)\}^{-1}\nonumber\\
~~~~~=\sum_{s,t,s',t'} a_{st}\left[\begin{matrix}R_1 & R_2 \cr \bar R_3& \bar R_4\end{matrix}
\right] \{\lambda_t^{(-)}(R_2,\bar R_3)\}^{-1} a_{s' t}\left[\begin{matrix}
 R_1&\bar R_3 \cr R_2 &\bar R_4  \end{matrix} \right]
\{\lambda_{s'}^{(-)}(R_1,\bar R_3)\}^{-1}\nonumber\\
~~~~\times a_{s't'}\left[\begin{matrix}\bar R_3 & R_1\cr R_2&\bar R_4 \end{matrix}\right]
\lambda_{t'}^{(+)}(R_1,R_2) a_{\tilde s t'}\left[ \begin{matrix} \bar R_3 & R_2 \cr
                                                   R_1&\bar R_4\end{matrix}\right]~.~
~~~~~~~~~~~~~
\end{eqnarray}
\subsubsection{Six Punctured $S^2$ boundaries}
We have obtained these identities by studying equivalence of three-balls with four-punctured
$S^2$ boundaries. We shall now look at a generalisation of Figure \ref {figs:equv}
for three-balls with six-punctured $S^2$ boundaries as depicted in Figure \ref {figs:sixp}
where the braiding operator ${\cal B}=g_6=(\{b_4^{(+)}\}^{-1} b_3^{(-)})(\{b_5^{(+)}\}^{-1}b_4^{(-)}
\{b_3^{(+)}\}^{-1}b_2^{(-)})$.\\
\begin{figure}[htbp]
\centering
	\input{sixpun.pspdftex}
\caption{Six-punctured boundary}
\label{figs:sixp}
\end{figure}
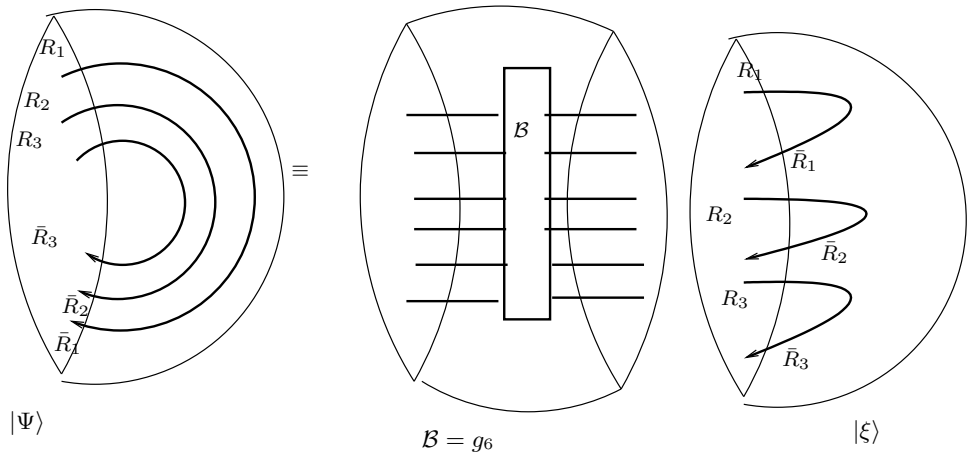
The Chern-Simons functional integral on these three-balls corresponds to a state in a space
of six-point correlator conformal blocks in $SU(N)_k$ Wess-Zumino conformal field theory.
They can be expanded in a convenient six point conformal block bases. Two such
basis states are drawn in Figure \ref {figs:sixpt}.
\begin{figure}[htbp]
\centering
	\input{sixpt.pspdftex}
\caption{Six point conformal block bases}
\label{figs:sixpt}
\end{figure}
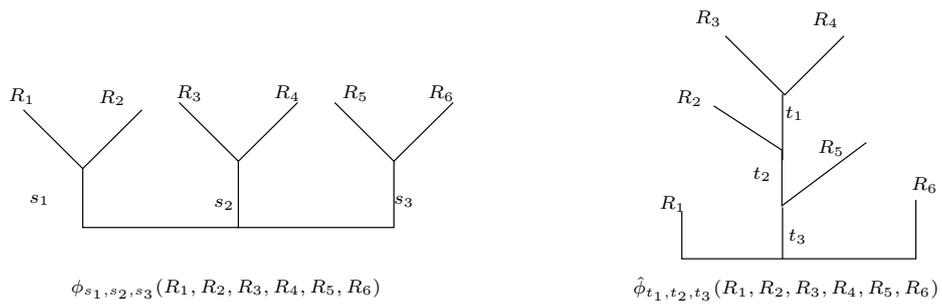
In terms of these six-point conformal block bases, we can relate the state $\vert \Psi \rangle$
and the state $\vert \xi \rangle$ as follows:
{\small
\begin{eqnarray}
\vert \Psi \rangle&=& \epsilon(R_1,R_2,R_3)\prod_{i=1}^3 \sqrt{dim_q R_i} 
\vert \hat {\phi}_{0,0,0}(R_1,R_2,R_3,\bar R_3,\bar R_2,\bar R_1)\rangle
={\cal B} \nu^{(1),(2)} \vert \xi \rangle\nonumber\\
&=&{\cal B} \prod_{i=1}^3 \sqrt{dim_q R_i} 
\vert {\phi}_{0,0,0}(R_1,\bar R_1, R_2,\bar R_2, R_3,\bar R_3) \rangle~,
\end{eqnarray}
}
where $\epsilon(R_1,R_2,R_3)=\pm 1= \epsilon(R_1,R_2)\epsilon(R_1,R_3) \epsilon(R_2,R_3)$ for symmetric and 
antisymmetric representations. 
Applying the braiding operator ${\cal B}=g_6$ on the six-point 
conformal block in the above equation, we obtain the following relation:
{\small
\begin{eqnarray}
\epsilon(R_1,R_2,R_3) \vert \hat {\phi}_{0,0,0}(R_1,R_2,R_3,\bar R_3, \bar R_2,\bar R_1)\rangle
=\sum
\lambda_{q_1}^{(-)}(\bar R_1,R_2)
\{\lambda_{s_1}^{(+)}(\bar R_1, \bar R_2)\}^{-1} ~~~~~~~~~~~~~\nonumber\\
\times \lambda_{\mu_1}^{(-)}(\bar R_1,R_3)
\{\lambda_{\nu_1}^{(+)}(\bar R_1, \bar R_3)\}^{-1}
 \lambda_{p_2}^{(-)}(\bar R_2,R_3)
\{\lambda_{\mu_2}^{(+)}(\bar R_2, \bar R_3)\}^{-1}
a_{t_1 0}\left[\begin{matrix}0&R_2\cr \bar R_2&0\end{matrix}\right]~~~~~~~~~~~~~\nonumber\\
\times
a_{0q_1}\left[\begin{matrix}R_1&\bar R_1\cr R_2&\bar t_1\end{matrix}\right]
a_{t_1 s_1}\left[\begin{matrix}p_1&\bar R_1\cr \bar R_2&0\end{matrix}\right]
~ a_{p_1 q_1}\left[\begin{matrix}R_1&R_2 \cr  \bar R_1 & \bar t_1\end{matrix}
\right] 
a_{t_2s_1}\left[\begin{matrix}p_1 & \bar R_2\cr\bar R_1&0\end{matrix}\right]
a_{0 \mu_1}\left[\begin{matrix}t_2&\bar R_1\cr R_3&\bar R_3\end{matrix}\right]~~~~~~~~~~~~~
\nonumber\\
\times
a_{t_2 p_2}\left[\begin{matrix}p_1&\bar R_2\cr R_3& {\bar \nu}_1\end{matrix}\right]
a_{\nu_1 \mu_1}\left[\begin{matrix}t_2&R_3\cr\bar R_1&\bar R_3\end{matrix}\right]
~a_{t_3 p_2}\left[\begin{matrix}p_1 & R_3 \cr \bar R_2&\bar {\nu}_1 \end{matrix}\right]
a_{\nu_1 \mu_2}\left[\begin{matrix}t_3&\bar R_2\cr\bar R_3&\bar R_1\end{matrix}\right]
a_{\nu_2 \mu_2}\left[\begin{matrix}t_3&\bar R_3\cr\bar R_2&\bar R_1\end{matrix}\right]
~~~~~~~~~~~~~\nonumber\\
\times a_{t_3 p_3}\left[\begin{matrix}p_1&R_3\cr \bar R_3&{\bar \nu}_2\end{matrix} \right]
\vert \phi_{p_1,p_3,\bar {\nu}_2}(R_1,R_2,R_3,\bar R_3,\bar R_2,\bar R_1)\rangle~.~~~~~~~~~~~~~~~~~~~~~~~~
\label {expre}
\end{eqnarray}}
We can simplify the RHS of the above expression using the property (\ref {sing}). Further,
the summation over index $q_1,\mu_1$ can be done using the identity (\ref {idy1}). 
The simplified equation suggests another identity for the Racah coefficients:
\begin{eqnarray}
 \sum_s
a_{ts}\left[\begin{matrix}R_1& R_2\cr \bar R_3 & \bar R_4 \end{matrix} \right]
\epsilon_s q^{\pm C_s \over 2}
a_{ls}
\left[\begin{matrix}R_1& \bar R_3\cr R_2&\bar R_4 \end{matrix} \right]~~~~~~~~~~~~~~~~~~~~~
~~~~~~~~~~~~~~~~~~~~~~~~~~~~~~~~~~~~~~~~~~~~~~~
~\nonumber\\
~~~~~~~~~~~~~~~~~~~~~~~~~~~=(\epsilon_t~\epsilon_l)^{-1} q^{\mp(C_t+C_l)\over 2}
\prod_{i=1}^4 (\epsilon_{R_i}q^{\pm(C_{R_i})\over 2})
a_{t l}\left[\begin{matrix}R_2&R_1\cr
\bar R_3&\bar R_4 \end{matrix} \right]~.~~~~~~~~~~~~~~~~~~\label {idy2}
\end{eqnarray}
This is a generalisation of identity (\ref {idy1}).
Using this identity, we can do the summation over index $p_2$ and $\mu_2$ and
further simplify the RHS of the expression (\ref {expre}).
The close similarity of these $SU(N)$ coefficients  to the $SU(2)$ Racah coefficient identities
suggests that all the
identities of $SU(2)$ quantum coefficients must be generalisable to the 
$SU(N)$ coefficients and hence we postulate the following identity 
{\small
\begin{eqnarray}
\sum_{l_1}a_{r_2 l_1}\left[\begin{matrix}r_1&R_3\cr
R_4&R_5\end{matrix}\right]
a_{r_1 l_2}\left[\begin{matrix}R_1&R_2\cr l_1&R_5\end{matrix}\right]
a_{\bar l_1 l_3}\left[\begin{matrix}\bar l_2&R_2\cr R_3&R_4\end{matrix}\right]=~~~~~~~~~~~~~~~~~~~
\nonumber\\
a_{r_2 l_2}\left[\begin{matrix}R_1 &l_3\cr R_4&R_5\end{matrix}\right]
a_{r_1l_3}\left[\begin{matrix}R_1&R_2\cr R_3&\bar r_2\end{matrix}\right]~,
\label {idy3}
\end{eqnarray}
}
where we have appropriately chosen conjugate representations 
which are consistent with the definition of the Racah matrix.
Again, using the above identity with $l_3=0$, we can do the
the summation over $\nu_1$ index in the simplified
RHS of eqn.(\ref {expre}). Finally, we see the eqn.(\ref {expre})
reducing to
\begin{equation}
\vert  \hat {\phi}_{0,0,0}(R_1,R_2,R_3,\bar R_3, \bar R_2,\bar R_1)\rangle=
a_{p_1 0} \left[\begin{matrix}R_1&R_2\cr \bar R_2&\bar R_1\end{matrix}\right]
\vert \phi_{p_1,0,p_1}(R_1,R_2,R_3,\bar R_3,\bar R_2,\bar R_1)\rangle~.
\end{equation}
Using the properties of the duality matrix, RHS can be seen to be LHS.
This elaborate exercise on the equivalence of two states corresponding
to six-punctured boundary confirms that the correctness of the identity
(\ref {idy3}). Armed with these identities, we try to determine
the Racah coefficients for some representations which we present
in the next section.
\section{$SU(N)$ quantum Racah coefficients}
\label{sec:SU(N)_quantum}
We shall use the properties and identities derived in the previous 
sections to obtain the duality matrix coefficients which 
will be useful for computing the non-torus knot and non-torus 
links.

For knots, all the strands carry same representation.
So for obtaining non-torus knot invariants, we have to evaluate two types of 
Racah matrices-namely.,
\begin{equation}
a_{\hat l \hat s}\left[\begin{matrix} R&\bar R\cr R&\bar R
\end{matrix}\right]~~,a_{l \hat s}\left[\begin{matrix} R& R\cr \bar R&\bar R
\end{matrix}\right]~,
\end{equation}
where first type can be shown to be a symmetric matrix 
from the properties of the Racah matrix. We could
evaluate the symmetric first type Racah coefficients for 
$R=\Yboxdim6pt\yng(1),\Yboxdim6pt\yng(2), \Yboxdim6pt\yng(1,1)$ using eqns.(\ref {ortho1},\ref{ortho2}) \cite{RamaDevi:1992np,Ramadevi:1996:PHD}.
However, for the second type Racah matrix we could only evaluate
the coefficients for the fundamental representation($R=\Yboxdim6pt\yng(1)$).

Similarly for the two-component links, we can place 
two representations $R_1,R_2$ on the component knots. In this case, 
we can have three types of Racah matrices as follows:
\begin{equation}
 a_{\hat l \hat t}\left[\begin{matrix}R_1&\bar R_1\cr   R_2&\bar R_2\end{matrix}\right]~,~
a_{l \hat s}\left[\begin{matrix}R_1& R_2\cr \bar  R_2&\bar R_1\end{matrix}\right]~,~
a_{l \hat t}\left[\begin{matrix}R_1&R_2\cr   \bar R_1&\bar R_2\end{matrix}\right]~.
\end{equation}
Now, we will present the Racah coefficients for some representations which will be
useful to compute the non-torus knot and link polynomials in $U(N)$ Chern-Simons
theory.
\subsection{Coefficients when $R_1$ is fundamental}
\begin{enumerate}
 \item
For the simplest fundamental representation $R=\Yboxdim6pt\yng(1)$, the two types
of Racah coefficient matrices are \cite{RamaDevi:1992np,Ramadevi:1996:PHD}:
\begin{displaymath}
$${$a_{ts}\left[
{
\begin{tabular}{cc} 
$R=\Yboxdim6pt\yng(1)$
& $\bar{R}=\Yboxdim6pt\young({\mdot})$
\\ 
 $R=\Yboxdim6pt\yng(1)$
& 
 $\bar{R}=\Yboxdim6pt\young({\mdot})$ 
\end{tabular}}
\right]=\frac{1}{dim_qR}$
$\left({ 
\begin{tabular}{c|c c}  
~
& $s=0$
& $\bar{R}=\Yboxdim6pt\young({\mdot}~)$
\\ 
\hline 
{$t=0$}
& $-1$
&{\scriptsize$\sqrt{[N-1][N+1]}$ }\\ 
  $\Yboxdim6pt\young({\mdot}~)$
& {\scriptsize$\sqrt{[N-1][N+1]}$} 
&$1$
\\ 
  \end{tabular} }\right)$}~,
$$
$${$a_{ts}\left[
{
\begin{tabular}{cc} 
$R=\Yboxdim6pt\yng(1)$
&  $R=\Yboxdim6pt\yng(1)$
\\ 
 $\bar{R}=\Yboxdim6pt\young({\mdot})$
&  $\bar{R}=\Yboxdim6pt\young({\mdot})$
\end{tabular}}
\right]=\frac{1}{dim_qR}$
$\left({\renewcommand{\arraystretch}{1.5}
\begin{tabular}{c|c c}  
{} 
& $s=0$
& $\Yboxdim6pt\young({\mdot}~)$ \\ 
\hline 
$t=\Yboxdim6pt\yng(1,1)$
& $-\sqrt{\frac{[N][N-1]}{[2]}}$
&$\sqrt{\frac{[N][N+1]}{[2]}}$ \\ 
  $\Yboxdim6pt\yng(2)$
& $\sqrt{\frac{[N][N+1]}{[2]}}$ 
&$\sqrt{\frac{[N][N-1]}{[2]}}$
\\ 
  \end{tabular} }\right)$}~,
$$
\end{displaymath}
where $dim_q(R=\Yboxdim6pt\yng(1))=[N]$.
\item Next, we look at Racah coefficient matrices where
 $R_1=\Yboxdim6pt\yng(1) \neq R_2$. This will be useful
for the computation of two-component links.
\begin{displaymath}
$${$a_{ts}\left[
{
\begin{tabular}{cc} 
$R_1=\Yboxdim6pt\yng(1)$
& $\bar{R}=\Yboxdim6pt\young({\mdot})$
\\ 
$R_2=\Yboxdim6pt\yng(2)$
&  $\bar{R}_2=\Yboxdim6pt\young({\mdot}{\mdot})$
\end{tabular}}
\right]=\frac{1}{K}$
$\left({ \renewcommand{\arraystretch}{1.5}
\begin{tabular}{c|c c}  
{} 
& $s=\Yboxdim6pt\yng(1)$
&$\Yboxdim6pt\young({\mdot}~~)$ \\ 
\hline 
$t=0$
&{\scriptsize$-\sqrt{[N]}$}
&{\scriptsize$\sqrt{\frac{[N-1][N][N+2]}{[2]}}$} \\ 
$\Yboxdim6pt\young({\mdot}~)$
& {\scriptsize$\sqrt{\frac{[N-1][N][N+2]}{[2]}}$} 
&{\scriptsize$\sqrt{[N]}$}
\\ 
  \end{tabular} }\right)$}~,
$$
\end{displaymath}
where $K=\sqrt{dim_{q}R_{1}dim_{q}R_{2}}$. Similarly, the second and third type
Racah matrix coefficients for $R_1=\Yboxdim6pt\yng(1), R_2=\Yboxdim6pt\yng(2)$ are
\begin{displaymath}
$${$a_{ts}\left[
{
\begin{tabular}{cc} 
$R_1=\Yboxdim6pt\yng(1)$
&$R_2=\Yboxdim6pt\yng(2)$
\\ 
$\bar{R}_1=\Yboxdim6pt\young(\mdot)$
& $\bar{R}_2=\Yboxdim6pt\young({\mdot}{\mdot})$
\end{tabular}}
\right]=$
$\left({ \renewcommand{\arraystretch}{1.7} 
\begin{tabular}{c|c c}  
{} 
& $s=\Yboxdim6pt\yng(1)$
&$\Yboxdim6pt\young({\mdot}~~)$ \\ 
\hline 
$t=\Yboxdim6pt\yng(2,1)$
& $-\sqrt{\frac{[N-1]}{[3][N+1]}}$
&$\sqrt{\frac{[2][N+2]}{[3][N+1]}}$ \\ 
 $\Yboxdim6pt\yng(3)$
& $\sqrt{\frac{[2][N+2]}{[3][N+1]}}$ 
&$\sqrt{\frac{[N-1]}{[3][N+1]}}$
~\\ 
  \end{tabular}}\right)$~,
}$$
\end{displaymath}
\begin{displaymath}
$${$a_{ts}\left[
\begin{array}{cc} 
R_{1}=\Yboxdim6pt\yng(1)
& R_{2}=\Yboxdim6pt\yng(2)
\\ 
\bar{R}_{2}=\Yboxdim6pt\young({\mdot}{\mdot})
& \bar{R}_{1}=\Yboxdim6pt\young(\mdot)
\end{array}\right]=\frac{1}{K}$
$\left({ \renewcommand{\arraystretch}{1.5} 
\begin{tabular}{c|c c}  
{} 
& $s=0$
&\Yboxdim6pt\young({\mdot}~) \\ 
\hline 
$t=\Yboxdim6pt\yng(2,1)$
& {\scriptsize$-\sqrt{\frac{[N-1][N][N+1]}{[3]}}$}
&{\scriptsize$\sqrt{\frac{[N][N+1][N+2]}{[2][3]}}$} \\ 
$\Yboxdim6pt\yng(3)$
& {\scriptsize$\sqrt{\frac{[N][N+1][N+2]}{[2][3]}}$} 
&{\scriptsize$\sqrt{\frac{[N-1][N][N+1]}{[3]}}$}
\\ 
  \end{tabular} }\right)$}~.
$$
\end{displaymath}
\item Interestingly, we could find the coefficients for $R_1=\Yboxdim6pt\yng(1), R_2=
\setlength{\unitlength}{1pt}
\begin{picture}(30,0)
\put(0,0){\Yboxdim6pt\yng(5)}
\put(12,7){$n$}
\put(11,9){\vector(-1,0){11}}
\put(18,9){\vector(1,0){10}}
\end{picture}$ 
using the identities. The corresponding conjugate representations are
 $\bar{R}_1=\Yboxdim6pt\young(\mdot), \bar{R}_2=
\setlength{\unitlength}{1pt}
\begin{picture}(30,12)
\put(0,0){\Yboxdim6pt\young({\mdot}{\mdot}{\mdot}{\mdot}{\mdot})}
\put(12,7){$n$}
\put(11,9){\vector(-1,0){11}}
\put(18,9){\vector(1,0){10}}
\end{picture}.$ 
The three types of Racah coefficients for this class of representations are
\begin{displaymath}
$${$a_{ts}\left[
\begin{array}{cc} 
R_{1}
& \bar{R}_{1}
\\ 
R_{2}
& \bar{R}_{2}
\end{array}
\right]=\frac{1}{K}$
$\left({\renewcommand{\arraystretch}{1.7}
\begin{tabular}{l|cc}  
 
& $s=\setlength{\unitlength}{1pt}
\begin{picture}(30,0)
\put(0,0){\Yboxdim6pt\yng(4)}
\put(7,7){\scriptsize{n-1}}
\put(6,9){\vector(-1,0){6}}
\put(18,9){\vector(1,0){5}}
\end{picture}$
& $\setlength{\unitlength}{1pt}
\begin{picture}(30,0)
\put(0,0){\Yboxdim6pt\young({\mdot}~~~~~)}
\put(10,7){\scriptsize{n+1}}
\put(9,9){\vector(-1,0){8}}
\put(26,9){\vector(1,0){9}}
\end{picture}$\\ 
\hline 
$t=0$
& $-\sqrt{\frac{[N][N+1]...[N+n-2]}{[n-1]!}}$
&$\sqrt{\frac{[N-1][N]...[N+n]}{[N+n-1][n]!}}$ \\ 
$\Yboxdim6pt\young({\mdot}~)$
& $\sqrt{\frac{[N-1][N]...[N+n]}{[N+n-1][n]!}}$ 
&$\sqrt{\frac{[N][N+1]...[N+n-2]}{[n-1]!}}$
\\ 
  \end{tabular} }\right)$}~,
$$
$${$a_{ts}\left[
\begin{array}{cc} 
R_{1}
& R_{2}
\\ 
\bar{R}_{1}
& \bar{R}_{2}
\end{array}
\right]=$
$\left({ \renewcommand{\arraystretch}{1.5} 
\begin{tabular}{c|c c}  
{} 
& $s=
\setlength{\unitlength}{1pt}
\begin{picture}(30,14)
\put(0,0){\Yboxdim6pt\yng(5,1)}
\put(12,13){\scriptsize{n}}
\put(11,15){\vector(-1,0){11}}
\put(18,15){\vector(1,0){10}}
\end{picture}$
&$\setlength{\unitlength}{1pt}
\begin{picture}(30,12)
\put(0,0){\Yboxdim6pt\young(~~~~~~)}
\put(10,7){\scriptsize{n+1}}
\put(9,9){\vector(-1,0){8}}
\put(26,9){\vector(1,0){9}}
\end{picture}$ \\ 
\hline 
$t=\setlength{\unitlength}{1pt}
\begin{picture}(30,0)
\put(0,0){\Yboxdim6pt\yng(4)}
\put(7,7){\scriptsize{n-1}}
\put(6,9){\vector(-1,0){6}}
\put(18,9){\vector(1,0){5}}
\end{picture}$
& $-\sqrt{\frac{[N-1]}{[n+1][N+n-1]}}$
&$\sqrt{\frac{[n][N+n]}{[n+1][N+n-1]}}$ \\ 
 $\setlength{\unitlength}{1pt}
\begin{picture}(30,0)
\put(0,0){\Yboxdim6pt\young({\mdot}~~~~~)}
\put(10,7){\scriptsize{n+1}}
\put(9,9){\vector(-1,0){8}}
\put(26,9){\vector(1,0){9}}
\end{picture}$
& $\sqrt{\frac{[n][N+n]}{[n+1][N+n-1]}}$ 
&$\sqrt{\frac{[N-1]}{[4][N+2]}}$
\\ 
  \end{tabular}}\right)$}~,
$$
$${$a_{ts}\left[
\begin{array}{cc} 
R_{1}
& R_{2}
\\ 
\bar{R}_{2}
& \bar{R}_{1}
\end{array}\right]=\frac{1}{K}$
$\left({\renewcommand{\arraystretch}{1.5} 
\begin{tabular}{c|c c}  
{} 
& $s=0$
&$\Yboxdim6pt\young({\mdot}~)$ \\ 
\hline 
$t=\setlength{\unitlength}{1pt}
\begin{picture}(30,20)
\put(0,0){\Yboxdim6pt\yng(5,1)}
\put(12,13){\scriptsize{n}}
\put(11,15){\vector(-1,0){11}}
\put(18,15){\vector(1,0){10}}
\end{picture}$
& {\scriptsize$-\sqrt{\frac{[N-1][N]...[N+n-1]}{[n+1][n-1]!}}$}
&{\scriptsize$\sqrt{\frac{[N][N+1]...[N+n]}{[n+1]!}}$} \\ 
$\setlength{\unitlength}{1pt}
\begin{picture}(30,12)
\put(0,0){\Yboxdim6pt\young(~~~~~~)}
\put(10,7){\scriptsize{n+1}}
\put(9,9){\vector(-1,0){8}}
\put(26,9){\vector(1,0){9}}
\end{picture}$
&{\scriptsize$\sqrt{\frac{[N][N+1]...[N+n]}{[n+1]!}}$}
&{\scriptsize$\sqrt{\frac{[N-1][N]...[N+n-1]}{[n+1][n-1]!}}$}
\\ 

  \end{tabular} }\right)~,$}$$
\end{displaymath}
where $K=\sqrt{dim_q R_1 dim_q R_2}$.
\item
Similar exercise could be done for 
$\begin{array}{cccc}
R_{1}= \Yboxdim6pt\yng(1),
&R_{2}=\Yboxdim6pt\yng(1,1),
\end{array}
$
and their conjugate representations are
$
\begin{array}{cccc}
\bar{R}_{1}=\Yboxdim6pt\young(\mdot),
&\bar{R}_{2}=~\setlength{\unitlength}{1pt}
\begin{picture}(22,0)
\put(-3,-10){\Yboxdim6pt\yng(1,1,1,1)}
\put(5,-1){\scriptsize$N-2$}
\put(7,5){\vector(0,1){8}}
\put(7,-3){\vector(0,-1){8}}
\end{picture}
\end{array}
$~. These representations give
\begin{displaymath}
$${$a_{ts}\left[
\begin{array}{cc} 
R_{1}
& \bar{R}_{1}
\\ 
R_{2}
& \bar{R}_{2}
\end{array}
\right]=\frac{1}{\sqrt{dim_{q}R_{1}\, dim_{q}R_{2}}}$
$\left({ \renewcommand{\arraystretch}{1.4}
\begin{tabular}{c|c c}  
{} 
& $s=\Yboxdim6pt\yng(1)$
&$\Yboxdim6pt\young({\mdot}~,:~)$ \\ 
\hline 
{$t=0$}
& {\scriptsize$\sqrt{[N]}$}
&{\scriptsize$\sqrt{\frac{[N-2][N][N+1]}{[2]}}$} \\ 
 $\Yboxdim6pt\young({\mdot}~)$
& {\scriptsize$\sqrt{\frac{[N-2][N][N+1]}{[2]}}$} 
&{\scriptsize$-\sqrt{[N]}$}
\\ 
  \end{tabular} }\right)$}~,
$$
$${$a_{ts}\left[
\begin{array}{cc} 
R_{1}
& R_{2}
\\ 
\bar{R}_{1}
&\bar{R}_{2}
\end{array}
\right]=$
$\left({ \renewcommand{\arraystretch}{2}
\begin{tabular}{c |c c}  
& $s=\Yboxdim6pt\yng(1)$
&$\Yboxdim6pt\young({\mdot}~,:~)$ \\ 
\hline 
$t=\Yboxdim6pt\yng(2,1)$
& $\sqrt{\frac{[N+1]}{[3][N-1]}}$
&$-\sqrt{\frac{[2][N-2]}{[3][N-1]}}$ \\ 
 \Yboxdim6pt\yng(1,1,1)
& $-\sqrt{\frac{[2][N-2]}{[3][N-1]}}$ 
&$-\sqrt{\frac{[N+1]}{[3][N-1]}}$
\\ 
  \end{tabular}}\right)$}~,
$$
$${$a_{ts}\left[
\begin{array}{cc} 
R_{1}
& R_{2}
\\ 
\bar{R}_{2}
& \bar{R}_{1}
\end{array}
\right]=\frac{1}{\sqrt{dim_{q}R_{1}\, dim_{q}R_{2}}}$
$\left({\renewcommand{\arraystretch}{1.8}
\begin{tabular}{c|c c}  
{} 
& {\scriptsize$s=0$}
&\Yboxdim6pt\young({\mdot}~) \\ 
\hline 
$t=\Yboxdim6pt\yng(2,1)$
& {\scriptsize$\sqrt{\frac{[N-1][N][N+1]}{[3]}}$}
&{\scriptsize$-\sqrt{\frac{[N-2][N][N-1]}{[2][3]}}$} \\ 
 \Yboxdim6pt\yng(1,1,1) 
& {\scriptsize$-\sqrt{\frac{[N-2][N][N-1]}{[2][3]}}$} 
&{\scriptsize$-\sqrt{\frac{[N-1][N][N+1]}{[3]}}$}
\\ 
  \end{tabular} }\right)$}~.
$$
\end{displaymath}
\item
These results can be generalised for $R_1=\Yboxdim6pt\yng(1)$ and
$R_{2}=\setlength{\unitlength}{1pt}
\begin{picture}(24,0)
\put(-3,-10){\Yboxdim6pt\yng(1,1,1,1)}
\put(5,-1){\scriptsize$n$}
\put(7,5){\vector(0,1){8}}
\put(7,-3){\vector(0,-1){8}}
\end{picture}$, whose conjugate representations are
$\bar{R}_{1}=\Yboxdim6pt\young(\mdot)$, 
$\bar{R}_{2}=~\setlength{\unitlength}{1pt}
\begin{picture}(30,16)
\put(-3,-10){\Yboxdim6pt\yng(1,1,1,1)}
\put(5,-1){\scriptsize$N-n$}
\put(7,5){\vector(0,1){8}}
\put(7,-3){\vector(0,-1){8}}
\end{picture}$as follows:  
\begin{displaymath}
$${$a_{ts}\left[
\begin{array}{cc} 
R_{1} 
& \bar{R}_{1}
\\ R_{2}
& \bar{R}_{2}
\end{array}
\right]=\frac{1}{K}$
$\left({ \renewcommand{\arraystretch}{2} 
\begin{tabular}{c|c c}  
{} 
& ${\setlength{\unitlength}{1pt}
\begin{picture}(22,20)
\put(-14,6){s=}
\put(0,0){\Yboxdim6pt\yng(1,1,1)}
\put(10,11){\vector(0,1){6}}
\put(8,6){\scriptsize$n-1$}
\put(10,6){\vector(0,-1){7}}
\end{picture}}$
& {\setlength{\unitlength}{1pt}
$$\begin{picture}(18,24)
\put(-12,0){\Yboxdim6pt\young({\mdot}~,:~,:~,:~)}
\put(2,9){\scriptsize$n$}
\put(4,15){\vector(0,1){8}}
\put(4,7){\vector(0,-1){8}}
\end{picture}$$}\\ 
\hline 
\scriptsize$t=0$
& $\sqrt{\frac{[N][N-1]...[N-n+2]}{[n-1]!}}$
&$\sqrt{\frac{[N+1][N]...[N-n]}{[N-n+1][n]!}}$ \\ 
\Yboxdim6pt\young({\mdot}~)
& $\sqrt{\frac{[N+1][N]...[N-n]}{[N-n+1][n]!}}$
&$-\sqrt{\frac{[N][N-1]...[N-n+2]}{[n-1]!}}$
\\ 
  \end{tabular} }\right)$}~,
$$
$${$a_{ts}\left[
\begin{array}{cc} 
R_{1}
& R_{2}
\\ 
\bar{R}_{1} 
& \bar{R}_{2}
\end{array}\right]=$
$\left({ \renewcommand{\arraystretch}{1.5}
\begin{tabular}{c |c c}  
{} 
& {\setlength{\unitlength}{1pt}
$$\begin{picture}(22,20)
\put(-14,6){s=}
\put(0,0){\Yboxdim6pt\yng(1,1,1)}
\put(10,11){\vector(0,1){6}}
\put(8,6){\scriptsize$n-1$}
\put(10,6){\vector(0,-1){7}}
\end{picture}$$}
& {\setlength{\unitlength}{1pt}
$$\begin{picture}(18,24)
\put(-12,0){\Yboxdim6pt\young({\mdot}~,:~,:~,:~)}
\put(2,9){\scriptsize$n$}
\put(4,15){\vector(0,1){8}}
\put(4,7){\vector(0,-1){8}}
\end{picture}$$}

\\
\hline 
{{\scriptsize$t=$}\setlength{\unitlength}{1pt}
$$\begin{picture}(22,20)
\put(0,-12){\Yboxdim6pt\yng(2,1,1,1)}
\put(10,-1){\vector(0,1){6}}
\put(8,-5){\scriptsize$n-1$}
\put(10,-5){\vector(0,-1){7}}
\end{picture}$$}
& $\sqrt{\frac{[N+1]}{[N+n][N-n+1]}}$
&$-\sqrt{\frac{[n][N-n]}{[n+1][N-n+1]}}$ \\ 
 {\setlength{\unitlength}{1pt}
$$\begin{picture}(18,24)
\put(4,-10){\Yboxdim6pt\yng(1,1,1,1)}
\put(12,-1){\scriptsize$n+1$}
\put(12,5){\vector(0,1){8}}
\put(12,-3){\vector(0,-1){8}}
\end{picture}$$} 
& $-\sqrt{\frac{[n][N-n]}{[n+1][N-n+1]}}$ 
&$-\sqrt{\frac{[n][N-n]}{[n+1][N-n+1]}}$
\\ 
  \end{tabular}}\right)$}~,
$$
$${$a_{ts}\left[
\begin{array}{cc} 
R_{1}
& R_{2}
\\ 
\bar{R}_{2}
& \bar{R}_{1}
\end{array}\right]=\frac{(-1)^n}{K}$
$\left({\renewcommand{\arraystretch}{1.5} 
\begin{tabular}{c|c c}  
{} 
& $s=0$
&$\Yboxdim6pt\young({\mdot}~)$ \\ 
\hline
{{\scriptsize$t=$}\setlength{\unitlength}{1pt}
$$\begin{picture}(22,20)
\put(0,-12){\Yboxdim6pt\yng(2,1,1,1)}
\put(10,-1){\vector(0,1){6}}
\put(8,-5){\scriptsize$n-1$}
\put(10,-5){\vector(0,-1){7}}
\end{picture}$$}
& {\scriptsize$\sqrt{\frac{[N+1][N]...[N-n+1]}{[n+1][n-1]!}}$}
&{\scriptsize$-\sqrt{\frac{[N][N-1]...[N-n]}{[n]!}}$} \\ 
~~~~{\setlength{\unitlength}{1pt}
$$\begin{picture}(18,24)
\put(-3,-10){\Yboxdim6pt\yng(1,1,1,1)}
\put(5,-1){\scriptsize$n+1$}
\put(7,5){\vector(0,1){8}}
\put(7,-3){\vector(0,-1){8}}
\end{picture}$$} 
& {\scriptsize$-\sqrt{\frac{[N][N-1]...[N-n]}{[n]!}}$} 
&{\scriptsize$-\sqrt{\frac{[N+1][N]...[N-n+1]}{[n+1][n-1]!}}$}
\\ 
  \end{tabular} }\right)$}~,
$$
\end{displaymath}
where $K=\sqrt{dim_q R_1 dim_q R_2}~.$
\end{enumerate}
\subsection{Coefficients for $R_1$ symmetric or antisymmetric representations}
\begin{enumerate}
\item For
$\begin{array}{cc}
R=\Yboxdim6pt\yng(2),
&\bar{R}=\Yboxdim6pt\young({.}{.})
\end{array}
$
\begin{displaymath}
$${$a_{ts}\left[
\renewcommand{\arraystretch}{1.5}
\begin{array}{cc} 
R& \bar{R}
\\ 
R
& \bar{R}
\end{array}\right]=\frac{1}{K}$
$\left({ 
\begin{tabular}{c|c c c}  
{} 
& $s=\overset{\tilde{\rho}_0}{0}$
& $\overset{\tilde{\rho}_1}{\Yboxdim6pt\young({\mdot}~)}$ 
&$\overset{\tilde{\rho}_2}{\Yboxdim6pt\young({\mdot}{\mdot}~~)}$ \\ 
\hline 
{$t=\tilde {\rho}_{0}$}
& {\scriptsize$\sqrt{dim_q\tilde {\rho}_{0}}$}
&{\scriptsize$-\sqrt{dim_q\tilde {\rho}_{1}}$} 
&{\scriptsize$\sqrt{dim_q\tilde {\rho}_{2}}$}
\\ 
 {$\tilde {\rho}_{1}$} 
& {\scriptsize$-\sqrt{dim_q\tilde {\rho}_{1}}$} 
&$\frac{dim_q\tilde {\rho}_{2}}{dim_qR-1}-1$
&{$\frac{\sqrt{{dim_q\tilde {\rho}_{1}}dim_q\tilde{\rho}_{2}}}{dim_qR-1}$}
\\ 
{$\tilde {\rho}_{2}$}
&{\scriptsize$\sqrt{dim_q\tilde {\rho}_{2}}$}
&{$\frac{\sqrt{{dim_q\tilde {\rho}_{1}}dim_q\tilde{\rho}_{2}}}{dim_qR-1}$}
&{$\frac{dim_q\tilde {\rho}_{1}}{dim_qR-1}-1$}\\
  \end{tabular} }\right)$}~,
$$
\end{displaymath}
where
\begin{eqnarray}
K=dim_{q}R=\frac{
[N][N+1]}{[2]}
,~dim_{q}\tilde{\rho}_{0}=1,~dim_{q}\tilde{\rho}_{1}=[N+1][N-1],&~&
~\nonumber\\
 dim_{q}\tilde{\rho}_{2}=\frac{[N-1][N]^{2}[N+3]}{[2]^{2}}~.~~~~~~~~~~~~~~~~~
~~~~~~~~~~~~~~~~~~~~~~~~~~~~
\end{eqnarray}
The second type Racah
matrix coefficients are
{\small
\begin{displaymath}
$${$a_{ts}\left[
\begin{array}{cc} 
R 
& R
\\ 
\bar{R}
& \bar{R}
\end{array}\right]=\frac{1}{K}$
$\left({ \renewcommand{\arraystretch}{1.5}
\begin{tabular}{l
|c c c}  
t
& $s=\overset{\tilde{\rho}_0}{0}$
& $\overset{\tilde{\rho}_1}{\Yboxdim6pt\young({\mdot}~)}$ 
&$\overset{\tilde{\rho}_2}{\Yboxdim6pt\young({\mdot}{\mdot}~~)}$ \\ 
\hline 
$\yng(2,2)~(\rho_1)$
& {\scriptsize$\sqrt{dim_q {\rho}_{1}}$}
&{\scriptsize$-\frac{[N][N+1]}{[2]}\sqrt{\frac{1}{[3]}}$} 
&{\scriptsize$\frac{[N]}{[2]}\sqrt{\frac{[N+3][N+1]}{[3]}}$}
\\ 
 $\yng(3,1)~(\rho_2)$
&{\scriptsize$-\sqrt{dim_q\rho_2}$}
&{\scriptsize$x\sqrt{\frac{[N][N+2]}{[4][2]}}$}
&{\scriptsize$y\sqrt{\frac{[N][N+1][N+2][N+3]}{[4][2]}}$}
\\ 
$\yng(4)~(\rho_3)$
&{\scriptsize$\sqrt{dim_q {\rho}_{3}}$}
&{\scriptsize$u\sqrt{\frac{[N-1][N][N+2][N+3]}{[4][3][2]}}$}
&{\scriptsize$v\sqrt{\frac{[N-1][N][N+1][N+2]}{[4][3][2]}}$}\\
  \end{tabular} }\right)$}~,
$$
\end{displaymath}
}
where the quantum dimensions of the representations are
\begin{eqnarray}
dim_{q}\rho_{1}=\frac{[N-1][N]^{2}[N+1]}{[2]^{2}[3]}, ~dim_{q}\rho_{2}=
\frac{[N-1][N][N+1][N+2]}{[4][2]}~,\nonumber\\
dim_{q}\rho_{3}=\frac{[N][N+1][N+2][N+3]}{[4][3][2]}~.~~~~~~~~~~~~~~~~~~~~~~~~~~~~~~~~~~~~~~~~~
\end{eqnarray}
The variables $x,y,u,v$ are 
{\small
\begin{equation}
x=\frac{[N+3][N]}{[N+2]}-[N-1],~y=\frac{[N]}{[N+2]},~u=\frac{[2][N+1]}{[N+2]},~
v=\frac{[N+1][2]}{[N+2]}-1~.\nonumber
\end{equation}
}
\item For
$R=\Yboxdim6pt\yng(1,1),
~\bar{R}={\setlength{\unitlength}{1pt}
\begin{picture}(20,20)
\put(0,0){\Yboxdim6pt\yng(1,1,1,1)}
\put(8,9){\tiny{$N-2$}}
\put(12,15){\vector(0,1){8}}
\put(12,7){\vector(0,-1){8}}
\end{picture}}$
\begin{displaymath}$$
$a_{ts}\left[{
\begin{tabular}{cc}  
$R$
&$\bar{R}$
\\
$R$
&$\bar{R}$
 \end{tabular} }\right]=\frac{1}{K}$
$\left({\renewcommand{\arraystretch}{1.5}
\begin{tabular}{c|c c c}  
{} 
& $s=\overset{\tilde{\rho}_0}{0}$
& $\overset{\tilde{\rho}_1}{\Yboxdim6pt\young({.}~)}$
&${\setlength{\unitlength}{1pt}
\begin{picture}(20,24)
\put(15,0){\Yboxdim6pt\yng(2,2,1,1)}
\put(-8,9){\tiny{$N-2$}}
\put(9,14){\vector(0,1){8}}
\put(9,8){\vector(0,-1){8}}
\put(15,26){$\tilde{\rho}_2$}
\end{picture}}$
\\ 
\hline 
{$t=\tilde {\rho}_{0}$}
& {\scriptsize$\sqrt{dim_q\tilde {\rho}_{0}}$}
&{\scriptsize$-\sqrt{dim_q\tilde {\rho}_{1}}$} 
&{\scriptsize$\sqrt{dim_q\tilde {\rho}_{2}}$}
\\ 
 {$\tilde {\rho}_{1}$} 
& {\scriptsize$-\sqrt{dim_q\tilde {\rho}_{1}}$} 
&$\frac{dim_q\tilde {\rho}_{2}}{dim_qR-1}-1$
&{$\frac{\sqrt{{dim_q\tilde {\rho}_{1}}dim_q\tilde{\rho}_{2}}}{dim_qR-1}$}
\\ 
{$\tilde {\rho}_{2}$}
&{\scriptsize$\sqrt{dim_q\tilde {\rho}_{2}}$}
&{$\frac{\sqrt{{dim_q\tilde {\rho}_{1}}dim_q\tilde{\rho}_{2}}}{dim_qR-1}$}
&{$\frac{dim_q\tilde {\rho}_{1}}{dim_qR-1}-1$}\\
  \end{tabular} }\right)$~,$$
\end{displaymath}
where
\begin{eqnarray}
K=dim_{q}R=\frac{[N][N-1]}{[2]},~
dim_{q}\tilde{\rho}_{0}=1, ~dim_{q}\tilde{\rho}_{1}=[N+1][N-1], ~\nonumber\\ 
dim_{q}\tilde{\rho}_{2}=\frac{[N+1][N]^{2}[N-3]}{[2]^{2}}~.~~~~~~~~~~~~~~~~~~~~~
~~~~~~~~~~~~~~~~~~~~~~~~~~
\end{eqnarray}
The second type Racah matrix coefficients are
\begin{displaymath}$$
$a_{ts}\left[{ 
\begin{tabular}{cc}  
{$R$} 
&{$R$}
\\
{$\bar{R}$}
&{$\bar{R}$}
\\
  \end{tabular} }\right]=\frac{1}{K}$
$\left({ \renewcommand{\arraystretch}{1.5}
\begin{tabular}{l|c c c}  
t 
& $s=\overset{\tilde{\rho}_0}{0}$
& $\overset{\tilde{\rho}_1}{\Yboxdim6pt\young({.}~)}$
&${\setlength{\unitlength}{1pt}
\begin{picture}(20,24)
\put(15,0){\Yboxdim6pt\yng(2,2,1,1)}
\put(-8,9){\tiny{$N-2$}}
\put(9,14){\vector(0,1){8}}
\put(9,8){\vector(0,-1){8}}
\put(15,26){$\tilde{\rho}_2$}
\end{picture}}$
\\ 
\hline 
$\Yboxdim6pt\yng(2,2)~(\rho_1)$
& {\scriptsize$\sqrt{dim_q {\rho}_{1}}$}
&{\scriptsize$\frac{[N][N-1]}{[2]}\sqrt{\frac{1}{[3]}}$} 
&{\scriptsize$\frac{[N]}{[2]}\sqrt{\frac{[N-3][N-1]}{[3]}}$}
\\ 
$\Yboxdim6pt\yng(2,1,1)~(\rho_2)$
&{\scriptsize$-\sqrt{dim_q\rho_2}$}
&{\scriptsize$x\sqrt{\frac{[N][N-2]}{[4][2]}}$}
&{\scriptsize$y\sqrt{\frac{[N][N-1][N-2][N-3]}{[4][2]}}$}
\\ 
$\Yboxdim6pt\yng(1,1,1,1)~(\rho_3)$
&{\scriptsize$\sqrt{dim_q {\rho}_{3}}$}
&{\scriptsize$-u\sqrt{\frac{[N+1][N][N-2][N-3]}{[4][3][2]}}$}
&{\scriptsize$v\sqrt{\frac{[N+1][N][N-1][N-2]}{[4][3][2]}}$}\\
  \end{tabular} }\right)$~,
$$\end{displaymath}
where 
\begin{eqnarray*}
dim_{q}\rho_{1}&=&\frac{[N-1][N]^{2}[N+1]}{[2]^{2}[3]},\\ 
dim_{q}\rho_{2}&=&\frac{[N+1][N][N-1][N-2]}{[4][2]}, \\
dim_{q}\rho_{3}&=&\frac{[N][N-1][N-2][N-3]}{[4][3][2]}~,
\end{eqnarray*}
and the variables are
\begin{eqnarray}
x=[N+1]-\frac{[N-3][N]}{[N-2]}, &~&y=
\frac{[N]}{[N-2]}, \nonumber\\
u=\frac{[2][N-1]}{[N-2]}, &~& v=\frac{[N-1][2]}{[N-2]}-1~.
\end{eqnarray}
\item For $R_1=\Yboxdim6pt\yng(2)\neq R_2=\Yboxdim6pt\yng(3)$,  which is will be useful for the computation of links,
the second type Racah coefficients are
{\small
\begin{displaymath}
$${$a_{ts}\left[
\renewcommand{\arraystretch}{1.1}
\begin{array}{cc} 
R_1 
& R_2
\\ 
\bar{R}_1
& \bar{R}_2
\end{array}\right]=$
$\left({\renewcommand{\arraystretch}{1.2} 
\begin{tabular}{c|c c c}  
{} 
t& $s=\overset{\tilde {\rho}_{0}}{0}$
&$\overset{\tilde {\rho}_{1}}{\Yboxdim6pt\young(.~~)}$
&$\overset{\tilde {\rho}_{1}}{\Yboxdim6pt\young({.}{.}~~~)}$\\ 
\hline 
{$\rho_0$}
& {\tiny$z_1\sqrt{\frac{[N-1][N][N+1][N+2]}{[4][3][2]}}$}
&{\tiny$-\frac{[2]}{[N+1]}\sqrt{\frac{[N][N+1]}{[4][3]}}$} 
&{\tiny$\frac{[2]}{[N+2]}\sqrt{\frac{[N+2][N+4]}{[4][2]}}$}
\\ 
 {$\rho_1$} 
&{\tiny$-z_1[2]\sqrt{\frac{[N-1][N+1][N+2][N+3]}{[2][3][5]}}$}
&{\tiny$z_2\sqrt{\frac{[N+1][N+3]}{[3][5]}}$}
&{\tiny$z_3\sqrt{\frac{[N][N+2][N+3][N+4]}{[2][5]}}$}
\\ 
{$\rho_2$}
&{\tiny$z_1[3]\sqrt{\frac{[N+1][N+2][N+3][N+4]}{[5][4][3][2]}}$}
&{\tiny$z_4[3]\sqrt{\frac{[N-1][N+1][N+3][N+4]}{[5][4][3]}}$}
&{\tiny$z_3\sqrt{\frac{[N-1][N][N+2][N+3]}{[5][4][2]}}$}\\
  \end{tabular} }\right)$~,}$$
\end{displaymath}
}where
$\begin{array}{ccc}
\rho_0=\Yboxdim6pt\yng(3,2),
&\rho_1=\Yboxdim6pt\yng(4,1),
&\rho_2=\Yboxdim6pt\yng(5)
\end{array}$ 
and the variables $z_i$'s are
\begin{eqnarray*}
z_1=\frac{[2]}{[N+1][N+2]},&~&z_2=z_1\left(\frac{[N+3][N+4]-\frac{[N][N-1]}{[3]}}
{\frac{[N]}{[3]}+[N+4]}\right),\\
z_3=\frac{[2]}{[N+2][N+3]},&~&z_4=\frac{[2]}{[N+1][N+3]}.~
\end{eqnarray*}

The  Racah coefficients of third type is given by
{\small
\begin{equation}
$${$a_{ts}\left[
\renewcommand{\arraystretch}{1.1}
\begin{array}{cc} 
R_1 
& R_2
\\ 
\bar{R}_2
& \bar{R}_1
\end{array}\right]=$
$\tfrac1{\sqrt{K}}\left({\renewcommand{\arraystretch}{1.2} 
\begin{tabular}{c|c c c}  
{} 
t& $s=0$
&$\overset{ {\tilde\rho}_{1}}{\Yboxdim6pt\young(.~)}$
&$\overset{ {\tilde\rho}_{2}}{\Yboxdim6pt\young({.}{.}~~)}$\\ 
\hline 
{$\rho_0$}
& {\tiny$\sqrt{\dim_q \rho_{0}}$}
&{\tiny$-\tfrac{[N][N+1]}{[3]}\sqrt{\frac{[N+3]}{[4]}}$} 
&{\tiny$\tfrac{[N]}{[3]}\sqrt{\frac{[N+1][N+3][N+4]}{[2][4]}}$}
\\ 
 {$\rho_1$} 
&{\tiny$-\sqrt{\dim_q \rho_{1}}$}
&{\tiny$-([N-2]+[N]-[N+4])\frac{[N+1]}{[2][3]}\sqrt{\frac{[N]}{[5]}}$}
&{\tiny$\tfrac{[2][N]}{[3]}\sqrt{\frac{[N][N+1][N+4]}{[2][5]}}$}
\\ 
{$\rho_1$}
&{\tiny$\sqrt{\dim_q \rho_{2}}$}
&{\tiny$[N + 1] \sqrt{\frac{[N-1] [N][N + 4]}{[4] [5]}}$}
&{\tiny$[N]\sqrt{\frac{[N-1] [N][N+1]}{[2][4] [5]}}$}\\
  \end{tabular} }\right)$~,}$$
\end{equation}
}
The quantum dimensions of the representations in terms of the $q$-numbers are
\begin{eqnarray}
\sqrt{K}=\sqrt{dim_qR_1dim_qR_2}=\frac{[N][N+1]}{[2]}\sqrt{\frac{[N+2]}{[3]}},~
dim_q\rho_0=\frac{[N-1][N]^2[N+1][N+2]}{[4][3][2]}, \nonumber\\
dim_q\rho_1=\frac{[N-1][N][N+1][N+2][N+3]}{[2][3][5]},~
dim_q\rho_2=\frac{[N][N+1][N+2][N+3][N+4]}{[2][3][4][5]}\nonumber
\end{eqnarray}

Using the identities, it should be possible to generalise these Racah matrices for 
$R_2=\setlength{\unitlength}{1pt}
\begin{picture}(35,12)
\put(0,0){\Yboxdim6pt\yng(6)}
\put(14,9){\vector(-1,0){14}}
\put(21,9){\vector(1,0){13}}
\put(15,7){$n$}
\end{picture},$
which are again $3\times 3$ matrices.
\item Equivalently, we could write the Racah matrix coefficients when
$R_1=\Yboxdim6pt\yng(1,1) \neq R_2$ where $R_2$ is totally antisymmetric $n$-th rank 
tensor (represented by $n$-vertical box). For $R_1=\Yboxdim6pt\yng(1,1)$ and $R_2=\Yboxdim6pt\yng(1,1,1)$, the
second type Racah coefficient matrix is
{\small
\begin{displaymath}
$${$a_{ts}\left[
\begin{array}{cc} 
R_1 
& R_2
\\ 
\bar{R}_1
& \bar{R}_2
\end{array}\right]=$
$\left({\renewcommand{\arraystretch}{1.4}
\begin{tabular}{l|c c c}  
{} 
t& $s=\overset{\tilde{\rho}_0}{\Yboxdim6pt\yng(1)}$
& $\overset{\tilde{\rho}_1}{\Yboxdim6pt\young(.~,:~)}$
&{\setlength{\unitlength}{1pt}
\begin{picture}(20,26)
\put(15,0){\Yboxdim6pt\yng(2,2,2,1)}
\put(-8,9){\tiny{$N-4$}}
\put(9,14){\vector(0,1){8}}
\put(9,8){\vector(0,-1){8}}
\put(15,27){$\tilde{\rho}_2$}

\end{picture} }\\ 
\hline 
{$\rho_0$}
& {\tiny$z_1\sqrt{\frac{[N+1][N][N-1][N-2]}{[4][3][2]}}$}
&{\tiny$\frac{[2]}{[N-1]}\sqrt{\frac{[N][N-1]}{[4][3]}}$} 
&{\tiny$\frac{[2]}{[N-2]}\sqrt{\frac{[N-2][N-4]}{[4][2]}}$}
\\ 
 {$\rho_1$} 
&{\tiny$-z_1[2]\sqrt{\frac{[N+1][N-1][N-2][N-3]}{[2][3][5]}}$}
&{\tiny$z_2\sqrt{\frac{[N-1][N-3]}{[3][5]}}$}
&{\tiny$z_3\sqrt{\frac{[N][N-2][N-3][N-4]}{[2][5]}}$}
\\ 
{$\rho_2$}
&{\tiny$z_1[3]\sqrt{\frac{[N-1][N-2][N-3][N-4]}{[5][4][3][2]}}$}
&{\tiny$-z_4[3]\sqrt{\frac{[N+1][N-1][N-3][N-4]}{[5][4][3]}}$}
&{\tiny$z_3\sqrt{\frac{[N+1][N][N-2][N-3]}{[5][4][2]}}$}\\
  \end{tabular} }\right)$~,
}$$
\end{displaymath}
}
where $\rho_0=\Yboxdim6pt\yng(2,2,1)$, $\rho_1=\Yboxdim6pt\yng(2,1,1,1)$, $\rho_2=\Yboxdim6pt\yng(1,1,1,1,1)$ 
and the variables are\\
\begin{eqnarray*}
\begin{array}{cccc}
z_1=\frac{[2]}{[N-1][N-2]},&~&
z_2=z_1\left(\frac{-[N-3][N-4]+\frac{[N][N+1]}{[3]}}{\frac{[N]}{[3]}+[N-4]}\right),\\
z_3=\frac{[2]}{[N-2][N-3]},&~&
z_4=\frac{[2]}{[N-1][N-3]}~.
\end{array}
\end{eqnarray*}

\end{enumerate}
Finally, with these data available, we  evaluate the polynomial form for the
non-torus knots and links in Figures
\ref{fig:plat}\&\ref{figs:plat} in \ref{sec:appenB} and \ref{sec:appenC} using the forumula in \ref{sec:appenA}.
 From these invariants, the reformulated invariants $f_R[\mathcal{K}],~f_{R_1,R_2}[\mathcal{L}]$ (\ref {findd})
are obtained and shown to obey  eqn.(\ref{fexp}). 

\section{Discussion and Conclusion}
In this paper, we have attempted a challenging problem of 
obtaining matrix elements of the duality matrix which has properties and 
identities similar to the quantum Racah coefficients. Particularly,
we derived these identities and properties by studying the
equivalence of states in the space of correlator conformal blocks
in the $SU(N)_k$ Wess-Zumino conformal field theory.
 
We have tabulated the Racah coefficients for some class of representations
which will be useful to compute non-torus knots and non-torus two component
links. We have presented the polynomial form for all the non-torus knots and non-torus links in Figures \ref{fig:plat}\&\ref{figs:plat} (see \ref{sec:appenB} and \ref{sec:appenC}) and obtained their reformulated invariants. These invariants
obey the conjectured form (\ref{fexp}) \cite{Ooguri:1999bv,Labastida:2000yw}
confirming  the correctness of our Racah coefficients in section \ref{sec:SU(N)_quantum}.

We believe that there must be a systematic way of writing a closed form 
expression similar to the expression obtained for $SU(2)$ quantum Racah 
coefficients \citep{Kirillov:1989,Kaul:1993hb}. There are papers in the literature addressing 
classical Racah and quantum Racah coefficients. Unfortunately, we do not see
such explicit coefficients in section 4 to compare. We hope to study
those papers which may help us to obtain a closed form expression for 
$SU(N)$ quantum Racah coefficients.

There are interesting recent developments relating torus knots to spectral
curve in the $B$-model topological strings \cite{Brini:2011wi}, Poincare polynomial
computation from refined Chern-Simons theory, Khovanov homology, fivebranes \cite{Witten:2011zz,Aganagic:2011sg},
and the polynomial invariants from counting of solutions 
in four-dimensional theories \cite{Gaiotto:2011nm} 
We hope to extend these recent works to non-torus links and report in future.

\vskip.5cm
\noindent
{\bf Acknowledgments}: P.R would like to thank the hospitality of Center for Quantum
spacetime, Sogang University where this work was completed.
This work was supported by the 
National Research Foundation of Korea(NRF) grant funded by the
Korea government(MEST) through the Center for Quantum Spacetime(CQUeST)
of Sogang University with grant number 2005-0049409. 
\vskip.5cm
\newpage

\appendix
\section{Formulae for $U(N)$ link invariants in terms of $SU(N)$ quantum Racah coefficients and braiding eigenvalues}
\label{sec:appenA}
 In this appendix we give the expression of $U(N)$ link invariant for
all the non torus knots  and links in Figure \ref{fig:plat} and Figure \ref{figs:plat}
in terms of $SU(N)$ quantum Racah coefficients and braiding eigenvalues. 
\subsection{Non Torus Knots}
\begin{flalign}
V_{R}^{\{U(N)\}}[4_{1};\,0] = & \sum_{s,t,s^{\prime}}\epsilon_{s}^{\bar{R},R}\,\sqrt{dim_{q}s}\,\epsilon_{s^{\prime}}^{R,R}\,\sqrt{dim_{q}s^{\prime}}\,(\lambda_{s}^{(-)}(\bar{R},\, R))^{2}\, a_{ts}\left[\begin{array}{cc}
R & \bar{R}\\
R & \bar{R}
\end{array}\right]&\nonumber\\
   &(\lambda_{t}^{(-)}(\bar{R},\, R))^{-1}\, a_{ts^{\prime}}\left[\begin{array}{cc}
\bar{R} & R\\
R & \bar{R}
\end{array}\right]\,(\lambda_{s^{\prime}}^{(+)}(R,\, R))^{-1}.&
\end{flalign}
\begin{flalign}
V_{R}^{\{U(N)\}}[5_{2};\,0] = & q^{(-5\kappa_{R}+\frac{5l^{2}}{2N})}\sum_{s,t,s^{\prime}}\epsilon_{s}^{R,R}\,\sqrt{dim_{q}s}\,\epsilon_{s^{\prime}}^{R,R}\,\sqrt{dim_{q}s^{\prime}}\,(\lambda_{s}^{(+)}(R,\, R))^{-2}\, a_{ts}\left[\begin{array}{cc}
\bar{R} & R\\
R & \bar{R}
\end{array}\right]&\nonumber \\
   & (\lambda_{t}^{(-)}(\bar{R},\, R))^{-2}\, a_{ts^{\prime}}\left[\begin{array}{cc}
R & \bar{R}\\
\bar{R} & R
\end{array}\right]\,(\lambda_{\bar{s^{\prime}}}^{(+)}(R,\, R))^{-1}.&
\end{flalign}
\begin{flalign}
V_{R}^{\{U(N)\}}[6{}_{1};\,0] = & q^{(-2\kappa_{R}+\frac{l^{2}}{N})}\sum_{s,t,s^{\prime}}\epsilon_{s}^{\bar{R},R}\,\sqrt{dim_{q}s}\,\epsilon_{s^{\prime}}^{R,R}\,\sqrt{dim_{q}s^{\prime}}\,(\lambda_{s}^{(-)}(\bar{R},\, R))^{2}\, a_{ts}\left[\begin{array}{cc}
R & \bar{R}\\
R & \bar{R}
\end{array}\right]&\nonumber \\
   & (\lambda_{t}^{(-)}(\bar{R},\, R))^{-3}\, a_{ts^{\prime}}\left[\begin{array}{cc}
\bar{R} & R\\
R & \bar{R}
\end{array}\right]\,(\lambda_{s^{\prime}}^{(+)}(R,\, R))^{-1}.&
\end{flalign}
\begin{flalign}
V_{R}^{\{U(N)\}}[6_{2};\,0] = & q^{(-2\kappa_{R}+\frac{l^{2}}{N})}\sum_{s,t,s^{\prime},u,v}\epsilon_{s}^{R,R}\,\sqrt{dim_{q}s}\,\epsilon_{v}^{\bar{R},R}\,\sqrt{dim_{q}v}\,\lambda_{s}^{(+)}(R,\, R)\, a_{ts}\left[\begin{array}{cc}
\bar{R} & R\\
R & \bar{R}
\end{array}\right]&\nonumber \\
   & \lambda_{t}^{(-)}(R,\,\bar{R})\, a_{ts^{\prime}}\left[\begin{array}{cc}
\bar{R} & R\\
\bar{R} & R
\end{array}\right]\,(\lambda_{s^{\prime}}^{(-)}(R,\,\bar{R}))^{-1}\, a_{us^{\prime}}\left[\begin{array}{cc}
\bar{R} & \bar{R}\\
R & R
\end{array}\right]&\nonumber \\
   & (\lambda_{\bar{u}}^{(+)}(R,\, R))^{-2}a_{uv}\left[\begin{array}{cc}
\bar{R} & \bar{R}\\
R & R
\end{array}\right]\,(\lambda_{v}^{(-)}(\bar{R},\, R))^{-1}.&
\end{flalign}
\begin{flalign}
V_{R}^{\{U(N)\}}[6_{3};\,0] = & \sum_{s,t,s^{\prime},u,v}\epsilon_{s}^{R,R}\,\sqrt{dim_{q}s}\,\epsilon_{v}^{\bar{R},R}\,\sqrt{dim_{q}v}\,(\lambda_{s}^{(+)}(R,\, R))^{-2}\, a_{ts}\left[\begin{array}{cc}
\bar{R} & R\\
R & \bar{R}
\end{array}\right]&\nonumber \\
   & (\lambda_{t}^{(-)}(R,\,\bar{R}))^{-1}\, a_{ts^{\prime}}\left[\begin{array}{cc}
\bar{R} & R\\
\bar{R} & R
\end{array}\right]\,\lambda_{s^{\prime}}^{(-)}(R,\,\bar{R})\, a_{us^{\prime}}\left[\begin{array}{cc}
\bar{R} & \bar{R}\\
R & R
\end{array}\right]&\nonumber \\
   & (\lambda_{u}^{(+)}(\bar{R},\,\bar{R}))a_{uv}\left[\begin{array}{cc}
\bar{R} & \bar{R}\\
R & R
\end{array}\right]\,(\lambda_{v}^{(-)}(\bar{R},\, R)).&
\end{flalign}
\begin{flalign}
V_{R}^{\{U(N)\}}[7_{2};\,0]=& q^{(-7\kappa_{R}+\frac{7l^{2}}{2N})}\sum_{s,t,s^{\prime}}\epsilon_{s}^{R,R}\,\sqrt{dim_{q}s}\,\epsilon_{s^{\prime}}^{R,R}\,\sqrt{dim_{q}s^{\prime}}\,(\lambda_{s}^{(+)}(R,\, R))^{-2}\, a_{ts}\left[\begin{array}{cc}
\bar{R} & R\\
R & \bar{R}
\end{array}\right]&\nonumber \\
  &(\lambda_{t}^{(-)}(R,\,\bar{R}))^{-4}\, a_{ts^{\prime}}\left[\begin{array}{cc}
\bar{R} & R\\
R & \bar{R}
\end{array}\right]\,(\lambda_{s^{\prime}}^{(+)}(R,\, R))^{-1}.&
\end{flalign}
\begin{flalign}
V_{R}^{\{U(N)\}}[7_{3};\,0] =&q^{(7\kappa_{R}-\frac{7l^{2}}{2N})} \sum_{s,t,s^{\prime}}\epsilon_{s}^{R,\bar{R}}\,\sqrt{dim_{q}s}\,\epsilon_{s^{\prime}}^{R,\bar{R}}\,\sqrt{dim_{q}s^{\prime}}\,(\lambda_{s}^{(-)}(R,\,\bar{R}))^{3}\, a_{ts}\left[\begin{array}{cc}
\bar{R} & \bar{R}\\
R & R
\end{array}\right]&\nonumber \\
   & (\lambda_{t}^{(+)}(\bar{R},\,\bar{R}))^{3}\, a_{ts^{\prime}}\left[\begin{array}{cc}
\bar{R} & \bar{R}\\
R & R
\end{array}\right]\,\lambda_{s^{\prime}}^{(-)}(R,\,\bar{R}).&
\end{flalign}
\begin{flalign}
V_{R}^{\{U(N)\}}[7_{4},;\,0] = & q^{(7\kappa_{R}-\frac{7l^{2}}{2N})}\sum_{s,t,s^{\prime},u,v}\epsilon_{s}^{R,R}\,\sqrt{dim_{q}s}\,\epsilon_{v}^{R,R}\,\sqrt{dim_{q}v}\,\lambda_{s}^{(+)}(R,\, R)\, a_{ts}\left[\begin{array}{cc}
\bar{R} & R\\
R & \bar{R}
\end{array}\right]&\nonumber \\
   & (\lambda_{t}^{(-)}(R,\,\bar{R}))^{2}\, a_{ts^{\prime}}\left[\begin{array}{cc}
R & \bar{R}\\
\bar{R} & R
\end{array}\right]\,\lambda_{s^{\prime}}^{(+)}(\bar{R},\,\bar{R})\, a_{us^{\prime}}\left[\begin{array}{cc}
R & \bar{R}\\
\bar{R} & R
\end{array}\right]&\nonumber \\
   & (\lambda_{u}^{(-)}(R,\,\bar{R}))^{2}\, a_{uv}\left[\begin{array}{cc}
\bar{R} & R\\
R & \bar{R}
\end{array}\right]\,\lambda_{v}^{(+)}(R,\, R).&
\end{flalign}
\begin{flalign}
V_{R}^{\{U(N)\}}[7_{5};\,0] =&q^{(-7\kappa_{R}+\frac{7l^{2}}{2N})}  \sum_{s,t,s^{\prime},u,v}\epsilon_{s}^{R,\bar{R}}\,\sqrt{dim_{q}s}\,\epsilon_{v}^{R,\bar{R}}\,\sqrt{dim_{q}v}\,(\lambda_{s}^{(-)}(R,\, \bar{R}))^{-1}\, a_{ts}\left[\begin{array}{cc}
\bar{R} & \bar{R}\\
R & R
\end{array}\right]&\nonumber \\
   & (\lambda_{\bar{t}}^{(+)}(R,\, R))^{-1}\, a_{ts^{\prime}}\left[\begin{array}{cc}
\bar{R} & \bar{R}\\
R & R
\end{array}\right]\,(\lambda_{s^{\prime}}^{(-)}(\bar{R},\, R))^{-2}\, a_{us^{\prime}}\left[\begin{array}{cc}
\bar{R} & \bar{R}\\
R & R
\end{array}\right]&\nonumber \\
   & (\lambda_{\bar{u}}^{(+)}(R,\, R))^{-2}\, a_{uv}\left[\begin{array}{cc}
\bar{R} & \bar{R}\\
R & R
\end{array}\right]\,(\lambda_{v}^{(-)}(\bar{R},\, R))^{-1}.&
\end{flalign}
\begin{flalign}
V_{R}^{\{U(N)\}}[7_{6};\,0]= & q^{(-3\kappa_{R}+\frac{3l^{2}}{2N})}\sum_{s,t,s^{\prime},u,v}\epsilon_{s}^{R,\bar{R}}\,\sqrt{dim_{q}s}\,\epsilon_{v}^{\bar{R},R}\,\sqrt{dim_{q}v}\,(\lambda_{s}^{(-)}(R,\,\bar{R}))^{-2}\, a_{ts}\left[\begin{array}{cc}
\bar{R} & R\\
\bar{R} & R
\end{array}\right]&\nonumber \\
   & (\lambda_{t}^{(-)}(\bar{R},\, R))^{2}\, a_{ts^{\prime}}\left[\begin{array}{cc}
\bar{R} & R\\
\bar{R} & R
\end{array}\right]\,(\lambda_{s^{\prime}}^{(-)}(R,\,\bar{R}))^{-1}\, a_{us^{\prime}}\left[\begin{array}{cc}
\bar{R} & \bar{R}\\
R & R
\end{array}\right]&\nonumber \\
   & (\lambda_{u}^{(+)}(\bar{R},\,\bar{R}))^{-1}\, a_{uv}\left[\begin{array}{cc}
\bar{R} & \bar{R}\\
R & R
\end{array}\right]\,(\lambda_{v}^{(-)}(\bar{R},\, R))^{-1}.&
\end{flalign}
\begin{flalign}
V_{R}^{\{U(N)\}}[7_{7};\,0] = & q^{(\kappa_{R}-\frac{l^{2}}{2N})}\sum_{s,t,s^{\prime},u,v,w,x}\epsilon_{s}^{R,R}\,\sqrt{dim_{q}s}\,\epsilon_{x}^{R,R}\,\sqrt{dim_{q}x}\,(\lambda_{s}^{(+)}(R,\, R))\, a_{ts}\left[\begin{array}{cc}
\bar{R} & R\\
R & \bar{R}
\end{array}\right]&\nonumber \\
   & (\lambda_{t}^{(-)}(\bar{R},\, R))\, a_{ts^{\prime}}\left[\begin{array}{cc}
R & \bar{R}\\
R & \bar{R}
\end{array}\right]\,(\lambda_{s^{\prime}}^{(-)}(\bar{R},\, R))^{-1}\, a_{us^{\prime}}\left[\begin{array}{cc}
R & R\\
\bar{R} & \bar{R}
\end{array}\right]&\nonumber \\
   & (\lambda_{\bar{u}}^{(+)}(\bar{R},\,\bar{R}))^{-1}\, a_{uv}\left[\begin{array}{cc}
R & R\\
\bar{R} & \bar{R}
\end{array}\right]\,(\lambda_{v}^{(-)}(R,\,\bar{R}))^{-1}a_{wv}\left[\begin{array}{cc}
R & \bar{R}\\
R & \bar{R}
\end{array}\right]&\nonumber \\
   & \lambda_{w}^{(-)}(R,\bar{R})\, a_{wx}\left[\begin{array}{cc}
\bar{R} & R\\
R & \bar{R}
\end{array}\right]\,\lambda_{x}^{(+)}(R,R).&
\end{flalign}
\begin{flalign}
V_{R}^{\{U(N)\}}[8{}_{1};\,0]= & q^{(-4\kappa_{R}+\frac{2l^{2}}{N})}\sum_{s,t,s^{\prime}}\epsilon_{s}^{\bar{R},R}\,\sqrt{dim_{q}s}\,\epsilon_{s^{\prime}}^{R,R}\,\sqrt{dim_{q}s^{\prime}}\,(\lambda_{s}^{(-)}(\bar{R},\, R))^{2}\, a_{ts}\left[\begin{array}{cc}
R & \bar{R}\\
R & \bar{R}
\end{array}\right]&\nonumber \\
   & (\lambda_{t}^{(-)}(\bar{R},\, R))^{-5}\, a_{ts^{\prime}}\left[\begin{array}{cc}
\bar{R} & R\\
R & \bar{R}
\end{array}\right]\,(\lambda_{s^{\prime}}^{(+)}(R,\, R))^{-1}.&
\end{flalign}
\begin{flalign}
V_{R}^{\{U(N)\}}[9_{2};\,0]= & q^{(-7\kappa_{R}+\frac{7l^{2}}{2N})}\sum_{s,t,s^{\prime}}\epsilon_{s}^{R,R}\,\sqrt{dim_{q}s}\,\epsilon_{s^{\prime}}^{R,R}\,\sqrt{dim_{q}s^{\prime}}\,(\lambda_{s}^{(+)}(R,\, R))^{-2}\, a_{ts}\left[\begin{array}{cc}
\bar{R} & R\\
R & \bar{R}
\end{array}\right]&\nonumber \\
   & (\lambda_{t}^{(-)}(\bar{R},\, R))^{-6}\, a_{ts^{\prime}}\left[\begin{array}{cc}
R & \bar{R}\\
\bar{R} & R
\end{array}\right]\,(\lambda_{s^{\prime}}^{(+)}(R,\, R))^{-1}.&
\end{flalign}
\begin{flalign}
V_{R}^{\{U(N)\}}[10{}_{1};\,0] = & q^{(-6\kappa_{R}+\frac{3l^{2}}{N})}\sum_{s,t,s^{\prime}}\epsilon_{s}^{\bar{R},R}\,\sqrt{dim_{q}s}\,\epsilon_{s^{\prime}}^{R,R}\,\sqrt{dim_{q}s^{\prime}}\,(\lambda_{s}^{(-)}(\bar{R},\, R))^{2}\, a_{ts}\left[\begin{array}{cc}
R & \bar{R}\\
R & \bar{R}
\end{array}\right]&\nonumber \\
   & (\lambda_{t}^{(-)}(\bar{R},\, R))^{-7}\, a_{ts^{\prime}}\left[\begin{array}{cc}
\bar{R} & R\\
R & \bar{R}
\end{array}\right]\,(\lambda_{s^{\prime}}^{(+)}(R,\, R))^{-1}.&
\end{flalign}

For framed knots $\mathcal{K}$ with framing number $p$, the invariants will be related to the zero-framed knot invariants as 
\begin{equation}
V_{R}^{\{U(N)\}}[\mathcal{K};\,p]=q^{p \kappa_R} V_{R}^{\{U(N)\}}[\mathcal{K};\,0]~.
\end{equation}

\subsection{Non Torus Links}
In the context of links, we can place different representations on the component knots. The invariants
are hence called multicolored links.
\begin{flalign}
V_{(R_{1},R_{2})}^{\{U(N)\}}[6_{2};\,0,0] = & q^{\frac{3l_{R_{1}}l_{R_{2}}}{N}}\sum_{s,t,s^{\prime}}\epsilon_{s}^{R_{1},R_{2}}\,\sqrt{dim_{q}s}\,\epsilon_{s^{\prime}}^{\bar{R}_{1},\bar{R}_{2}}\,\sqrt{dim_{q}s^{\prime}}\,(\lambda_{s}^{(+)}(R_{1},\, R_{2}))^{-3}\, a_{t\bar{s}}\left[\begin{array}{cc}
R_{2} & \bar{R}_{1}\\
\bar{R}_{2} & R_{1}
\end{array}\right]&\nonumber \\
   & (\lambda_{\bar{t}}^{(-)}(\bar{R}_{1},\, R_{2}))^{-2}\, a_{t\bar{s^{\prime}}}\left[\begin{array}{cc}
\bar{R}_{1} & R_{2}\\
R_{1} & \bar{R}_{2}
\end{array}\right]\,(\lambda_{s^{\prime}}^{(+)}(\bar{R}_{1},\,\bar{R}_{2}))^{-1}.&
\end{flalign}
\begin{flalign}
V_{(R_{1},R_{2})}^{\{U(N)\}}[6_{3};\,0,0] = & q^{(-2\kappa_{R_{2}}+\frac{l_{R_{2}}^{2}}{N}-\frac{2l_{R_{1}}l_{R_{2}}}{N})}\sum_{s,t,s^{\prime},u,v}\epsilon_{s}^{R_{1},R_{2}}\,\sqrt{dim_{q}s}\,\epsilon_{v}^{\bar{R}_{1},\bar{R}_{2}}\,\sqrt{dim_{q}v}\,\lambda_{s}^{(+)}(R_{1},\, R_{2})&\nonumber \\
   & a_{ts}\left[\begin{array}{cc}
\bar{R}_{1} & R_{2}\\
R_{1} & \bar{R}_{2}
\end{array}\right]\,\lambda_{\bar{t}}^{(-)}(R_{1},\,\bar{R}_{2})\, a_{ts^{\prime}}\left[\begin{array}{cc}
\bar{R}_{1} & R_{2}\\
\bar{R}_{2} & R_{1}
\end{array}\right]\,(\lambda_{s^{\prime}}^{(-)}(R_{2},\,\bar{R}_{2}))^{-2}&\nonumber \\
   & a_{us^{\prime}}\left[\begin{array}{cc}
\bar{R}_{1} & R_{2}\\
\bar{R}_{2} & R_{1}
\end{array}\right]\,\lambda_{u}^{(-)}(\bar{R}_{1},\, R_{2})\, a_{uv}\left[\begin{array}{cc}
R_{2} & \bar{R}_{1}\\
\bar{R}_{2} & R_{1}
\end{array}\right]\,\lambda_{v}^{(+)}(\bar{R}_{1},\,\bar{R}_{2}).&
\end{flalign}
\begin{flalign}
V_{(R_{1},R_{2})}^{\{U(N)\}}[7_{1};\,0,0] = & q^{(-\kappa_{R_{2}}+\frac{l_{R_{2}}^{2}}{2N}+\frac{l_{R_{1}}l_{R_{2}}}{N})}\sum_{s,t,s^{\prime},u,v}\epsilon_{s}^{R_{1},R_{2}}\,\sqrt{dim_{q}s}\,\epsilon_{v}^{\bar{R}_{1},R_{2}}\,\sqrt{dim_{q}v}\,(\lambda_{s}^{(+)}(R_{1},\, R_{2}))&\nonumber \\
  & a_{ts}\left[\begin{array}{cc}
\bar{R}_{1} & R_{2}\\
R_{1} & \bar{R}_{2}
\end{array}\right]\,\lambda_{\bar{t}}^{(-)}(R_{1},\,\bar{R}_{2})\, a_{ts^{\prime}}\left[\begin{array}{cc}
\bar{R}_{1} & R_{2}\\
\bar{R}_{2} & R_{1}
\end{array}\right]\,(\lambda_{s^{\prime}}^{(-)}(R_{2},\,\bar{R}_{2}))^{-1}&\nonumber \\
   & a_{us^{\prime}}\left[\begin{array}{cc}
\bar{R}_{1} & \bar{R}_{2}\\
R_{2} & R_{1}
\end{array}\right]\,\lambda_{u}^{(+)}(\bar{R}_{1},\,\bar{R}_{2})^{-3}\, a_{uv}\left[\begin{array}{cc}
\bar{R}_{2} & \bar{R}_{1}\\
R_{2} & R_{1}
\end{array}\right]\,(\lambda_{v}^{(-)}(\bar{R}_{1},\, R_{2}))^{-1}.&
\end{flalign}
\begin{flalign}
V_{(R_{1},R_{2})}^{\{U(N)\}}[7_{2};\,0,0]= & q^{(-\kappa_{R_{2}}+\frac{l_{R_{2}}^{2}}{2N}-\frac{l_{R_{1}}l_{R_{2}}}{N})}\sum_{s,t,s^{\prime},u,v}\epsilon_{s}^{R_{1},R_{2}}\,\sqrt{dim_{q}s}\,\epsilon_{v}^{\bar{R}_{1},R_{2}}\,\sqrt{dim_{q}v}\,(\lambda_{s}^{(+)}(R_{1},\, R_{2}))^{-2}&\nonumber \\
   & a_{ts}\left[\begin{array}{cc}
\bar{R}_{1} & R_{1}\\
R_{2} & \bar{R}_{2}
\end{array}\right]\,(\lambda_{\bar{t}}^{(-)}(R_{2},\,\bar{R}_{2}))^{-1}\, a_{ts^{\prime}}\left[\begin{array}{cc}
\bar{R}_{1} & R_{1}\\
\bar{R}_{2} & R_{2}
\end{array}\right]\,(\lambda_{s^{\prime}}^{(-)}(R_{1},\,\bar{R}_{2}))&\nonumber \\
   & a_{us^{\prime}}\left[\begin{array}{cc}
\bar{R}_{1} & \bar{R}_{2}\\
R_{1} & R_{2}
\end{array}\right]\,\lambda_{\bar{u}}^{(+)}(R_{1},\, R_{2})^{2}\, a_{uv}\left[\begin{array}{cc}
\bar{R}_{2} & \bar{R}_{1}\\
R_{2} & R_{1}
\end{array}\right]\,\lambda_{v}^{(-)}(\bar{R}_{1},\, R_{2}).&
\end{flalign}
\begin{flalign}
V_{(R_{1},R_{2})}^{\{U(N)\}}[7_{3};\,0,0] = & q^{(-3\kappa_{R_{2}}+\frac{3l_{R_{2}}^{2}}{2N})}\sum_{s,t,s^{\prime},u,v}\epsilon_{s}^{R_{1},R_{2}}\,\sqrt{dim_{q}s}\,\epsilon_{v}^{\bar{R}_{1},R_{2}}\,\sqrt{dim_{q}v}\,\lambda_{s}^{(+)}(R_{1},\, R_{2})&\nonumber \\
  & a_{ts}\left[\begin{array}{cc}
\bar{R}_{1} & R_{2}\\
R_{1} & \bar{R}_{2}
\end{array}\right]\,\lambda_{\bar{t}}^{(-)}(R_{1},\,\bar{R}_{2})\, a_{ts^{\prime}}\left[\begin{array}{cc}
\bar{R}_{1} & R_{2}\\
\bar{R}_{2} & R_{1}
\end{array}\right]\,(\lambda_{s^{\prime}}^{(-)}(R_{2},\,\bar{R}_{2}))^{-3}&\nonumber \\
   & a_{us^{\prime}}\left[\begin{array}{cc}
\bar{R}_{1} & \bar{R}_{2}\\
R_{2} & R_{1}
\end{array}\right]\,(\lambda_{u}^{(+)}(\bar{R}_{1},\,\bar{R}_{2}))^{-1}\, a_{uv}\left[\begin{array}{cc}
\bar{R}_{2} & \bar{R}_{1}\\
R_{2} & R_{1}
\end{array}\right]&\nonumber \\
   & (\lambda_{v}^{(-)}(\bar{R}_{1},\, R_{2}))^{-1}.&
\end{flalign}

Including the framing numbers $p_1,p_2$ on the component knots of these two-component torus links ${\cal L}$, the  framed multicolored invariant will be
\begin{equation}
V_{(R_{1},R_{2})}^{\{U(N)\}}[{\cal L} ;\,p_1,p_2]=q^{(p_1 \kappa_{R_1}+ p_2 \kappa_{R_2})}~ 
V_{(R_{1},R_{2})}^{\{U(N)\}}[{\cal L};\,0,0]~.
\end{equation}

\section{Knot  Polynomials }
\label{sec:appenB}
\Yboxdim5pt
In this appendix we  present  the polynomial form  of  the $U(N)$ link invariant for non torus knots
in Figure \ref{fig:plat} for representation  whose Young tableau diagrams are $\Yboxdim6pt\yng(1)$, $\Yboxdim6pt\yng(2)$ and $\Yboxdim6pt\yng(1,1)$. The polynomial corresponding to  representation $\yng(1)$ is proportional to HOMFLY-PT  polynomial $P(\lambda,t)[\mathcal{K}]$ \cite{Freyd:1985dx,PT}
upto unknot $U$  normalisation:
\Yboxdim4pt
\begin{equation}
P(\lambda,q )[\mathcal{K}] ={V_{\yng(1)}^{U(N)}[\mathcal{K};0]\over V_{\yng(1)}^{U(N)}[U]}= {(q^{1/2}-q^{-1/2})\over (\lambda^{1/2}-\lambda^{-1/2})}V_{\yng(1)}^{U(N)}[\mathcal{K};0]~.
\end{equation}

We list them so that we can directly use  them in the computation of 
reformulated invariants in \ref{sec:appenD}.

\begin{enumerate}

\item
For fundamental representation $R=\Yboxdim6pt\yng(1)$ placed on the knot, the $U(N)$ knot polynomials are
\Yboxdim4pt
\begin{center}
\begin{tabular}{c p{11cm}}
$V_{\yng(1)}^{U(N)}[4_{1}]=$ &$\frac{(\lambda-1)}{\lambda^{3/2}(q-1)\sqrt{q}}\big[-\lambda-\lambda q^{2}+\left(\lambda^{2}+\lambda+1\right)q\bigl]$
\end{tabular}  
\end{center}
\begin{center}
\begin{tabular}{c p{11cm}}
$V_{\yng(1)}^{U(N)}[5_{2}] =$ &$\frac{1}{(-1+q)\sqrt{q}\lambda^{7/2}}\bigl[q-q\lambda^{3}+\lambda(-1+\lambda^{2})+q^{2}\lambda(-1+\lambda^{2})\bigl]$
\end{tabular}  
\end{center}
\begin{center}
\begin{tabular}{c p{11cm}}
$V_{\yng(1)}^{U(N)}[6_{1}] =$ &$ \frac{1}{\lambda^{5/2}(q-1)\sqrt{q}}\bigl[-\lambda^{3}+\lambda+(\lambda-\lambda^{3})q^{2}+(\lambda^{4}+\lambda^{3}-\lambda^{2}-1)q\bigl]$
\end{tabular}  
\end{center}
\begin{center}
\begin{tabular}{c p{11cm}}
$V_{\yng(1)}^{U(N)}[6_{2}]  =$ & $\frac{(-1+\lambda)}{(-1+q)q^{3/2}\lambda^{5/2}}\bigl[-\lambda-q^{4}\lambda-q^{2}(1+2\lambda)+q(1+\lambda+\lambda^{2})+q^{3}(1+\lambda+\lambda^{2})\bigl]$
\end{tabular}  
\end{center}
\begin{center}
\begin{tabular}{c p{11cm}}
$V_{\yng(1)}^{U(N)}[6_{3}]=$ & $-\frac{(-1+\lambda)}{(-1+q)q^{3/2}\lambda^{3/2}}\bigl[-\lambda-q^{4}\lambda+q(1+\lambda+\lambda^{2})+q^{3}(1+\lambda+\lambda^{2})-q^{2}(1+3\lambda+\lambda^{2})\bigl]$
\end{tabular}  
\end{center}
 \begin{center}
\begin{tabular}{c p{11cm}} 
$V_{\yng(1)}^{U(N)}[7_{2}]=$ & $\frac{1}{(-1+q)\sqrt{q}\lambda^{9/2}}\bigl[\lambda(-1+\lambda^{3})+q^{2}\lambda(-1+\lambda^{3})-q(-1-\lambda^{2}+\lambda^{3}+\lambda^{4})\bigl]$
\end{tabular}  
\end{center}
\begin{center}
\begin{tabular}{c p{11cm}}
$V_{\yng(1)}^{U(N)}[7_{3}] =$ & $\frac{\lambda^{3/2}}{(-1+q)q^{3/2}}\bigl[-1+q+\lambda^{2}-q\lambda^{3}+q^{4}(-1+\lambda^{2})+q^{2}(-1-\lambda+2\lambda^{2})-q^{3}(-1+\lambda^{3})\bigl]$
\end{tabular}  
\end{center}
\begin{center}
\begin{tabular}{c p{11cm}}
$V_{\yng(1)}^{U(N)}[7_{4}]=$ & $\frac{(-1+\lambda)\lambda^{1/2}}{(-1+q)q^{1/2}}\bigl[(1+\lambda)^{2}+q^{2}(1+\lambda)^{2}-q(2+2\lambda+2\lambda^{2}+\lambda^{3})\bigl]$
\end{tabular}  
\end{center}
\begin{center}
\begin{tabular}{c p{11cm}}
$V_{\yng(1)}^{U(N)}[7_{5}]  =$ & $\ensuremath{\frac{(-1+\lambda)}{(-1+q)q^{3/2}\lambda^{9/2}}\bigl[\lambda(1+\lambda)+q^{4}\lambda(1+\lambda)-q(1+\lambda)^{2}}-q^{3}(1+\lambda)^{2}+q^{2}(1+2\lambda+2\lambda^{2})\bigl]$
\end{tabular}  
\end{center}
\begin{center}
\begin{tabular}{c p{11cm}}
$V_{\yng(1)}^{U(N)}[7_{6}]  =$ & $-\frac{(-1+\lambda)}{(-1+q)q^{3/2}\lambda^{7/2}}\bigl[\lambda^{2}+q^{4}\lambda^{2}-q\lambda(2+2\lambda+\lambda^{2})-q^{3}\lambda(2+2\lambda+\lambda^{2})+q^{2}(1+2\lambda+3\lambda^{2}+\lambda^{3})\bigl]$
\end{tabular}  
\end{center}
\begin{center}
\begin{tabular}{c p{11cm}}
$V_{\yng(1)}^{U(N)}[7_{7}]=$ & $\frac{(-1+\lambda)}{(-1+q)q^{3/2}\lambda^{3/2}}\bigl[\lambda+q^{4}\lambda-q(1+2\lambda+2\lambda^{2})-q^{3}(1+2\lambda+2\lambda^{2})+q^{2}(2+4\lambda+2\lambda^{2}+\lambda^{3})\bigl]$
\end{tabular}  
\end{center}
\begin{center}
\begin{tabular}{c p{11cm}} 
$V_{\yng(1)}^{U(N)}[8_{1}]=$ &$ \frac{1}{(q-1)\sqrt{q}\lambda^{7/2}}\bigl[-\lambda^{4}+\lambda+q^{2}(\lambda-\lambda^{4})+q(\lambda^{5}+\lambda^{4}-\lambda^{2}-1)\bigl]$
\end{tabular}  
\end{center}
\begin{center}
\begin{tabular}{c p{11cm}}
$V_{\yng(1)}^{U(N)}[9_{1}]=$
&$\frac{1}{(q-1)\sqrt{q}\lambda^{11/2}}\bigl[\lambda(\lambda^{4}-1)q^{2}-(\lambda^{5}+\lambda^{4}-\lambda^{2}-1)q+\lambda(\lambda^{4}-1)\bigl]$
\end{tabular}  
\end{center}
\begin{center}
\begin{tabular}{c p{11cm}}
$V_{\yng(1)}^{U(N)}[10_{1}]=$
& $\frac{1}{(q-1)\sqrt{q}\lambda^{9/2}}\bigl[-\lambda^{5}+\lambda+q^{2}(\lambda-\lambda^{5})+q(\lambda^{6}+\lambda^{5}-\lambda^{2}-1)\bigl]$
\end{tabular}  
\end{center}

\renewcommand{\baselinestretch}{1.5}\selectfont
\item For symmetric second rank representation $R=\Yboxdim6pt\yng(2)$ , the knot polynomials are
\Yboxdim4pt
\begin{center}
\begin{tabular}{c p{11cm}}
$V_{\yng(2)}^{U(N)}[4_{1}]=$ &$\frac{(-1+\lambda ) (-1+q \lambda ) }{(-1+q)^2 q^2 (1+q) \lambda ^3}\bigl[(-1+\lambda ) \lambda +3 q^3 \lambda ^2-q^6 (-1+\lambda ) \lambda ^2+q^4 \lambda  (-1+\lambda ^2)+q^5 \lambda ^2 (-1-\lambda +\lambda ^2)-q (-1+\lambda +\lambda ^2)+q^2 (\lambda -\lambda ^3)\bigl]$
\end{tabular}  
\end{center}
\begin{center}
\begin{tabular}{c p{11cm}}
$V_{\yng(2)}^{U(N)}[5_{2}]=$ &$\frac{1}{(-1+q)^2 q^5 (1+q) \lambda ^7}\bigl[q (-1+\lambda )^2-(-1+\lambda )^2 \lambda +q^9 \lambda ^4 (-1+\lambda ^2)+q^3 (-1+\lambda )^2 \lambda ^2 (1+\lambda +\lambda ^2)+q^6 \lambda ^4 (-1-\lambda +2 \lambda ^2)+q^5 \lambda ^2 (-1+\lambda -\lambda ^2+\lambda ^3)-q^8 \lambda ^3 (-1+\lambda -\lambda ^2+\lambda ^3)+q^4 \lambda  (-1+\lambda +\lambda ^2-\lambda ^5)+q^7 (\lambda ^3-\lambda ^6)\bigl]$
\end{tabular}  
\end{center}
\begin{center}
\begin{tabular}{c p{11cm}}
$V_{\yng(2)}^{U(N)}[6_{1}]=$ &$\frac{(-1+\lambda ) (-1+q \lambda ) }{(-1+q)^2 q^4 (1+q) \lambda ^5}\bigl[(-1+\lambda ) \lambda +q (-1+\lambda )^2 (1+\lambda )+q^2 (-1+\lambda )^2 \lambda  (1+\lambda )+q^7 \lambda ^4 (-3+\lambda ^2)+2 q^6 \lambda ^3 (-1+\lambda ^2)-2 q^3 \lambda ^2 (-1+\lambda +\lambda ^2)+q^5 \lambda ^2 (-1+2 \lambda +4 \lambda ^2)-q^4 \lambda  (1-\lambda -3 \lambda ^2+2 \lambda ^3+\lambda ^4)+q^8 (\lambda ^3-\lambda ^5)\bigl]$
\end{tabular}  
\end{center}
\begin{center}
\begin{tabular}{c p{11cm}}
$V_{\yng(2)}^{U(N)}[6_{2}]=$ &$\frac{(-1+\lambda ) (-1+q \lambda ) }{(-1+q)^2 q^6 (1+q) \lambda ^5}\bigl[q+(-1+\lambda ) \lambda -q \lambda ^2-q^9 (-4+\lambda ) \lambda ^2-q^{12} (-1+\lambda ) \lambda ^2-q^2 (-1+\lambda )^2 (1+\lambda )+q^3 \lambda  (-3+2 \lambda )-2 q^4 (-1+\lambda ^2)+q^{10} \lambda  (-1+\lambda ^2)+q^{11} \lambda ^2 (-1-\lambda +\lambda ^2)+q^6 (-1-3 \lambda +4 \lambda ^2)+q^8 \lambda  (2-3 \lambda ^2+\lambda ^3)+q^7 (1-2 \lambda -3 \lambda ^2+\lambda ^4)+q^5 (-1+3 \lambda +2 \lambda ^2-2 \lambda ^3+\lambda ^4)\bigl]$
\end{tabular}  
\end{center}
\begin{center}
\begin{tabular}{c p{11cm}}
$V_{\yng(2)}^{U(N)}[6_{3}]=$ &$-\frac{(-1+\lambda ) (-1+q \lambda ) }{(-1+q)^2 q^5 (1+q) \lambda ^3}\bigl[-(-1+\lambda ) \lambda +q^{12} (-1+\lambda ) \lambda ^2+q^3 (1+3 \lambda -4 \lambda ^2)+q (-1+\lambda ^2)+q^9 \lambda ^2 (-4+3 \lambda +\lambda ^2)+q^7 (-1+4 \lambda +\lambda ^2-4 \lambda ^3)+q^5 \lambda  (-4+\lambda +4 \lambda ^2-\lambda ^3)-q^4 (2-3 \lambda -3 \lambda ^2+\lambda ^3)+q^{10} \lambda  (1-\lambda -2 \lambda ^2+\lambda ^3)+q^2 (1-2 \lambda -\lambda ^2+\lambda ^3)-q^8 \lambda (1-3 \lambda -3 \lambda ^2+2 \lambda ^3)+q^{11} (\lambda ^2-\lambda ^4)+q^6 (1+\lambda -9 \lambda ^2+\lambda ^3+\lambda ^4)\bigl]$
\end{tabular}  
\end{center}
\begin{center}
\begin{tabular}{c p{11cm}}
$V_{\yng(2)}^{U(N)}[7_{2}]=$ &$\frac{1}{(-1+q)^2 q^7 (1+q) \lambda ^9}\bigl[q (-1+\lambda )^2-(-1+\lambda )^2 \lambda -q^4 (-1+\lambda )^2 \lambda +q^3 (-1+\lambda )^2 \lambda ^2+q^5 (-1+\lambda )^2 \lambda ^4 (1+\lambda +\lambda ^2)+q^{11} \lambda ^5 (-1+\lambda ^3)+q^8 \lambda ^5 (-1-\lambda +2 \lambda ^3)+q^7 \lambda ^3 (-1+\lambda -\lambda ^2+\lambda ^4)-q^{10} \lambda ^4 (-1+\lambda -\lambda ^2+\lambda ^4)-q^6 \lambda ^3 (1-2 \lambda +\lambda ^2-2 \lambda ^3+\lambda ^4+\lambda ^5)+q^9 (\lambda ^4+\lambda ^6-\lambda ^7-\lambda ^8)]$
\end{tabular}  
\end{center}
\begin{center}
\begin{tabular}{c p{11cm}}
$V_{\yng(2)}^{U(N)}[7_{3}]=$ &$\frac{(-1+\lambda ) \lambda ^3 (-1+q \lambda ) }{(-1+q)^2 q^3 (1+q)}\bigl[1+q (-1+\lambda )-q^{12} (-1+\lambda )+\lambda -q^{14} (-1+\lambda ) \lambda ^2+q^{13} (-1+\lambda )^2 \lambda  (1+\lambda )-q^2 (1+2 \lambda )-q^4 \lambda  (-4+\lambda ^2)-2 q^3 (-1+\lambda ^2)+q^5 (-2-\lambda +3 \lambda ^2)+q^8 (-1+\lambda +4 \lambda ^2-2 \lambda ^3)+q^9 (1-2 \lambda +\lambda ^3)+q^6 (1-3 \lambda -2 \lambda ^2+\lambda ^3)+q^{10} \lambda  (2-2 \lambda -\lambda ^2+\lambda ^3)+q^7 (1+3 \lambda -3 \lambda ^2-2 \lambda ^3+\lambda ^4)+q^{11} (-1+2 \lambda ^2-2 \lambda ^3+\lambda ^4)\bigl]$
\end{tabular}  
\end{center}
\begin{center}
\begin{tabular}{c p{11cm}}
$V_{\yng(2)}^{U(N)}[7_{4}]=$ &$\frac{(-1+\lambda ) \lambda  (-1+q \lambda )}{(-1+q)^2 q (1+q)}\bigl[-q^{10} (-1+\lambda ) \lambda ^4+(1+\lambda )^2+2 q (-1+\lambda ^2)+q^2 (-1-6 \lambda -\lambda ^2+\lambda ^3)+q^9 \lambda ^3 (2-2 \lambda -\lambda ^2+\lambda ^3)-q^3 (-4-2 \lambda +5 \lambda ^2+\lambda ^3)+q^8 \lambda ^2 (3-2 \lambda -3 \lambda ^2+2 \lambda ^3)+q^5 (-2-4 \lambda +6 \lambda ^2+3 \lambda ^3-3 \lambda ^4)+q^7 \lambda  (2-2 \lambda -4 \lambda ^2+3 \lambda ^3+\lambda ^4)-q^4 (1-6 \lambda -4 \lambda ^2+4 \lambda ^3+\lambda ^4)+q^6 (1-2 \lambda -4 \lambda ^2+5 \lambda ^3+\lambda ^4-\lambda ^5)\bigl]$
\end{tabular}  
\end{center}
\begin{center}
\begin{tabular}{c p{11cm}}
$V_{\yng(2)}^{U(N)}[7_{5}]=$ &$\frac{(-1+\lambda ) (-1+q \lambda )}{(-1+q)^2 q^9 (1+q) \lambda ^9}\bigl[(-1+\lambda ) \lambda -q^{13} (-1+\lambda ) \lambda ^3+q^2 (-1+\lambda )^3 (1+\lambda )+q^{14} \lambda ^3 (1+\lambda )-q^{12} \lambda ^3 (3+\lambda )+3 q^{11} \lambda ^2 (-1+\lambda ^2)+q^{10} \lambda  (-1+6 \lambda ^2)-q^3 \lambda  (3-4 \lambda +\lambda ^3)+q (1-2 \lambda ^2+\lambda ^3)+q^8 \lambda  (3-5 \lambda -5 \lambda ^2+3 \lambda ^3)+q^9 (\lambda +5 \lambda ^2-3 \lambda ^3-3 \lambda ^4)+q^4 (2-\lambda -5 \lambda ^2+3 \lambda ^3+\lambda ^4)+q^7 (1-3 \lambda -5 \lambda ^2+6 \lambda ^3+\lambda ^4)-q^6 (1+2 \lambda -7 \lambda ^2+2 \lambda ^4)+q^5 (-1+5 \lambda -\lambda ^2-5 \lambda ^3+2 \lambda ^4)\bigl]$
\end{tabular}  
\end{center}
\begin{center}
\begin{tabular}{c p{11cm}}
$V_{\yng(2)}^{U(N)}[7_{6}]=$ &$\frac{(-1+\lambda ) (-1+q \lambda )}{(-1+q)^2 q^6 (1+q) \lambda ^7}\bigl[(-1+\lambda )^2 \lambda ^2-q^{12} (-1+\lambda ) \lambda ^4+q^{11} \lambda ^4 (-2+\lambda ^2)+q \lambda  (-2+3 \lambda +\lambda ^2-2 \lambda ^3)-q^{10} \lambda ^3 (2-3 \lambda ^2+\lambda ^3)-q^9 \lambda ^3 (-1-7 \lambda +2 \lambda ^2+\lambda ^3)+q^7 \lambda ^2 (3-7 \lambda -7 \lambda ^2+4 \lambda ^3)+q^6 \lambda ^2 (-4-6 \lambda +10 \lambda ^2+\lambda ^3-\lambda ^4)+q^4 \lambda  (-1+8 \lambda -\lambda ^2-7 \lambda ^3+\lambda ^4)+q^3 \lambda  (2-8 \lambda ^2+4 \lambda ^3+\lambda ^4)+q^8 \lambda ^2 (1+5 \lambda -4 \lambda ^2-4 \lambda ^3+2 \lambda ^4)+q^5 \lambda  (-2-2 \lambda +10 \lambda ^2+2 \lambda ^3-4 \lambda ^4+\lambda ^5)-q^2 (-1+\lambda +4 \lambda ^2-3 \lambda ^3-2 \lambda ^4+\lambda ^5)\bigl]$
\end{tabular}  
\end{center}
\begin{center}
\begin{tabular}{c p{11cm}}
$V_{\yng(2)}^{U(N)}[7_{7}]=$ &$\frac{(-1+q \lambda )}{(-1+q)^2 q^5 (1+q) \lambda ^3} \bigl[(-1+\lambda )^2 \lambda +q^{12} (-1+\lambda )^3 \lambda ^2-q (-1+\lambda )^2 (1+2 \lambda )-2 q^{11} (-1+\lambda )^2 \lambda ^2 (-1-\lambda +\lambda ^2)+q^3 (-1+\lambda )^2 (1+7 \lambda +2 \lambda ^2)+q^5 (-1+\lambda )^2 (1-9 \lambda -8 \lambda ^2+2 \lambda ^3)+q^9 (-1+\lambda )^2 \lambda  (-1-8 \lambda -\lambda ^2+2 \lambda ^3)+q^2 (2-5 \lambda +\lambda ^2+4 \lambda ^3-2 \lambda ^4)-q^7 (-1+\lambda )^2 (1-5 \lambda -11 \lambda ^2+2 \lambda ^4)+q^8 \lambda  (-3+11 \lambda -16 \lambda ^3+8 \lambda ^4)+q^6 (2-18 \lambda ^2+15 \lambda ^3+5 \lambda ^4-4 \lambda ^5)+q^4 (-4+6 \lambda +8 \lambda ^2-12 \lambda ^3+\lambda ^4+\lambda ^5)+q^{10} \lambda  (1-2 \lambda -4 \lambda ^2+8 \lambda ^3-\lambda ^4-3 \lambda ^5+\lambda ^6)\bigl]$
\end{tabular}  
\end{center}
\begin{center}
\begin{tabular}{c p{11cm}}
$V_{\yng(2)}^{U(N)}[8_{1}]=$ &$\frac{1}{(-1+q)^2 q^6 (1+q) \lambda ^7}\bigl[q (-1+\lambda )^2-(-1+\lambda )^2 \lambda -q^4 (-1+\lambda )^2 \lambda +q^3 (-1+\lambda )^2 \lambda ^2+q^5 \lambda ^5 (-1+\lambda ^3)+q^{11} \lambda ^5 (-1+\lambda +\lambda ^3-\lambda ^4)-q^6 \lambda ^4 (-1+\lambda -\lambda ^2+\lambda ^4)+q^9 \lambda ^5 (-1+\lambda -\lambda ^2+\lambda ^4)+q^8 \lambda ^4 (-1-\lambda +2 \lambda ^4)+q^{10} \lambda ^4 (1-\lambda +2 \lambda ^2-\lambda ^3-\lambda ^4-\lambda ^5+\lambda ^6)+q^7 (\lambda ^4+\lambda ^6-\lambda ^8-\lambda ^9)\bigl]$
\end{tabular}  
\end{center}
\begin{center}
\begin{tabular}{c p{11cm}}
$V_{\yng(2)}^{U(N)}[9_{2}]=$ & $\frac{1}{(-1+q)^2 q^9 (1+q) \lambda ^{11}}\bigl[q (-1+\lambda )^2-(-1+\lambda )^2 \lambda -q^4 (-1+\lambda )^2 \lambda +q^3 (-1+\lambda )^2 \lambda ^2+q^{13} \lambda ^6 (-1+\lambda ^4)+q^{10} \lambda ^6 (-1-\lambda +2 \lambda ^4)+q^9 \lambda ^4 (-1+\lambda -\lambda ^2+\lambda ^5)-q^{12} \lambda ^5 (-1+\lambda -\lambda ^2+\lambda ^5)+q^7 \lambda ^5 (1-\lambda -\lambda ^4+\lambda ^5)-q^8 \lambda ^4 (1-2 \lambda +\lambda ^2-\lambda ^3-\lambda ^4+\lambda ^5+\lambda ^6)+q^{11} (\lambda ^5+\lambda ^7-\lambda ^9-\lambda ^{10})\bigl]$
\end{tabular}  
\end{center}
\begin{center}
\begin{tabular}{c p{11cm}}
$V_{\yng(2)}^{U(N)}[10_{2}]=$ &$\frac{1}{(-1+q)^2 q^8 (1+q) \lambda ^9}\bigl[q (-1+\lambda )^2-(-1+\lambda )^2 \lambda -q^4 (-1+\lambda )^2 \lambda +q^3 (-1+\lambda )^2 \lambda ^2+q^7 \lambda ^6 (-1+\lambda ^4)+q^{13} \lambda ^6 (-1+\lambda +\lambda ^4-\lambda ^5)-q^8 \lambda ^5 (-1+\lambda -\lambda ^2+\lambda ^5)+q^{11} \lambda ^6 (-1+\lambda -\lambda ^2+\lambda ^5)+q^{10} \lambda ^5 (-1-\lambda +2 \lambda ^5)-q^9 \lambda ^5 (-1-\lambda ^2+\lambda ^5+\lambda ^6)+q^{12} \lambda ^5 (1-\lambda +2 \lambda ^2-\lambda ^3-\lambda ^5-\lambda ^6+\lambda ^7)\bigl]$
\end{tabular}  
\end{center}
\end{enumerate}
\renewcommand{\baselinestretch}{1}\selectfont

There seems to be a symmetry transformation on the polynomial variables which gives the 
$U(N)$ invariants of knots carrying antisymmetric second rank tensor representation $R=\Yboxdim6pt\yng(1,1)$.
The symmetry relation for these non-torus knots (see also eqn.(7) in \cite{Itoyama:2012fq})
\begin{align}
\Yboxdim4pt
V_{\yng(2)}^{U(N)}[\mathcal{K}](q^{-1},\lambda)=V_{\yng(1,1)}^{U(N)}[\mathcal{K}](q,\lambda). \label{symm}
\end{align}
We  checked  our polynomials for knots $6_2,6_3,7_2,7_3$  with the results obtained in Ref.\cite{Itoyama:2012qt} using character expansion approach.
Before we use these polynomial invariants in verifying Ooguri-Vafa conjecture, we shall enumerate
the multi-colored link polynomials in the following appendix.
\section{Link Polynomials}
\label{sec:appenC}

In this appendix we list the $U(N)$ link invariant for non torus links
given in Figure \ref{figs:plat} for representations $R_1,R_2\in   \Big{\{}\Yboxdim6pt\yng(1), \Yboxdim6pt\yng(2), \Yboxdim6pt\yng(1,1)\Big{\}}$.

\renewcommand{\baselinestretch}{1.5}\selectfont
\begin{enumerate}
\item
For $R_1=\Yboxdim6pt\yng(1)$, $R_2=\Yboxdim6pt\yng(1)$    : 
\Yboxdim4pt
\begin{center}
\begin{tabular}{c p{11cm}}
$V_{(\yng(1),\,\yng(1))}^{U(N)}[6_{2}]=$ &$\frac{1}{(-1+q)^2 q \lambda }\bigl[\lambda(-1+\lambda ^2)+q^4 \lambda  (-1+\lambda ^2)+q(1+\lambda -2 \lambda ^3)+q^3 (1+\lambda -2 \lambda ^3)+q^2 (-1-2 \lambda +\lambda ^2+2 \lambda ^3)\bigl]$
\end{tabular}  
\end{center}
\begin{center}
\begin{tabular}{c p{11cm}}
$V_{(\yng(1),\,\yng(1))}^{U(N)}[6_{3}]=$ &$\frac{(-1+\lambda ) }{(-1+q)^2 q \lambda ^3}\bigl[\lambda +q^4 \lambda -q (1+3 \lambda +2 \lambda ^2)-q^3 (1+3 \lambda +2 \lambda ^2)+q^2 (2+4 \lambda +3 \lambda ^2+\lambda ^3)\bigl]$
\end{tabular}  
\end{center}
\begin{center}
\begin{tabular}{c p{11cm}}
$V_{(\yng(1),\,\yng(1))}^{U(N)}[7_{1}]=$ &$\frac{(-1+\lambda ) }{(-1+q)^2 q^2 \lambda ^2}\bigl[-\lambda -q^6 \lambda +q (1+\lambda )^2+q^5 (1+\lambda )^2-q^2 \left(2+3 \lambda +\lambda ^2\right)+q^3 \left(2+3 \lambda +\lambda ^2\right)-q^4 \left(2+3 \lambda +\lambda ^2\right)\bigl]$
\end{tabular}  
\end{center}
\begin{center}
\begin{tabular}{c p{11cm}}
$V_{(\yng(1),\,\yng(1))}^{U(N)}[7_{2}]=$ &$\frac{(-1+\lambda ) }{(-1+q)^2 q^2 \lambda ^2}\bigl(-\lambda -q^6 \lambda +q (1+\lambda )^2-2 q^2 (1+\lambda )^2-2 q^4 (1+\lambda )^2+q^5 (1+\lambda )^2+q^3 \left(2+5 \lambda +3 \lambda ^2\right)\bigl]$
\end{tabular}  
\end{center}
\begin{center}
\begin{tabular}{c p{11cm}}
$V_{(\yng(1),\,\yng(1))}^{U(N)}[7_{3}]=$ &$\frac{(-1+\lambda ) }{(-1+q)^2 q \lambda ^3}\bigl[-\lambda  (1+\lambda )-q^4 \lambda  (1+\lambda )+q (1+\lambda )^3+q^3 (1+\lambda )^3-q^2 \left(2+4 \lambda +5 \lambda ^2+\lambda ^3\right)\bigl]$
\end{tabular}  
\end{center}
\Yboxdim5pt
\item
For $R_1=\yng(1)$, $R_2=\yng(2)$    : 
\Yboxdim4pt
\begin{center}
\begin{tabular}{c p{11cm}}
$V_{(\yng(1),\,\yng(2))}^{U(N)}[6_{2}]=$ &$\frac{(-1+\lambda ) }{(-1+q)^3 \sqrt{q} (1+q) \lambda ^{3/2}}\bigl[q-\lambda -q^7 \lambda ^3+q^8 \lambda ^3+q^2 \lambda  (1+\lambda +\lambda ^2)-q^6 \lambda  (1+\lambda +\lambda ^2)+q^5 \lambda  (1+2 \lambda +2 \lambda ^2)-q^4 (-1+\lambda ^3)-q^3 (1+2 \lambda +\lambda ^2+\lambda ^3)\bigl]$
\end{tabular}  
\end{center}
\begin{center}
\begin{tabular}{c p{11cm}}
$V_{(\yng(1),\,\yng(2))}^{U(N)}[6_{3}]=$ &$\frac{(-1+\lambda ) }{(-1+q)^3 q^{5/2} (1+q) \lambda ^{7/2}}\bigl[-\lambda +q^7 \lambda ^2-q^2 (1+\lambda )+q^5 \lambda ^3 (1+\lambda )+q (1+\lambda )^2-q^6 \lambda  (1+\lambda +2 \lambda ^2)+q^4 (1+3 \lambda +2 \lambda ^2+\lambda ^3)-q^3 (1+2 \lambda +2 \lambda ^2+2 \lambda ^3)\bigl]$
\end{tabular}  
\end{center}
\begin{center}
\begin{tabular}{c p{11cm}}
$V_{(\yng(1),\,\yng(2))}^{U(N)}[7_{1}]=$ &$\frac{1}{(-1+q)^3 q^{5/2} (1+q) \lambda ^{5/2}}\bigl[q+q^2 (-1+\lambda )+(-1+\lambda ) \lambda -q^{10} (-1+\lambda ) \lambda ^2-q \lambda ^3-q^8 (-1+\lambda ) \lambda ^3+q^7 (1+\lambda -2 \lambda ^3)-2 q^4 (-1+\lambda ^3)+q^9 \lambda  (-1+\lambda ^3)+q^5 (-1+\lambda +\lambda ^3-\lambda ^4)+q^3 (-1-\lambda +\lambda ^2+\lambda ^4)+q^6 (-1-\lambda +\lambda ^3+\lambda ^4)\bigl]$
\end{tabular}  
\end{center}
\begin{center}
\begin{tabular}{c p{11cm}}
$V_{(\yng(1),\,\yng(2))}^{U(N)}[7_{2}]=$ &$-\frac{(-1+q \lambda ) }{(-1+q)^3 q^{7/2} (1+q) \lambda ^{5/2}}\bigl[(q+(-1+\lambda ) \lambda +q^9 (-1+\lambda ) \lambda -q \lambda ^2+q^5 (-1+3 \lambda +\lambda ^2-3 \lambda ^3)+q^8 (\lambda -\lambda ^3)-q^2 (1-2 \lambda +\lambda ^3)+q^4 (2+\lambda -4 \lambda ^2+\lambda ^3)+q^7 (1-2 \lambda ^2+\lambda ^3)+q^3 (-1-3 \lambda +3 \lambda ^2+\lambda ^3)+q^6 (-1-3 \lambda +3 \lambda ^2+\lambda ^3)\bigl]$
\end{tabular}  
\end{center}
\begin{center}
\begin{tabular}{c p{11cm}}
$V_{(\yng(1),\,\yng(2))}^{U(N)}[7_{3}]=$ &$\frac{1}{(-1+q)^3 q^{3/2} (1+q) \lambda ^{7/2}}\bigl[q^2 (-1+\lambda ^2)+\lambda  (-1+\lambda ^2)+q^4 (1+2 \lambda -3 \lambda ^4)+q^7 (\lambda ^2-\lambda ^4)-q (-1-\lambda +\lambda ^3+\lambda ^4)+q^6 \lambda  (-1-\lambda +\lambda ^3+\lambda ^4)+q^5 (\lambda ^3-\lambda ^5)+q^3 (-1-\lambda -2 \lambda ^2+2 \lambda ^3+\lambda ^4+\lambda ^5)\bigl]$
\end{tabular}  
\end{center}

\renewcommand{\baselinestretch}{1}\selectfont
Interchanging $R_1,R_2$ on the two components of the link,  gives the same polynomial:
\begin{equation}
V_{(\Yboxdim4pt\yng(1),\Yboxdim4pt\yng(2))}^{U(N)}[\mathcal{L}]=V_{(\Yboxdim4pt\yng(2),\Yboxdim4pt\yng(1))}^{U(N)}[\mathcal{L}]~,
\end{equation}
and replacing the second rank symmetric representation $\yng(2)$ by antisymmetric 
representation $\yng(1,1)$, the link polynomials are related as follows:

\begin{align}
\Yvcentermath1
V_{(\yng(1),\yng(2))}^{U(N)}[\mathcal{L}](q^{-1},\lambda)=-V_{\left(\yng(1),\yng(1,1)\right)}^{U(N)}[\mathcal{L}](q,\lambda)
\end{align}

\renewcommand{\baselinestretch}{1.5}\selectfont
\item
For $R_1=\Yboxdim6pt\yng(2)$, $R_2=\Yboxdim6pt\yng(2)$   : 
\Yboxdim4pt
\begin{center}
\begin{tabular}{c p{11cm}}
$V_{(\yng(2),\,\yng(2))}^{U(N)}[6_{2}]=$ &$\frac{1}{(-1+q)^4 q (1+q)^2 \lambda ^2}\big[-q^2 (-1+\lambda )^2-(-1+\lambda )^2 \lambda +q (-1+\lambda )^2 (1+\lambda )+q^{15} \lambda ^4 (-1+\lambda ^2)-q^{13} \lambda ^4 (-2+\lambda +\lambda ^2)+q^3 (-1+\lambda )^2 \lambda  (-2+\lambda +\lambda ^2+\lambda ^3)+q^{12} \lambda ^3 (-2-\lambda -2 \lambda ^2+5 \lambda ^3)+q^{11} \lambda ^2 (-1+\lambda -2 \lambda ^2+4 \lambda ^3-2 \lambda ^4)+q^{10} \lambda  (-1+2 \lambda +\lambda ^2+3 \lambda ^3-5 \lambda ^5)+q^8 \lambda  (1-4 \lambda +\lambda ^2-3 \lambda ^3+4 \lambda ^4+\lambda ^5)+q^5 (-1+4 \lambda -4 \lambda ^2-\lambda ^4+2 \lambda ^5)+q^9 \lambda  (1+2 \lambda -3 \lambda ^2-5 \lambda ^4+5 \lambda ^5)+q^7 (1-4 \lambda +\lambda ^2+2 \lambda ^3+3 \lambda ^4+\lambda ^5-4 \lambda ^6)+q^{14} (\lambda ^3+\lambda ^5-2 \lambda ^6)+2 q^4 (1-2 \lambda +\lambda ^3+\lambda ^5-\lambda ^6)+q^6 (-1+\lambda +5 \lambda ^2-4 \lambda ^3+\lambda ^4-5 \lambda ^5+3 \lambda ^6)\bigl]$
\end{tabular}  
\end{center}
\begin{center}
\begin{tabular}{c p{11cm}}
$V_{(\yng(2),\,\yng(2))}^{U(N)}[6_{3}]=$ &$\frac{(-1+\lambda ) (-1+q \lambda )}{(-1+q)^4 q^7 (1+q)^2 \lambda ^6}\bigl[(-1+\lambda ) \lambda +q^{12} (-1+\lambda )^2 \lambda ^2+q (1+\lambda -3 \lambda ^2)+q^2 (-2+4 \lambda +\lambda ^2-2 \lambda ^3)+q^{11} \lambda ^2 (-3+2 \lambda +3 \lambda ^2-2 \lambda ^3)+q^3 (-1-5 \lambda +9 \lambda ^2+3 \lambda ^3)+q^8 \lambda  (2-13 \lambda -5 \lambda ^2+10 \lambda ^3)+q^6 (-2+\lambda +19 \lambda ^2-2 \lambda ^3-6 \lambda ^4)+q^4 (4-5 \lambda -11 \lambda ^2+5 \lambda ^3+\lambda ^4)+q^5 (-1+9 \lambda -8 \lambda ^2-12 \lambda ^3+2 \lambda ^4)+q^9 \lambda  (2+7 \lambda -10 \lambda ^2-3 \lambda ^3+3 \lambda ^4)+q^{10} \lambda  (-1+2 \lambda +6 \lambda ^2-6 \lambda ^3-\lambda ^4+\lambda ^5)-q^7 (-1+7 \lambda +2 \lambda ^2-17 \lambda ^3+\lambda ^4+2 \lambda ^5)\bigl]$
\end{tabular}  
\end{center}
\begin{center}
\begin{tabular}{c p{11cm}}
$V_{(\yng(2),\,\yng(2))}^{U(N)}[7_{1}]=$ &$\frac{(-1+\lambda ) (-1+q \lambda ) }{(-1+q)^4 q^6 (1+q)^2 \lambda ^4}\bigl[(-1+\lambda ) \lambda -q^{18} (-1+\lambda ) \lambda ^2+q (1+\lambda -2 \lambda ^2)+q^{17} \lambda ^2 (-2+\lambda ^2)+q^3 \lambda  (-4+3 \lambda +\lambda ^2)+q^2 (-2+2 \lambda +\lambda ^2-\lambda ^3)+q^7 (5+5 \lambda -7 \lambda ^2+\lambda ^3)+q^4 (3+2 \lambda -5 \lambda ^2+\lambda ^3)-q^{15} \lambda  (-1-6 \lambda +2 \lambda ^2+\lambda ^3)+q^{14} \lambda  (3-5 \lambda -4 \lambda ^2+2 \lambda ^3)+q^{13} (1-5 \lambda -7 \lambda ^2+5 \lambda ^3)+q^{12} (-2-3 \lambda +11 \lambda ^2+2 \lambda ^3-2 \lambda ^4)+q^6 (1-8 \lambda +3 \lambda ^2+\lambda ^3-\lambda ^4)+q^{11} (-1+10 \lambda +2 \lambda ^2-6 \lambda ^3+\lambda ^4)-q^{16} (\lambda -3 \lambda ^3+\lambda ^4)-q^9 (2+11 \lambda -5 \lambda ^2-3 \lambda ^3+\lambda ^4)+q^5 (-4+3 \lambda +2 \lambda ^2-3 \lambda ^3+\lambda ^4)+q^8 (-5+7 \lambda +6 \lambda ^2-3 \lambda ^3+\lambda ^4)+q^{10} (5-\lambda -12 \lambda ^2+\lambda ^3+\lambda ^4)\bigl]$
\end{tabular}  
\end{center}
\begin{center}
\begin{tabular}{c p{11cm}}
$V_{(\yng(2),\,\yng(2))}^{U(N)}[7_{2}]=$ &$\frac{(-1+\lambda ) (-1+q \lambda ) }{(-1+q)^4 q^8 (1+q)^2 \lambda ^4}\bigl[((-1+\lambda ) \lambda -q^{18} (-1+\lambda ) \lambda ^2+q (1+\lambda -2 \lambda ^2)+q^{17} \lambda ^2 (-2+\lambda +\lambda ^2)+q^{15} (\lambda +4 \lambda ^2-6 \lambda ^3)+q^{16} \lambda  (-1+\lambda +2 \lambda ^2-2 \lambda ^3)+q^2 (-2+3 \lambda +\lambda ^2-\lambda ^3)+q^3 (-1-6 \lambda +5 \lambda ^2+\lambda ^3)+q^4(5-2 \lambda -9 \lambda ^2+2 \lambda ^3)+q^{14} \lambda  (2-8 \lambda +2 \lambda ^2+5 \lambda ^3)+q^{11} \lambda  (6-19 \lambda -4 \lambda ^2+9 \lambda ^3)+q^7 (5-13 \lambda -12 \lambda ^2+12 \lambda ^3)+q^9 (-4+3 \lambda +25 \lambda ^2-9 \lambda ^3-5 \lambda ^4)+q^{13} (1-5 \lambda +3 \lambda ^2+10 \lambda ^3-5 \lambda ^4)+q^{12} (-2+3 \lambda +12 \lambda ^2-14 \lambda ^3-3 \lambda ^4)+q^{10} (3-12 \lambda +20 \lambda ^3-3 \lambda ^4)+q^6 (-5-5 \lambda +19 \lambda ^2+\lambda ^3-2 \lambda ^4)+q^5 (-2+13 \lambda -2 \lambda ^2-6 \lambda ^3+\lambda ^4)+q^8 (1+13 \lambda -17 \lambda ^2-12 \lambda ^3+5 \lambda ^4)\bigl]$
\end{tabular}  
\end{center}
\begin{center}
\begin{tabular}{c p{11cm}}
$V_{(\yng(2),\,\yng(2))}^{U(N)}[7_{3}]=$ &$\frac{(-1+\lambda ) (-1+q \lambda )}{(-1+q)^4 q^6 (1+q)^2 \lambda ^6}\bigl[(-1+\lambda ) \lambda -q^{12} \lambda ^3 (4-4 \lambda -3 \lambda ^2+\lambda ^3)+q (1+\lambda -3 \lambda ^2+\lambda ^3)+q^{13} \lambda ^3 (-1-3 \lambda +\lambda ^2+\lambda ^3)+q^2 (-2+4 \lambda -3 \lambda ^3+\lambda ^4)-q^{11} \lambda ^2 (2-8 \lambda -8 \lambda ^2+5 \lambda ^3+\lambda ^4)-q^3 (1+5 \lambda -9 \lambda ^2+3 \lambda ^4)+q^9 \lambda  (2+3 \lambda -20 \lambda ^2-4 \lambda ^3+7 \lambda ^4)+q^{14} (\lambda ^3-\lambda ^5)+q^4 (4-5 \lambda -8 \lambda ^2+11 \lambda ^3+\lambda ^4-\lambda ^5)-q^8 \lambda  (-2+13 \lambda -3 \lambda ^2-22 \lambda ^3+\lambda ^4+\lambda ^5)+q^5 (-1+9 \lambda -10 \lambda ^2-10 \lambda ^3+9 \lambda ^4+\lambda ^5)+q^6 (-2+\lambda +16 \lambda ^2-13 \lambda ^3-12 \lambda ^4+2 \lambda ^5)+q^{10} \lambda  (-1+4 \lambda +5 \lambda ^2-15 \lambda ^3-3 \lambda ^4+2 \lambda ^5)+q^7 (1-7 \lambda +3 \lambda ^2+22 \lambda ^3-7 \lambda ^4-5 \lambda ^5+\lambda ^6)\bigl]$
\end{tabular}  
\end{center}
\end{enumerate}
\renewcommand{\baselinestretch}{1}\selectfont

Changing both the rank two symmetric representation $\Yboxdim6pt\yng(2)$ by 
antisymmetric representation $\Yboxdim6pt\yng(1,1)$, we find the following relation 
between the link polynomials:
\Yboxdim4pt
\begin{align}
\Yvcentermath1
V_{\left(\yng(2),\yng(2)\right)}^{U(N)}[\mathcal{L}](q,\lambda)=V_{\left(\yng(1,1),\yng(1,1)\right)}^{U(N)}[\mathcal{L}](q^{-1},\lambda).
\end{align} 

With these polynomial invariants available for the non-torus knots and links in Figures \ref{fig:plat}, \ref{figs:plat},
we are in a position to verify Ooguri-Vafa \cite{Ooguri:1999bv} and Labastida-Marino-Vafa 
\cite{Labastida:2000yw} conjectures.

\section{Reformulated link invariants}
\label{sec:appenD}

In this appendix we explicitly write  the reformulated link invariant for the  non torus
knots and links in Figure \ref{fig:plat} and Figure \ref{figs:plat}.
Rewriting the most general form of reformulated invariants $f_R[\mathcal{K}]$   and $f_{R_1,R_2}[{\cal L}]$ (see eqn.(\ref{findd})) 
for representations $\Yboxdim6pt\yng(1)$ and $\Yboxdim6pt\yng(2)$ on the component
knots, the expression  for knots are
\Yboxdim4pt
\begin{eqnarray}
f_{\yng(1)}[\mathcal{K}] & = & V_{\yng(1)}[\mathcal{K}]\label{eqn:fsingle}\\
f_{\yng(2)}[\mathcal{K}] & = & V_{\yng(2)}[\mathcal{K}]-\frac{1}{2}\left(V_{\yng(1)}[\mathcal{K}]^{2}+V_{\yng(1)}^{(2)}[\mathcal{K}]\right)\label{eqn:ftwohor}\\
f_{\yng(1,1)}[\mathcal{K}] & = & V_{\yng(1,1)}[\mathcal{K}]-\frac{1}{2}\left(V_{\yng(1)}[\mathcal{K}]^{2}-V_{\yng(1)}^{(2)}[\mathcal{K}]\right)\label{eqn:ftwover},
\end{eqnarray}
where we have suppressed $U(N)$ superscript on the knot invariants $(V_R[\mathcal{K}]\equiv
V_R^{U(N)}[\mathcal{K}](q,\lambda))$. Further,  we use the notation $V_{R}^{(n)}[\mathcal{K}]\equiv V_{R}[\mathcal{K}](q^n,\lambda^n)$. 

Ooguri-Vafa conjecture \cite{Ooguri:1999bv} states  that the  reformulated invariants for knots should have the following structure

\begin{align}
f_{R}(q,\lambda)=\sum_{s,Q}\frac{N_{Q,R,s}}{q^{1/2}-q^{-1/2}}a^{Q}q^{s},
\label{OVconj}
\end{align}
where $N_{Q,R,s}$ are integer and $Q$ and $s$ are, in general,
half integers. Clearly for$ \Yboxdim5ptR=\yng(1)$ (fundamental representation), the polynomial structure in \ref{sec:appenB} for 
$V_{\yng(1)}^{U(N)}[\mathcal{K}]$ (\ref{eqn:fsingle}) obeys eqn.(\ref{OVconj}). We will verify for $R=\Yboxdim6pt\yng(2)$ and $\Yboxdim6pt\yng(1,1)$ in 
the following subsection. 

The reformulated invariants in terms of two-component link invariants 
has the following form for $R_1,R_2 \in \Big{\{}\Yboxdim6pt\yng(1),\Yboxdim6pt\yng(2), \Yboxdim6pt\yng(1,1)\Big{\}}$:
\Yboxdim4pt
\begin{eqnarray}
f_{(\yng(1),\,\yng(1))}[\mathcal{L}] & = & V_{(\yng(1),\yng(1))}[\mathcal{L}]-V_{\yng(1)}[\mathcal{K}_{1}]V_{\yng(1)}[\mathcal{K}_{2}]\\
f_{(\yng(1),\,\yng(2))}[\mathcal{L}] & = & V_{(\yng(1),\yng(2))}[\mathcal{L}]-V_{(\yng(1),\yng(1))}[\mathcal{L}]V_{\yng(1)}[\mathcal{K}_{2}]-V_{\yng(1)}[\mathcal{K}_{1}]V_{\yng(2)}[\mathcal{K}_{2}]\nonumber \\
 &  & +V_{\yng(1)}[\mathcal{K}_{1}]V_{\yng(1)}[\mathcal{K}_{2}]^{2}\label{eqn:fonetwohor}\\
f_{\left(\yng(1),\,\yng(1,1)\right)}[\mathcal{L}] & = & V_{\left(\yng(1),\yng(1,1)\right)}[\mathcal{L}]-V_{(\yng(1),\yng(1))}[\mathcal{L}]V_{\yng(1)}[\mathcal{K}_{2}]-V_{\yng(1)}[\mathcal{K}_{1}]V_{\yng(1,1)}[\mathcal{K}_{2}]\nonumber \\
 &  & +V_{\yng(1)}[\mathcal{K}_{1}]V_{\yng(1)}[\mathcal{K}_{2}]^{2}\label{eqn:fonetwover}\\
f_{(\yng(2),\,\yng(1))}[\mathcal{L}] & = & V_{(\yng(2),\yng(1))}[\mathcal{L}]-V_{(\yng(1),\yng(1))}[\mathcal{L}]V_{\yng(1)}[\mathcal{K}_{1}]-V_{\yng(2)}[\mathcal{K}_{1}]V_{\yng(1)}[\mathcal{K}_{2}]\nonumber \\
 &  & +V_{\yng(1)}[\mathcal{K}_{1}]^{2}V_{\yng(1)}[\mathcal{K}_{2}]\\
f_{\left(\yng(1,1),\,\yng(1)\right)}[\mathcal{L}] & = & V_{\left(\yng(1,1),\yng(1)\right)}[\mathcal{L}]-V_{(\yng(1),\yng(1))}[\mathcal{L}]V_{\yng(1)}[\mathcal{K}_{1}]-V_{\yng(1,1)}[\mathcal{K}_{1}]V_{\yng(1)}[\mathcal{K}_{2}]\nonumber \\
 &  & +V_{\yng(1)}[\mathcal{K}_{1}]^{2}V_{\yng(1)}[\mathcal{K}_{2}]
\end{eqnarray}
\begin{eqnarray}
f_{(\yng(2),\,\yng(2))}[\mathcal{L}] & = & V_{(\yng(2),\yng(2))}[\mathcal{L}]-V_{\yng(2)}(\mathcal{K}_{1})V_{\yng(2)}[\mathcal{K}_{2}]-V_{(\yng(2),\yng(1))}[\mathcal{L}]V_{\yng(1)}[\mathcal{K}_{2}]\nonumber \\
 &  & -V_{(\yng(1),\yng(2))}[\mathcal{L}]V_{\yng(1)}[\mathcal{K}_{1}]-\frac{1}{2}V_{(\yng(1),\yng(1))}[\mathcal{L}]^{2}+2V_{(\yng(1),\yng(1))}[\mathcal{L}]V_{\yng(1)}[\mathcal{K}_{1}]V_{\yng(1)}[\mathcal{K}_{2}]\nonumber \\
 &  & +V_{\yng(1)}[\mathcal{K}_{1}]^{2}V_{\yng(2)}[\mathcal{K}_{2}]+V_{\yng(2)}[\mathcal{K}_{1}]V_{\yng(1)}[\mathcal{K}_{2}]^{2}-\frac{3}{2}V_{(\yng(1))}[\mathcal{K}_{1}]^{2}V_{(\yng(1))}[\mathcal{K}_{2}]^{2}\nonumber \\
 &  & -\frac{1}{2}V_{(\yng(1),\yng(1))}^{(2)}[\mathcal{L}]+\frac{1}{2}V_{\yng(1)}^{(2)}[\mathcal{K}_{1}]V_{\yng(1)}^{(2)}[\mathcal{K}_{2}]
\end{eqnarray}
\begin{eqnarray}
f_{\left(\yng(1,1),\,\yng(1,1)\right)}[\mathcal{L}] & = & V_{\left(\yng(1,1),\yng(1,1)\right)}[\mathcal{L}]-V_{\yng(1,1)}[\mathcal{K}_{1}]V_{\yng(1,1)}[\mathcal{K}_{2}]-V_{\left(\yng(1,1),\yng(1)\right)}[\mathcal{L}]V_{\yng(1)}[\mathcal{K}_{2}]\nonumber \\
 &  & -V_{\left(\yng(1),\yng(1,1)\right)}[\mathcal{L}]V_{\yng(1)}[\mathcal{K}_{1}]-\frac{1}{2}V_{(\yng(1),\yng(1))}[\mathcal{L}]^{2}+2V_{(\yng(1),\yng(1))}[\mathcal{L}]V_{\yng(1)}[\mathcal{K}_{1}]V_{\yng(1)}[\mathcal{K}_{2}]\nonumber \\
 &  & +V_{\yng(1)}[\mathcal{K}_{1}]^{2}V_{\yng(1,1)}[\mathcal{K}_{2}]+V_{\yng(1,1)}[\mathcal{K}_{1}]V_{\yng(1)}[\mathcal{K}_{2}]^{2}-\frac{3}{2}V_{(\yng(1))}[\mathcal{K}_{1}]^{2}V_{(\yng(1))}[\mathcal{K}_{2}]^{2}\nonumber \\
 &  & -\frac{1}{2}V_{(\yng(1),\yng(1))}^{(2)}[\mathcal{L}]+\frac{1}{2}V_{\yng(1)}^{(2)}[\mathcal{K}_{1}]V_{\yng(1)}^{(2)}[\mathcal{K}_{2}]
\end{eqnarray}

Here the components knots $\mathcal{K}_1$ and $\mathcal{K}_2$ are unknots for the non-torus links in Figure \ref{figs:plat}.
The generalisation of Ooguri-Vafa  conjecture for links was proposed in \cite{Labastida:2000yw}  which 
states that reformulated invariants for  $r$-component link should have the following structure
  \begin{align}
f_{(R_1, R_2, \ldots R_r)} (q, \lambda)= (q^{1/2}- q^{-1/2})^{r-2}
\sum_{Q,s} N_{(R_1, \ldots R_r),Q,s} q^s \lambda^Q~, \label{reformulated}
\end{align}
where $N_{(R_1,\cdots R_r),Q,s}$ are integer and $Q$ and $s$ are half integers.

We can see below that all the reformulated invariants we calculate indeed satisfy the conjecture.
\subsection{Reformulated invariant for knots }

We have already seen in appendix \ref{sec:appenB}}, $V_{\Yboxdim4pt\yng(1)}[\mathcal{K}]$ has the Ooguri-Vafa form given in eqn.(\ref{OVconj}). 

For the symmetric second rank tensor $R=\Yboxdim6pt\yng(2)$ placed on the knot, $f_{\Yboxdim4pt\yng(2)}[\mathcal{K}]$ are: 
\renewcommand{\baselinestretch}{1.5}\selectfont
\Yboxdim4pt
\begin{center}
\begin{tabular}{c p{11cm}}
$f_{\yng(2)}[4_{1}]=$ &$\frac{(-1+\lambda )^2 }{(-1+q) q^2 \lambda ^3}\left[-q+\lambda -q^5 \lambda ^3+q^4 \lambda ^4+q^3 \lambda  (1+\lambda )-q^2 \lambda ^2 (1+\lambda )\right]$
\end{tabular}  
\end{center}
\begin{center}
\begin{tabular}{c p{11cm}}
$f_{\yng(2)}[5_{2}]=$ &$-\frac{(1-q+q^2) (-1+\lambda )^2 }{(-1+q) q^5 \lambda ^7}\bigl[q (-1+\lambda )+q^2 (-1+\lambda )+\lambda +q^4 \lambda (1+\lambda +\lambda ^2)-q^3 (1+\lambda ^2+\lambda ^3+\lambda ^4)\bigl]$
\end{tabular}  
\end{center}
\begin{center}
\begin{tabular}{c p{11cm}}
$f_{\yng(2)}[6_{1}]=$ &$-\frac{(-1+\lambda )^2 }{(-1+q) q^4 \lambda ^5}\bigl[q-\lambda -q^2 \lambda +q^7 \lambda ^4 (1+\lambda )+q^3 (1+\lambda ^2)-q^5 \lambda  (1+\lambda +\lambda ^2)^2+q^4 \lambda(-1+\lambda +2 \lambda ^2+2 \lambda ^3+\lambda ^4)+q^6 (\lambda ^2+\lambda ^3+\lambda ^4-\lambda ^5-\lambda ^6\bigl]$
\end{tabular}  
\end{center}
\begin{center}
\begin{tabular}{c p{11cm}}
$f_{\yng(2)}[6_{2}]=$ &$-\frac{(1-q+q^2) (-1+\lambda )^2 }{(-1+q) q^6 \lambda ^5}\bigl[-\lambda -q^5 \lambda +q^2 (-1+\lambda ) \lambda +q^9 \lambda ^3-q^8 (-1+\lambda ) \lambda ^3+q^6 \lambda ^2 (1+\lambda )+q (1+\lambda ^2)-q^3 \lambda  (2+\lambda ^2)+q^4 (1+2 \lambda ^2)-q^7 \lambda  (1+\lambda +\lambda ^3)\bigl]$  
\end{tabular}
\end{center}
\begin{center}
\begin{tabular}{c p{11cm}}
$f_{\yng(2)}[6_{3}]=$ &$-\frac{(1-q+q^2) (-1+\lambda )^2 }{(-1+q) q^5 \lambda ^3}\bigl[-\lambda +q^2 \lambda ^2-q^7 \lambda ^2+q^9 \lambda ^3+q (1+\lambda ^2)+q^3 (-1-\lambda +\lambda ^2)+q^6 \lambda ^2 (-1+\lambda +\lambda ^2)-q^4 \lambda  (2+2 \lambda +2 \lambda ^2+\lambda ^3)+q^5 (1+2 \lambda +2 \lambda ^2+2 \lambda ^3)-q^8 (\lambda ^2+\lambda ^4)\bigl]$  
\end{tabular}
\end{center}
\begin{center}
\begin{tabular}{c p{11cm}}
$f_{\yng(2)}[7_{2}]=$ &$-\frac{(-1+\lambda )^2 }{(-1+q) q^7 \lambda ^9}\bigl[q-\lambda -q^2 \lambda -2 q^4 \lambda +q^3 (1+\lambda ^2)-q^8 \lambda  (1+\lambda +2 \lambda ^2+2 \lambda ^3+\lambda ^4)-q^6 \lambda  (2+\lambda +3 \lambda ^2+3 \lambda ^3+3 \lambda ^4+\lambda ^5)+q^5 (1+\lambda ^2+\lambda ^4+\lambda ^5+\lambda ^6)+q^7 (1+\lambda +3 \lambda ^2+3 \lambda ^3+4 \lambda ^4+3 \lambda ^5+\lambda ^6)\bigl]$  
\end{tabular}
\end{center}
\begin{center}
\begin{tabular}{c p{11cm}}
$f_{\yng(2)}[7_{3}]=$ &$\frac{(-1+\lambda )^2 \lambda ^3 }{(-1+q) q^2}\bigl[-2 q^{10} \lambda ^3-q^{12} \lambda ^3+q^{11} \lambda ^4-\lambda  (1+\lambda +\lambda ^2)+q^6 \lambda  (-1-\lambda -5 \lambda ^2+\lambda ^3)+q^4 \lambda  (-1-\lambda -4 \lambda ^2+\lambda ^3)-2 q^2 (1+2 \lambda +2 \lambda ^2+2 \lambda ^3)+q^5 \lambda  (1+2 \lambda -\lambda ^2+3 \lambda ^3)+q^8 (-1-\lambda -\lambda ^2-3 \lambda ^3+\lambda ^4)+q (1+3 \lambda +3 \lambda ^2+\lambda ^3+\lambda ^4)+q^3 (2+3 \lambda +4 \lambda ^2+2 \lambda ^4)+q^9 (1+\lambda +\lambda ^2-\lambda ^3+2 \lambda ^4)+q^7 (1+2 \lambda +2 \lambda ^2-\lambda ^3+2 \lambda ^4)\bigl]$  
\end{tabular}
\end{center}
\begin{center}
\begin{tabular}{c p{11cm}}
$f_{\yng(2)}[7_{4}]=$ &$\frac{(-1+\lambda )^2 \lambda  }{(-1+q) q}\bigl[-q^9 \lambda ^5+\lambda ^2 (1+\lambda )-q^7 \lambda ^3 (1+\lambda )+q^8 \lambda ^4 (-1+\lambda ^2)+q^6 \lambda ^2(1+\lambda +\lambda ^2+\lambda ^3+\lambda ^4)-q \lambda (2+4 \lambda +4 \lambda ^2+3 \lambda ^3+\lambda ^4)+q^5(\lambda +\lambda ^2-2 \lambda ^4-\lambda ^5)-q^3 (2+5 \lambda +5 \lambda ^2+5 \lambda ^3+4 \lambda ^4+\lambda ^5)+q^4 (1+\lambda +2 \lambda ^2+\lambda ^3-\lambda ^4+\lambda ^5+\lambda ^6)+q^2(1+5 \lambda +8 \lambda ^2+7 \lambda ^3+3 \lambda ^4+2 \lambda ^5+\lambda ^6)\bigl]$  
\end{tabular}
\end{center}
\begin{center}
\begin{tabular}{c p{11cm}}
$f_{\yng(2)}[7_{5}]=$ &$\frac{(1-q+q^2) (-1+\lambda )^2 }{(-1+q) q^9 \lambda ^9}\bigl[-q+\lambda -q^7 (-1+\lambda ) \lambda ^3+q^8 \lambda  (1+\lambda )^3+q^{10} \lambda  (1+2 \lambda +2 \lambda ^2)+q^4 (-1+\lambda +\lambda ^2+2 \lambda ^3)+q^6 (\lambda ^3-2 \lambda ^4)+q^3 (\lambda -\lambda ^4)-q^5 (1+\lambda +\lambda ^2+\lambda ^4)-q^9 (1+2 \lambda +3 \lambda ^2+3 \lambda ^3+2 \lambda ^4)\bigl]$  
\end{tabular}
\end{center}
\begin{center}
\begin{tabular}{c p{11cm}}
$f_{\yng(2)}[7_{6}]=$ &$-\frac{(-1+\lambda )^2}{(-1+q) q^6 \lambda ^7}\bigl[-(-1+\lambda ) \lambda ^2+q^{11} \lambda ^5-q^{10} \lambda ^5 (1+\lambda )+q^9 \lambda ^3 (-2-2 \lambda -\lambda ^2+\lambda ^3)+q \lambda  (-2+\lambda +\lambda ^2+\lambda ^3)-q^3 \lambda (1-3 \lambda +\lambda ^2+\lambda ^4)-q^2(-1+\lambda ^2+3 \lambda ^3+\lambda ^4)+q^8 \lambda ^2 (2+4 \lambda +5 \lambda ^2+4 \lambda ^3+\lambda ^4)-q^5 \lambda (3+4 \lambda +8 \lambda ^2+6 \lambda ^3+2 \lambda ^4)-q^7 \lambda (2+2 \lambda +7 \lambda ^2+6 \lambda ^3+4 \lambda ^4+2 \lambda ^5)+q^4 (1-\lambda +4 \lambda ^2+3 \lambda ^3+5 \lambda ^4+2 \lambda ^5)+q^6 (1+\lambda +6 \lambda ^2+6 \lambda ^3+7 \lambda ^4+2 \lambda ^5+\lambda ^6)\bigl]$  
\end{tabular}
\end{center}
\begin{center}
\begin{tabular}{c p{11cm}}
$f_{\yng(2)}[7_{7}]=$ &$\frac{(-1+\lambda )^2 }{(-1+q) q^5 \lambda ^3}\bigl[\lambda +2 q^3 \lambda +q^{11} (-1+\lambda ) \lambda ^3-q (1+\lambda )^2+q^2 (2+\lambda)-q^8 \lambda ^2 (-1+\lambda -2 \lambda ^2+\lambda ^3)+q^9 \lambda ^2 (-1-2 \lambda -\lambda ^2-\lambda ^3+\lambda ^4)-q^5 \lambda  (7+12 \lambda +12 \lambda ^2+5 \lambda ^3+2 \lambda ^4)+q^4 (-2+5 \lambda ^2+5 \lambda ^3+4 \lambda ^4)+q^{10} (\lambda ^2+\lambda ^3+2 \lambda ^4-2 \lambda ^5)+q^6 (2+8 \lambda +11 \lambda ^2+9 \lambda ^3+4 \lambda ^4-\lambda ^5)+q^7 (-1-3 \lambda -4 \lambda ^2-3 \lambda ^3+3 \lambda ^4-\lambda ^5+\lambda ^6)\bigl]$  
\end{tabular}
\end{center}
\begin{center}
\begin{tabular}{c p{11cm}}
$f_{\yng(2)}[8_{1}]=$ &$-\frac{(-1+\lambda )^2 }{(-1+q) q^6 \lambda ^7}\bigl[q-\lambda -q^2 \lambda -2 q^4 \lambda +q^3 (1+\lambda ^2)+q^5 (1+\lambda ^2)+q^9 \lambda ^5 (1+\lambda +\lambda ^2)-q^8 \lambda ^2 (-1-2 \lambda -2 \lambda ^2-2 \lambda ^3+\lambda ^5+\lambda ^6)+q^6 \lambda  (-1+\lambda +2 \lambda ^2+3 \lambda ^3+3 \lambda ^4+2 \lambda ^5+\lambda ^6)-q^7 \lambda  (1+2 \lambda +4 \lambda ^2+5 \lambda ^3+4 \lambda ^4+3 \lambda ^5+\lambda ^6)\bigl]$  
\end{tabular}
\end{center}
\begin{center}
\begin{tabular}{c p{11cm}}
$f_{\yng(2)}[9_{2}]=$ &$-\frac{(-1+\lambda )^2}{(-1+q) q^9 \lambda ^{11}} \bigl[q-\lambda -q^2 \lambda -2 q^4 \lambda -2 q^6 \lambda +q^3 (1+\lambda ^2)+q^5(1+\lambda ^2)-q^{10} \lambda  (1+\lambda +2 \lambda ^2+3 \lambda ^3+3 \lambda ^4+2 \lambda ^5+\lambda ^6)-q^8 \lambda  (2+\lambda +2 \lambda ^2+4 \lambda ^3+4 \lambda ^4+4 \lambda ^5+3 \lambda ^6+\lambda ^7)+q^7 (1+\lambda ^2+\lambda ^5+\lambda ^6+\lambda ^7+\lambda ^8)+q^9(1+\lambda +3 \lambda ^2+4 \lambda ^3+5 \lambda ^4+6 \lambda ^5+5 \lambda ^6+3 \lambda ^7+\lambda ^8)\bigl]$  
\end{tabular}
\end{center}
\begin{center}
\begin{tabular}{c p{11cm}}
$f_{\yng(2)}[10_{1}]=$ &$-\frac{(-1+\lambda )^2}{(-1+q) q^8 \lambda ^9} \bigl[q-\lambda -q^2 \lambda -2 q^4 \lambda -2 q^6 \lambda +q^3 (1+\lambda ^2)+q^5 (1+\lambda ^2)+q^7 (1+\lambda ^2)+q^{11} \lambda ^6 (1+\lambda +\lambda ^2+\lambda ^3)+q^{10} \lambda ^2 (1+2 \lambda +3 \lambda ^2+3 \lambda ^3+3 \lambda ^4+\lambda ^5-\lambda ^7-\lambda ^8)+q^8 \lambda  (-1+\lambda +2 \lambda ^2+3 \lambda ^3+4 \lambda ^4+4 \lambda ^5+3 \lambda ^6+2 \lambda ^7+\lambda ^8)-q^9 \lambda  (1+2 \lambda +4 \lambda ^2+6 \lambda ^3+7 \lambda ^4+6 \lambda ^5+5 \lambda ^6+3 \lambda ^7+\lambda ^8\bigl]$  
\end{tabular}
\end{center}

\renewcommand{\baselinestretch}{1}\selectfont

Changing the symmetric representation by antisymmetry representation, we find  the
following relation between the reformulated invariants:

\begin{align}
\Yboxdim4pt
f_{\yng(2)}[\mathcal{K}](q^{-1},\lambda)=f_{\yng(1,1)}[\mathcal{K}](q,\lambda).
\end{align}

\subsection{Reformulated invariant for links}

\begin{enumerate}

\item
For $R_1=\Yboxdim6pt\yng(1)$, $R_2=\Yboxdim6pt\yng(1)$    : 

\begin{center}
\begin{tabular}{c p{11cm}}
\Yboxdim4pt
$f_{(\yng(1),\yng(1))}[6_{2}]=$ &$\frac{(-1+\lambda ) }{q \lambda }\bigl[-q+\lambda +q^2 \lambda +\lambda ^2+q^2 \lambda ^2\bigl]$  
\end{tabular}
\end{center}

\begin{center}
\begin{tabular}{c p{11cm}}
$f_{(\yng(1),\yng(1))}[6_{3}]=$ &$\frac{(-1+\lambda ) }{q \lambda ^3}\bigl[\lambda +q^2 \lambda -q \left(1+\lambda +2 \lambda ^2\right)\bigl]$  
\end{tabular}
\end{center}

\begin{center}
\begin{tabular}{c p{11cm}}
$f_{(\yng(1),\yng(1))}[7_{1}]=$ &$\frac{(-1+\lambda ) }{q^2 \lambda ^2}\bigl[-\lambda -q^4 \lambda +q^2 (-2+\lambda ) \lambda +q (1+\lambda ^2)+q^3 (1+\lambda ^2)\bigl]$  
\end{tabular}
\end{center}

\begin{center}
\begin{tabular}{c p{11cm}}
$f_{(\yng(1),\yng(1))}[7_{2}]=$ &$\frac{(-1+\lambda ) }{q^2 \lambda ^2}\bigl[(-\lambda -3 q^2 \lambda -q^4 \lambda +q (1+\lambda ^2)+q^3(1+\lambda ^2)\bigl]$  
\end{tabular}
\end{center}

\begin{center}
\begin{tabular}{c p{11cm}}
$f_{(\yng(1),\yng(1))}[7_{3}]=$ &$\frac{\left(-1+\lambda ^2\right) }{q \lambda ^3}\bigl[q-\lambda -q^2 \lambda +q \lambda ^2\bigl]$  
\end{tabular}
\end{center}

\item
For $R_1=\Yboxdim6pt\yng(1)$, $R_2=\Yboxdim6pt\yng(2)$    : 
\Yboxdim4pt

\begin{center}
\begin{tabular}{c p{11cm}}
$f_{(\yng(1),\yng(2))}[6_{2}]=$ &$ \frac{(-1+\lambda ) }{\sqrt{q} \sqrt{\lambda }}\bigl[\lambda ^2+q \lambda ^2+q^3 \lambda ^2+q^4 \lambda ^2+q^2 \left(-1-\lambda +\lambda ^2\right)\bigl]$
\end{tabular}
\end{center}

\begin{center}
\begin{tabular}{c p{11cm}}
$f_{(\yng(1),\yng(2))}[6_{3}]=$ &$\frac{1}{q^{5/2} \lambda ^{7/2}}(1+q) (-1+\lambda ) \bigl[\lambda +q^2 \lambda -q \left(1+\lambda +\lambda ^2\right)\bigl]$  
\end{tabular}
\end{center}

\begin{center}
\begin{tabular}{c p{11cm}}
$f_{(\yng(1),\yng(2))}[7_{1}]=$ &$\frac{(-1+\lambda ) }{q^{5/2} \lambda ^{5/2}}\bigl[q-\lambda -q^6 \lambda ^2+q^3 (-2+\lambda ) \lambda ^2+q^2 \lambda ^3+q^4 \lambda  (1-\lambda +\lambda ^2)+q^5 \lambda  (1-\lambda +\lambda ^2)\bigl]$  
\end{tabular}
\end{center}

\begin{center}
\begin{tabular}{c p{11cm}}
$f_{(\yng(1),\yng(2))}[7_{2}]=$ &$\frac{(-1+\lambda )}{q^{7/2} \lambda ^{5/2}} \bigl[(q^4-\lambda -3 q^3 \lambda -q^6 \lambda ^2+q^5 \lambda ^3+q (1-\lambda +\lambda ^2)+q^2 (1-\lambda +\lambda ^2)\bigl]$  
\end{tabular}
\end{center}

\begin{center}
\begin{tabular}{c p{11cm}}
$f_{(\yng(1),\yng(2))}[7_{3}]=$ &$\frac{\left(-1+\lambda ^2\right) }{q^{3/2} \lambda ^{7/2}}\bigl[q-\lambda -q^3 \lambda ^2+q^2 \lambda ^3\bigl]$  
\end{tabular}
\end{center}

We have checked that $f_{(\yng(1),\yng(2))}[L]=f_{(\yng(2),\yng(1))}[L]$ for these links. We also have the symmetry relation 
\begin{align}
f_{(\yng(1),\yng(2))}[\mathcal{L}](q^{-1},\lambda)=-f_{\left(\yng(1),\yng(1,1)\right)}[\mathcal{L}](q,\lambda).
\end{align}

\item
For $R_1=\Yboxdim6pt\yng(2)$, $R_2=\Yboxdim6pt\yng(2)$    : 
\Yboxdim4pt
\renewcommand{\baselinestretch}{1.5}\selectfont
\begin{center}
\begin{tabular}{c p{11cm}}
$f_{(\yng(2),\yng(2))}[6_{2}]=$ &$\frac{1}{q^2 \lambda }\bigl[q (-1+\lambda )^2 \lambda +q^9 (-1+\lambda )^2 \lambda ^2+\lambda ^3-\lambda ^5+q^7 (-1+\lambda )^2 \lambda ^2 (3+\lambda )+q^{10} \lambda ^3 (-1+\lambda ^2)+q^8 \lambda ^2 (2-3 \lambda +\lambda ^2)+q^5 (-1+\lambda )^2 (-1-2 \lambda +2 \lambda ^2)+q^3 (-1+\lambda )^2 (1+\lambda +2 \lambda ^2)+q^4 (-1+\lambda +4 \lambda ^2-5 \lambda ^3+\lambda ^4)+q^6 \lambda  (-1+5 \lambda -8 \lambda ^2+3 \lambda ^3+\lambda ^4)+q^2 (-1+\lambda ^2+\lambda ^4-\lambda ^5)\bigl]$
\end{tabular}
\end{center}

\begin{center}
\begin{tabular}{c p{11cm}}
$f_{(\yng(2),\yng(2))}[6_{3}]=$ &$\frac{1}{q^7 \lambda ^6}\bigl[-(-1+\lambda )^2 \lambda +q^2 (-1+\lambda )^2 \lambda ^2-q^9 (-1+\lambda ) \lambda ^3+q^5 (-1+\lambda )^2 \lambda ^2 (2+\lambda ^2)-q^4 (-1+\lambda )^2 \lambda  (1-\lambda +2 \lambda ^2)+q^8 (-1+\lambda )^2 \lambda (1+\lambda +2 \lambda ^2)+q^3 \lambda ^2 (3-4 \lambda +3 \lambda ^2-2 \lambda ^3)+q^6 (-1+\lambda )^2 \lambda  (1+3 \lambda +2 \lambda ^2+2 \lambda ^3)+q (1-2 \lambda +\lambda ^2-\lambda ^3+\lambda ^4)-q^7 (1+2 \lambda ^2-2 \lambda ^3-2 \lambda ^4+\lambda ^6)\bigl]$  
\end{tabular}
\end{center}

\begin{center}
\begin{tabular}{c p{11cm}}
$f_{(\yng(2),\yng(2))}[7_{1}]=$ &$\frac{1}{q^6 \lambda ^4}\bigl[-(-1+\lambda )^2 \lambda +q^{11} (-1+\lambda )^4 \lambda ^2-q^{13} (-1+\lambda )^2 \lambda ^3+q (-1+\lambda )^2 (1+\lambda ^2)+3 q^9 (-1+\lambda )^2 \lambda ^2(1-\lambda +\lambda ^2)+q^3 (-1+\lambda )^2 (1-\lambda +2 \lambda ^2)+q^2 \lambda  (-2+4 \lambda -3 \lambda ^2+\lambda ^3)+q^{12} \lambda ^2 (1-2 \lambda +3 \lambda ^2-3 \lambda ^3+\lambda ^4)-q^6 (1+\lambda -\lambda ^2-2 \lambda ^3+\lambda ^4)+q^7 (-1+\lambda )^2 (1+2 \lambda +2 \lambda ^2-\lambda ^3+\lambda ^4)+q^8 \lambda ^2 (2-4 \lambda +7 \lambda ^2-7 \lambda ^3+2 \lambda ^4)-q^5 \lambda (-1+\lambda +2 \lambda ^3-3 \lambda ^4+\lambda ^5)+q^{10} \lambda  (-1+2 \lambda -7 \lambda ^2+11 \lambda ^3-7 \lambda ^4+2 \lambda ^5)-q^4 (-1+3 \lambda -4 \lambda ^2+\lambda ^3+\lambda ^4-\lambda ^5+\lambda ^6)\bigl]$  
\end{tabular}
\end{center}

\begin{center}
\begin{tabular}{c p{11cm}}
$f_{(\yng(2),\yng(2))}[7_{2}]=$ &$-\frac{1}{q^8 \lambda ^4}\bigl[(-1+\lambda )^2 \lambda +q^{13} (-1+\lambda )^2 \lambda ^3-q (-1+\lambda )^2 (1+\lambda ^2)+q^3 (-1+\lambda )^2 \lambda  (3-2 \lambda +\lambda ^2)+q^{11} (-1+\lambda )^2 \lambda  (-1-\lambda +\lambda ^2)+q^5 (-1+\lambda )^2 (-1+2 \lambda -5 \lambda ^2+\lambda ^3)-q^2 \lambda  (-1+4 \lambda -4 \lambda ^2+\lambda ^3)-q^{12} \lambda ^3 (-1+2 \lambda -2 \lambda ^2+\lambda ^3)-q^7 (-1+\lambda )^2 (1+\lambda +5 \lambda ^2+\lambda ^3)+q^{10} \lambda ^2 (1-2 \lambda -2 \lambda ^2+3 \lambda ^3)+q^6 \lambda  (3-10 \lambda +9 \lambda ^2-5 \lambda ^3+3 \lambda ^4)+q^4 (-1+6 \lambda -11 \lambda ^2+12 \lambda ^3-7 \lambda ^4+\lambda ^5)-q^9 \lambda  (3-3 \lambda -2 \lambda ^3+\lambda ^4+\lambda ^5)+q^8 (1+\lambda +2 \lambda ^2-5 \lambda ^3+\lambda ^4-\lambda ^5+\lambda ^6)\bigl]$  
\end{tabular}
\end{center}

\begin{center}
\begin{tabular}{c p{11cm}}
$f_{(\yng(2),\yng(2))}[7_{3}]=$ &$-\frac{(-1+\lambda )^2 }{q^6 \lambda ^6}\bigl[(-q+\lambda -q^5 \lambda +q^9 \lambda ^4 (1+\lambda )+q^4 \lambda  (1+\lambda )^2+q^6 \lambda ^2 (1+\lambda )^2 (1+\lambda ^2)-q^3 \lambda ^2 (1+\lambda +\lambda ^2)-q^7 \lambda  (1+2 \lambda +3 \lambda ^2+3 \lambda ^3+3 \lambda ^4+\lambda ^5)+q^8 (\lambda ^2+\lambda ^3+2 \lambda ^4-\lambda ^6)\bigl]$  
\end{tabular}
\end{center}

\renewcommand{\baselinestretch}{1}\selectfont

Changing both the rank two symmetric representation $\Yboxdim6pt\yng(2)$ by 
antisymmetric representation $\Yboxdim6pt\yng(1,1)$, we find  the
following relation between the reformulated invariants:
\Yboxdim4pt
\begin{align}
\Yvcentermath1
f_{\left(\yng(2),\yng(2)\right)}[\mathcal{L}](q,\lambda)=f_{\left(\yng(1,1),\yng(1,1)\right)}[\mathcal{L}](q^{-1},\lambda).
\end{align} 
\end{enumerate}

\end{document}

%% file: glue.pspdftex
\begin{picture}(0,0)%
\includegraphics{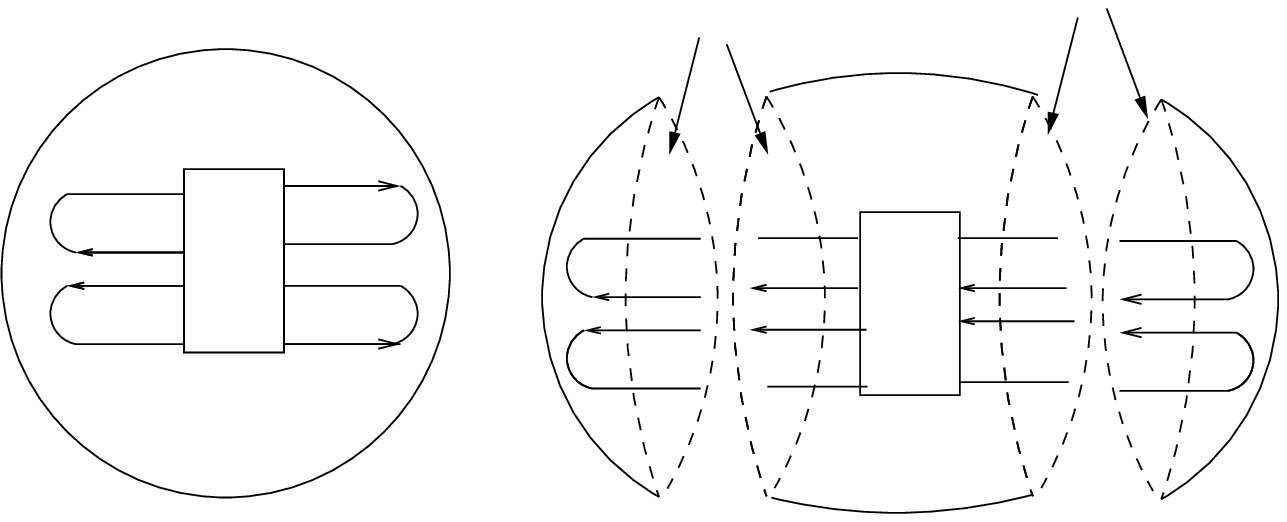}%
\end{picture}%
\setlength{\unitlength}{3947sp}%
\begingroup\makeatletter\ifx\SetFigFont\undefined%
\gdef\SetFigFont#1#2#3#4#5{%
  \reset@font\fontsize{#1}{#2pt}%
  \fontfamily{#3}\fontseries{#4}\fontshape{#5}%
  \selectfont}%
\fi\endgroup%
\begin{picture}(6144,2931)(262,-3448)
\put(586,-1754){\makebox(0,0)[lb]{\smash{{\SetFigFont{7}{8.4}{\rmdefault}{\mddefault}{\updefault}{\color[rgb]{0,0,0}$R_1$}%
}}}}
\put(507,-1475){\makebox(0,0)[lb]{\smash{{\SetFigFont{7}{8.4}{\rmdefault}{\mddefault}{\updefault}{\color[rgb]{0,0,0}$\bar R_1$}%
}}}}
\put(546,-2074){\makebox(0,0)[lb]{\smash{{\SetFigFont{7}{8.4}{\rmdefault}{\mddefault}{\updefault}{\color[rgb]{0,0,0}$R_2$}%
}}}}
\put(586,-2354){\makebox(0,0)[lb]{\smash{{\SetFigFont{7}{8.4}{\rmdefault}{\mddefault}{\updefault}{\color[rgb]{0,0,0}$\bar R_2$}%
}}}}
\put(1700,-2360){\makebox(0,0)[lb]{\smash{{\SetFigFont{7}{8.4}{\rmdefault}{\mddefault}{\updefault}{\color[rgb]{0,0,0}$R_1$}%
}}}}
\put(1705,-1423){\makebox(0,0)[lb]{\smash{{\SetFigFont{7}{8.4}{\rmdefault}{\mddefault}{\updefault}{\color[rgb]{0,0,0}$R_2$}%
}}}}
\put(1263,-1922){\makebox(0,0)[lb]{\smash{{\SetFigFont{9}{10.8}{\rmdefault}{\mddefault}{\updefault}{\color[rgb]{0,0,0}$\cal B$}%
}}}}
\put(6277,-1699){\makebox(0,0)[rb]{\smash{{\SetFigFont{7}{8.4}{\rmdefault}{\mddefault}{\updefault}{\color[rgb]{0,0,0}$R_2$}%
}}}}
\put(6197,-1979){\makebox(0,0)[rb]{\smash{{\SetFigFont{7}{8.4}{\rmdefault}{\mddefault}{\updefault}{\color[rgb]{0,0,0}$\bar R_2$}%
}}}}
\put(6197,-2578){\makebox(0,0)[rb]{\smash{{\SetFigFont{7}{8.4}{\rmdefault}{\mddefault}{\updefault}{\color[rgb]{0,0,0}$R_1$}%
}}}}
\put(6237,-2299){\makebox(0,0)[rb]{\smash{{\SetFigFont{7}{8.4}{\rmdefault}{\mddefault}{\updefault}{\color[rgb]{0,0,0}$\bar R_1$}%
}}}}
\put(3409,-662){\makebox(0,0)[lb]{\smash{{\SetFigFont{9}{10.8}{\rmdefault}{\mddefault}{\updefault}{\color[rgb]{0,0,0}oppositely oriented $S^2$ boundary}%
}}}}
\put(4358,-3221){\makebox(0,0)[lb]{\smash{{\SetFigFont{9}{10.8}{\rmdefault}{\mddefault}{\updefault}{\color[rgb]{0,0,0}$B_2$}%
}}}}
\put(3487,-982){\makebox(0,0)[lb]{\smash{{\SetFigFont{9}{10.8}{\rmdefault}{\mddefault}{\updefault}{\color[rgb]{0,0,0}1}%
}}}}
\put(3862,-916){\makebox(0,0)[lb]{\smash{{\SetFigFont{9}{10.8}{\rmdefault}{\mddefault}{\updefault}{\color[rgb]{0,0,0}1}%
}}}}
\put(5681,-883){\makebox(0,0)[lb]{\smash{{\SetFigFont{9}{10.8}{\rmdefault}{\mddefault}{\updefault}{\color[rgb]{0,0,0}2}%
}}}}
\put(5273,-982){\makebox(0,0)[lb]{\smash{{\SetFigFont{9}{10.8}{\rmdefault}{\mddefault}{\updefault}{\color[rgb]{0,0,0}2}%
}}}}
\put(3112,-3187){\makebox(0,0)[lb]{\smash{{\SetFigFont{9}{10.8}{\rmdefault}{\mddefault}{\updefault}{\color[rgb]{0,0,0}$B_1$}%
}}}}
\put(5781,-3187){\makebox(0,0)[lb]{\smash{{\SetFigFont{9}{10.8}{\rmdefault}{\mddefault}{\updefault}{\color[rgb]{0,0,0}$B_3$}%
}}}}
\put(3066,-1968){\makebox(0,0)[lb]{\smash{{\SetFigFont{7}{8.4}{\rmdefault}{\mddefault}{\updefault}{\color[rgb]{0,0,0}$R_1$}%
}}}}
\put(2986,-1688){\makebox(0,0)[lb]{\smash{{\SetFigFont{7}{8.4}{\rmdefault}{\mddefault}{\updefault}{\color[rgb]{0,0,0}$\bar R_1$}%
}}}}
\put(3026,-2288){\makebox(0,0)[lb]{\smash{{\SetFigFont{7}{8.4}{\rmdefault}{\mddefault}{\updefault}{\color[rgb]{0,0,0}$R_2$}%
}}}}
\put(3066,-2567){\makebox(0,0)[lb]{\smash{{\SetFigFont{7}{8.4}{\rmdefault}{\mddefault}{\updefault}{\color[rgb]{0,0,0}$\bar R_2$}%
}}}}
\put(3089,-3364){\makebox(0,0)[lb]{\smash{{\SetFigFont{9}{10.8}{\rmdefault}{\mddefault}{\updefault}{\color[rgb]{0,0,0}$\langle \Psi_0^{(1)}\vert$}%
}}}}
\put(2465,-1954){\makebox(0,0)[lb]{\smash{{\SetFigFont{9}{10.8}{\rmdefault}{\mddefault}{\updefault}{\color[rgb]{0,0,0}$\equiv$}%
}}}}
\put(1005,-676){\makebox(0,0)[lb]{\smash{{\SetFigFont{12}{14.4}{\rmdefault}{\mddefault}{\updefault}{\color[rgb]{0,0,0}$S^3$}%
}}}}
\put(5693,-3387){\makebox(0,0)[lb]{\smash{{\SetFigFont{9}{10.8}{\rmdefault}{\mddefault}{\updefault}{\color[rgb]{0,0,0}$\vert \chi_0^{(2)}\rangle$}%
}}}}
\put(4398,-3379){\makebox(0,0)[lb]{\smash{{\SetFigFont{9}{10.8}{\rmdefault}{\mddefault}{\updefault}{\color[rgb]{0,0,0}${\cal B} \nu^{(1),(2)}$}%
}}}}
\put(4511,-2066){\makebox(0,0)[lb]{\smash{{\SetFigFont{9}{10.8}{\rmdefault}{\mddefault}{\updefault}{\color[rgb]{0,0,0}$\cal B$}%
}}}}
\end{picture}%

%% file: unknot.pspdftex
\begin{picture}(0,0)%
\includegraphics{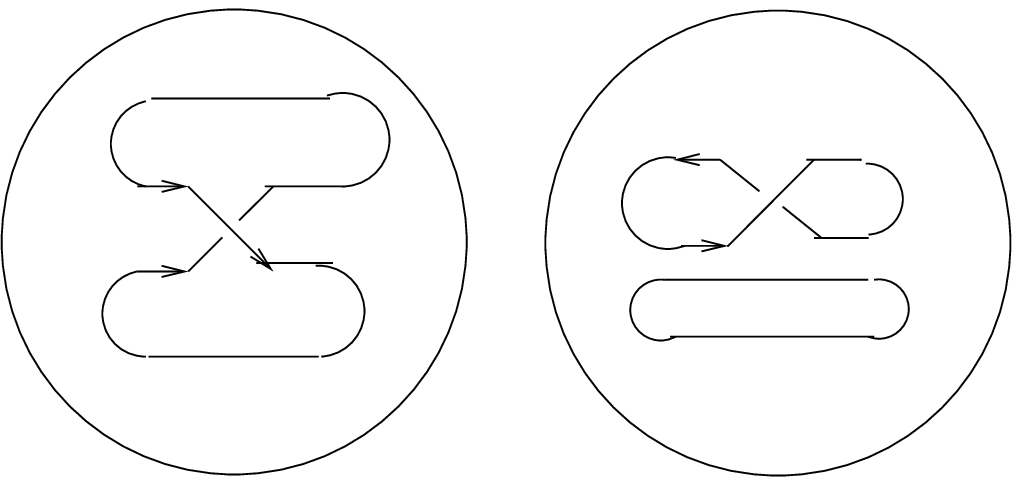}%
\end{picture}%
\setlength{\unitlength}{4144sp}%
\begingroup\makeatletter\ifx\SetFigFont\undefined%
\gdef\SetFigFont#1#2#3#4#5{%
  \reset@font\fontsize{#1}{#2pt}%
  \fontfamily{#3}\fontseries{#4}\fontshape{#5}%
  \selectfont}%
\fi\endgroup%
\begin{picture}(4628,2145)(912,-1765)
\put(1523, 52){\rotatebox{360.0}{\makebox(0,0)[lb]{\smash{{\SetFigFont{10}{12.0}{\rmdefault}{\mddefault}{\updefault}{\color[rgb]{0,0,0}$\bar R$}%
}}}}}
\put(1747,-1136){\rotatebox{90.0}{\makebox(0,0)[lb]{\smash{{\SetFigFont{10}{12.0}{\rmdefault}{\mddefault}{\updefault}{\color[rgb]{0,0,0}$R$}%
}}}}}
\put(1747,-1564){\rotatebox{90.0}{\makebox(0,0)[lb]{\smash{{\SetFigFont{10}{12.0}{\rmdefault}{\mddefault}{\updefault}{\color[rgb]{0,0,0}$\bar R$}%
}}}}}
\put(1747,-397){\rotatebox{90.0}{\makebox(0,0)[lb]{\smash{{\SetFigFont{10}{12.0}{\rmdefault}{\mddefault}{\updefault}{\color[rgb]{0,0,0}$R$}%
}}}}}
\put(3963,-643){\makebox(0,0)[lb]{\smash{{\SetFigFont{10}{12.0}{\rmdefault}{\mddefault}{\updefault}{\color[rgb]{0,0,0}$R$}%
}}}}
\put(3769,-254){\makebox(0,0)[lb]{\smash{{\SetFigFont{10}{12.0}{\rmdefault}{\mddefault}{\updefault}{\color[rgb]{0,0,0}$\bar R$}%
}}}}
\put(4236,-1265){\makebox(0,0)[lb]{\smash{{\SetFigFont{10}{12.0}{\rmdefault}{\mddefault}{\updefault}{\color[rgb]{0,0,0}$0$}%
}}}}
\put(3079,-700){\makebox(0,0)[lb]{\smash{{\SetFigFont{10}{12.0}{\rmdefault}{\mddefault}{\updefault}{\color[rgb]{0,0,0}$\equiv$}%
}}}}
\end{picture}%

%% file: equiva.pspdftex
\begin{picture}(0,0)%
\includegraphics{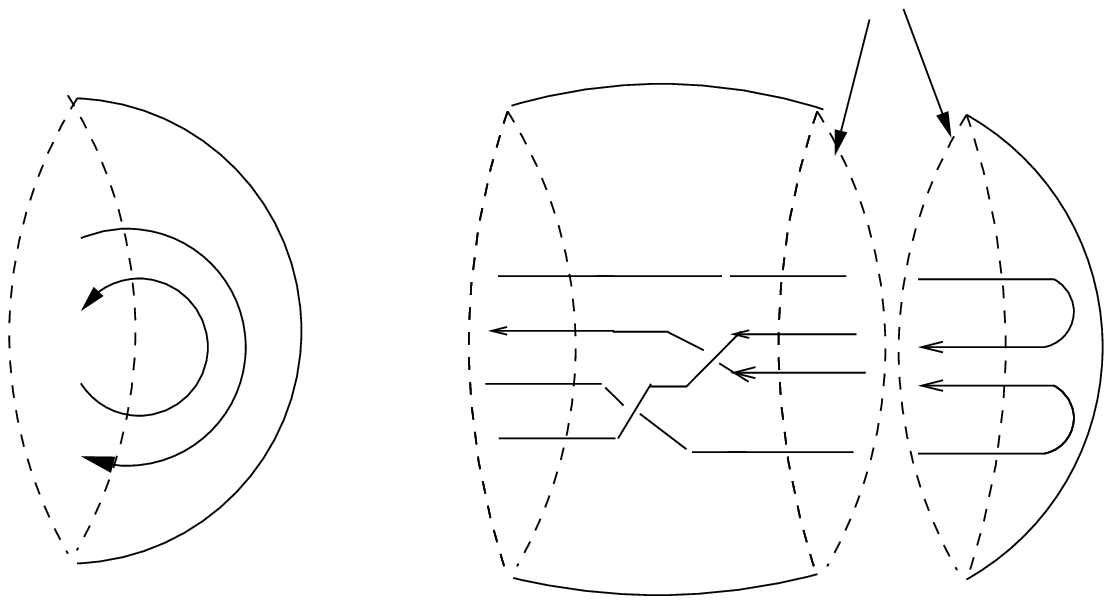}%
\end{picture}%
\setlength{\unitlength}{3947sp}%
\begingroup\makeatletter\ifx\SetFigFont\undefined%
\gdef\SetFigFont#1#2#3#4#5{%
  \reset@font\fontsize{#1}{#2pt}%
  \fontfamily{#3}\fontseries{#4}\fontshape{#5}%
  \selectfont}%
\fi\endgroup%
\begin{picture}(5790,3385)(473,-2935)
\put(3305,  3){\makebox(0,0)[lb]{\smash{{\SetFigFont{10}{12.0}{\rmdefault}{\mddefault}{\updefault}{\color[rgb]{0,0,0}1}%
}}}}
\put(5421, 41){\makebox(0,0)[lb]{\smash{{\SetFigFont{10}{12.0}{\rmdefault}{\mddefault}{\updefault}{\color[rgb]{0,0,0}2}%
}}}}
\put(4947,-72){\makebox(0,0)[lb]{\smash{{\SetFigFont{10}{12.0}{\rmdefault}{\mddefault}{\updefault}{\color[rgb]{0,0,0}2}%
}}}}
\put(3929,-2861){\makebox(0,0)[lb]{\smash{{\SetFigFont{10}{12.0}{\rmdefault}{\mddefault}{\updefault}{\color[rgb]{0,0,0}${\cal B} \nu^{(1),(2)}$}%
}}}}
\put(6114,-906){\makebox(0,0)[rb]{\smash{{\SetFigFont{9}{10.8}{\rmdefault}{\mddefault}{\updefault}{\color[rgb]{0,0,0}$R_2$}%
}}}}
\put(6021,-1929){\makebox(0,0)[rb]{\smash{{\SetFigFont{9}{10.8}{\rmdefault}{\mddefault}{\updefault}{\color[rgb]{0,0,0}$R_1$}%
}}}}
\put(6068,-1604){\makebox(0,0)[rb]{\smash{{\SetFigFont{9}{10.8}{\rmdefault}{\mddefault}{\updefault}{\color[rgb]{0,0,0}$\bar R_1$}%
}}}}
\put(5435,-2869){\makebox(0,0)[lb]{\smash{{\SetFigFont{10}{12.0}{\rmdefault}{\mddefault}{\updefault}{\color[rgb]{0,0,0}$\vert \chi_0^{(2)}\rangle$}%
}}}}
\put(1091,-154){\makebox(0,0)[lb]{\smash{{\SetFigFont{10}{12.0}{\rmdefault}{\mddefault}{\updefault}{\color[rgb]{0,0,0}1}%
}}}}
\put(2779,299){\makebox(0,0)[lb]{\smash{{\SetFigFont{10}{12.0}{\rmdefault}{\mddefault}{\updefault}{\color[rgb]{0,0,0}oppositely oriented $S^2$ boundary}%
}}}}
\put(2541,-1374){\makebox(0,0)[lb]{\smash{{\SetFigFont{10}{12.0}{\rmdefault}{\mddefault}{\updefault}{\color[rgb]{0,0,0}$\equiv$}%
}}}}
\put(1351,-668){\makebox(0,0)[lb]{\smash{{\SetFigFont{10}{12.0}{\rmdefault}{\mddefault}{\updefault}{\color[rgb]{0,0,0}$R_2$}%
}}}}
\put(3392,-826){\makebox(0,0)[lb]{\smash{{\SetFigFont{10}{12.0}{\rmdefault}{\mddefault}{\updefault}{\color[rgb]{0,0,0}${\cal B}=(b_2^{(+)})^{-1}b_3^{(-)}$}%
}}}}
\put(488,-2861){\makebox(0,0)[lb]{\smash{{\SetFigFont{10}{12.0}{\rmdefault}{\mddefault}{\updefault}{\color[rgb]{0,0,0}$\vert \hat{\Psi}_0^{(1)}\rangle$}%
}}}}
\put(1128,-1189){\makebox(0,0)[lb]{\smash{{\SetFigFont{10}{12.0}{\rmdefault}{\mddefault}{\updefault}{\color[rgb]{0,0,0}$\bar R_1$}%
}}}}
\end{picture}%

%% file: ybid.pspdftex
\begin{picture}(0,0)%
\includegraphics{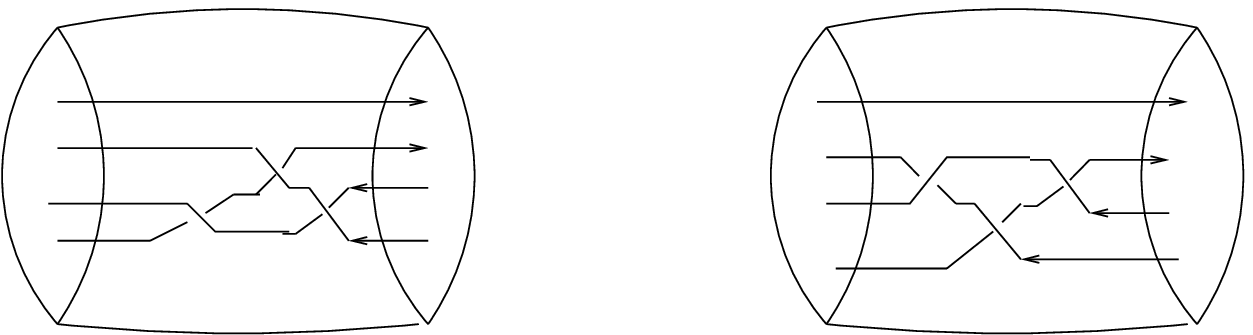}%
\end{picture}%
\setlength{\unitlength}{3947sp}%
\begingroup\makeatletter\ifx\SetFigFont\undefined%
\gdef\SetFigFont#1#2#3#4#5{%
  \reset@font\fontsize{#1}{#2pt}%
  \fontfamily{#3}\fontseries{#4}\fontshape{#5}%
  \selectfont}%
\fi\endgroup%
\begin{picture}(5976,1574)(1311,-1745)
\put(3277,-1425){\makebox(0,0)[lb]{\smash{{\SetFigFont{7}{8.4}{\rmdefault}{\mddefault}{\updefault}{\color[rgb]{0,0,0}$R_1$}%
}}}}
\put(3233,-1159){\makebox(0,0)[lb]{\smash{{\SetFigFont{7}{8.4}{\rmdefault}{\mddefault}{\updefault}{\color[rgb]{0,0,0}$R_2$}%
}}}}
\put(6968,-1248){\makebox(0,0)[lb]{\smash{{\SetFigFont{7}{8.4}{\rmdefault}{\mddefault}{\updefault}{\color[rgb]{0,0,0}$R_2$}%
}}}}
\put(7012,-1425){\makebox(0,0)[lb]{\smash{{\SetFigFont{7}{8.4}{\rmdefault}{\mddefault}{\updefault}{\color[rgb]{0,0,0}$R_1$}%
}}}}
\put(3321,-758){\makebox(0,0)[lb]{\smash{{\SetFigFont{7}{8.4}{\rmdefault}{\mddefault}{\updefault}{\color[rgb]{0,0,0}$\bar R_4$}%
}}}}
\put(3321,-981){\makebox(0,0)[lb]{\smash{{\SetFigFont{7}{8.4}{\rmdefault}{\mddefault}{\updefault}{\color[rgb]{0,0,0}$\bar R_3$}%
}}}}
\put(6923,-1026){\makebox(0,0)[lb]{\smash{{\SetFigFont{7}{8.4}{\rmdefault}{\mddefault}{\updefault}{\color[rgb]{0,0,0}$\bar R_3$}%
}}}}
\put(6923,-758){\makebox(0,0)[lb]{\smash{{\SetFigFont{7}{8.4}{\rmdefault}{\mddefault}{\updefault}{\color[rgb]{0,0,0}$\bar R_4$}%
}}}}
\put(4201,-961){\makebox(0,0)[lb]{\smash{{\SetFigFont{7}{8.4}{\rmdefault}{\mddefault}{\updefault}{\color[rgb]{0,0,0}$\equiv$}%
}}}}
\end{picture}%

%% file: sixpun.pspdftex
\begin{picture}(0,0)%
\includegraphics{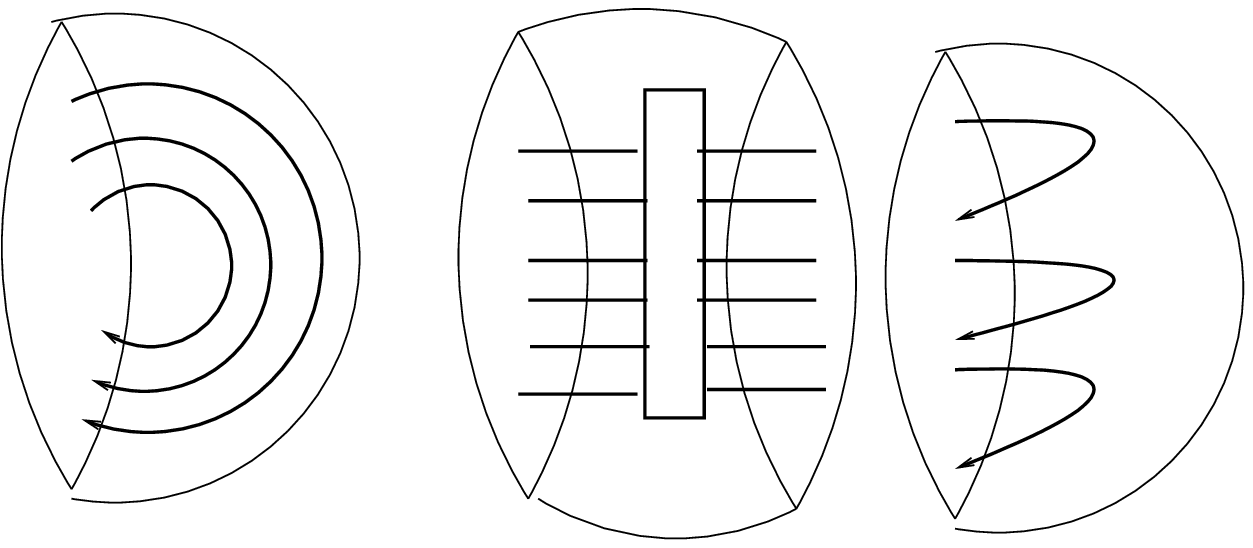}%
\end{picture}%
\setlength{\unitlength}{3947sp}%
\begingroup\makeatletter\ifx\SetFigFont\undefined%
\gdef\SetFigFont#1#2#3#4#5{%
  \reset@font\fontsize{#1}{#2pt}%
  \fontfamily{#3}\fontseries{#4}\fontshape{#5}%
  \selectfont}%
\fi\endgroup%
\begin{picture}(5977,2832)(939,-2691)
\put(1044,-503){\makebox(0,0)[lb]{\smash{{\SetFigFont{8}{9.6}{\rmdefault}{\mddefault}{\updefault}{\color[rgb]{0,0,0}$R_2$}%
}}}}
\put(1283,-1791){\makebox(0,0)[lb]{\smash{{\SetFigFont{8}{9.6}{\rmdefault}{\mddefault}{\updefault}{\color[rgb]{0,0,0}$\bar R_2$}%
}}}}
\put(2712,-933){\makebox(0,0)[lb]{\smash{{\SetFigFont{8}{9.6}{\rmdefault}{\mddefault}{\updefault}{\color[rgb]{0,0,0}$\equiv$}%
}}}}
\put(1092,-1361){\makebox(0,0)[lb]{\smash{{\SetFigFont{8}{9.6}{\rmdefault}{\mddefault}{\updefault}{\color[rgb]{0,0,0}$\bar R_3$}%
}}}}
\put(996,-742){\makebox(0,0)[lb]{\smash{{\SetFigFont{8}{9.6}{\rmdefault}{\mddefault}{\updefault}{\color[rgb]{0,0,0}$R_3$}%
}}}}
\put(1140,-170){\makebox(0,0)[lb]{\smash{{\SetFigFont{8}{9.6}{\rmdefault}{\mddefault}{\updefault}{\color[rgb]{0,0,0}$R_1$}%
}}}}
\put(1235,-2028){\makebox(0,0)[lb]{\smash{{\SetFigFont{8}{9.6}{\rmdefault}{\mddefault}{\updefault}{\color[rgb]{0,0,0}$\bar R_1$}%
}}}}
\put(5477,-312){\makebox(0,0)[lb]{\smash{{\SetFigFont{8}{9.6}{\rmdefault}{\mddefault}{\updefault}{\color[rgb]{0,0,0}$R_1$}%
}}}}
\put(5810,-885){\makebox(0,0)[lb]{\smash{{\SetFigFont{8}{9.6}{\rmdefault}{\mddefault}{\updefault}{\color[rgb]{0,0,0}$\bar R_1$}%
}}}}
\put(5286,-1218){\makebox(0,0)[lb]{\smash{{\SetFigFont{8}{9.6}{\rmdefault}{\mddefault}{\updefault}{\color[rgb]{0,0,0}$R_2$}%
}}}}
\put(5382,-1743){\makebox(0,0)[lb]{\smash{{\SetFigFont{8}{9.6}{\rmdefault}{\mddefault}{\updefault}{\color[rgb]{0,0,0}$R_3$}%
}}}}
\put(5762,-2124){\makebox(0,0)[lb]{\smash{{\SetFigFont{8}{9.6}{\rmdefault}{\mddefault}{\updefault}{\color[rgb]{0,0,0}$\bar R_3$}%
}}}}
\put(6001,-1457){\makebox(0,0)[lb]{\smash{{\SetFigFont{8}{9.6}{\rmdefault}{\mddefault}{\updefault}{\color[rgb]{0,0,0}$\bar R_2$}%
}}}}
\put(4093,-691){\makebox(0,0)[lb]{\smash{{\SetFigFont{8}{9.6}{\rmdefault}{\mddefault}{\updefault}{\color[rgb]{0,0,0}$\cal B$}%
}}}}
\put(6201,-2572){\makebox(0,0)[lb]{\smash{{\SetFigFont{9}{10.8}{\rmdefault}{\mddefault}{\updefault}{\color[rgb]{0,0,0}$\vert \xi \rangle$}%
}}}}
\put(958,-2515){\makebox(0,0)[lb]{\smash{{\SetFigFont{9}{10.8}{\rmdefault}{\mddefault}{\updefault}{\color[rgb]{0,0,0}$\vert \Psi \rangle$}%
}}}}
\put(3523,-2629){\makebox(0,0)[lb]{\smash{{\SetFigFont{9}{10.8}{\rmdefault}{\mddefault}{\updefault}{\color[rgb]{0,0,0}${\cal B}=g_6$}%
}}}}
\end{picture}%

%% file: sixpt.pspdftex
\begin{picture}(0,0)%
\includegraphics{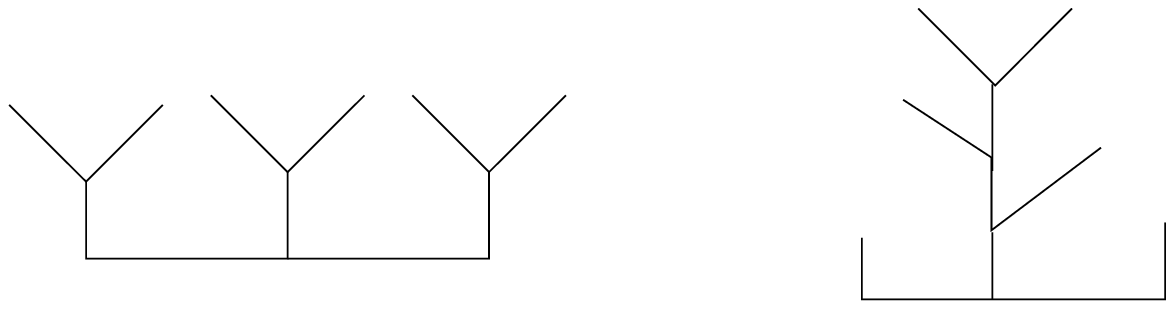}%
\end{picture}%
\setlength{\unitlength}{3947sp}%
\begingroup\makeatletter\ifx\SetFigFont\undefined%
\gdef\SetFigFont#1#2#3#4#5{%
  \reset@font\fontsize{#1}{#2pt}%
  \fontfamily{#3}\fontseries{#4}\fontshape{#5}%
  \selectfont}%
\fi\endgroup%
\begin{picture}(5666,1843)(155,-1702)
\put(170,-443){\makebox(0,0)[lb]{\smash{{\SetFigFont{7}{8.4}{\rmdefault}{\mddefault}{\updefault}{\color[rgb]{0,0,0}$R_1$}%
}}}}
\put(731,-452){\makebox(0,0)[lb]{\smash{{\SetFigFont{7}{8.4}{\rmdefault}{\mddefault}{\updefault}{\color[rgb]{0,0,0}$R_2$}%
}}}}
\put(1210,-424){\makebox(0,0)[lb]{\smash{{\SetFigFont{7}{8.4}{\rmdefault}{\mddefault}{\updefault}{\color[rgb]{0,0,0}$R_3$}%
}}}}
\put(1819,-424){\makebox(0,0)[lb]{\smash{{\SetFigFont{7}{8.4}{\rmdefault}{\mddefault}{\updefault}{\color[rgb]{0,0,0}$R_4$}%
}}}}
\put(2242,-424){\makebox(0,0)[lb]{\smash{{\SetFigFont{7}{8.4}{\rmdefault}{\mddefault}{\updefault}{\color[rgb]{0,0,0}$R_5$}%
}}}}
\put(2776,-424){\makebox(0,0)[lb]{\smash{{\SetFigFont{7}{8.4}{\rmdefault}{\mddefault}{\updefault}{\color[rgb]{0,0,0}$R_6$}%
}}}}
\put(2564,-1087){\makebox(0,0)[lb]{\smash{{\SetFigFont{7}{8.4}{\rmdefault}{\mddefault}{\updefault}{\color[rgb]{0,0,0}$s_3$}%
}}}}
\put(298,-1078){\makebox(0,0)[lb]{\smash{{\SetFigFont{7}{8.4}{\rmdefault}{\mddefault}{\updefault}{\color[rgb]{0,0,0}$s_1$}%
}}}}
\put(1445,-1105){\makebox(0,0)[lb]{\smash{{\SetFigFont{7}{8.4}{\rmdefault}{\mddefault}{\updefault}{\color[rgb]{0,0,0}$s_2$}%
}}}}
\put(556,-1640){\makebox(0,0)[lb]{\smash{{\SetFigFont{7}{8.4}{\rmdefault}{\mddefault}{\updefault}{\color[rgb]{0,0,0}$\phi_{s_1,s_2,s_3}(R_1,R_2,R_3,R_4,R_5,R_6)$}%
}}}}
\put(4437, 42){\makebox(0,0)[lb]{\smash{{\SetFigFont{7}{8.4}{\rmdefault}{\mddefault}{\updefault}{\color[rgb]{0,0,0}$R_3$}%
}}}}
\put(5165, 34){\makebox(0,0)[lb]{\smash{{\SetFigFont{7}{8.4}{\rmdefault}{\mddefault}{\updefault}{\color[rgb]{0,0,0}$R_4$}%
}}}}
\put(4317,-455){\makebox(0,0)[lb]{\smash{{\SetFigFont{7}{8.4}{\rmdefault}{\mddefault}{\updefault}{\color[rgb]{0,0,0}$R_2$}%
}}}}
\put(5201,-767){\makebox(0,0)[lb]{\smash{{\SetFigFont{7}{8.4}{\rmdefault}{\mddefault}{\updefault}{\color[rgb]{0,0,0}$R_5$}%
}}}}
\put(5782,-998){\makebox(0,0)[lb]{\smash{{\SetFigFont{7}{8.4}{\rmdefault}{\mddefault}{\updefault}{\color[rgb]{0,0,0}$R_6$}%
}}}}
\put(4216,-1117){\makebox(0,0)[lb]{\smash{{\SetFigFont{7}{8.4}{\rmdefault}{\mddefault}{\updefault}{\color[rgb]{0,0,0}$R_1$}%
}}}}
\put(4999,-546){\makebox(0,0)[lb]{\smash{{\SetFigFont{7}{8.4}{\rmdefault}{\mddefault}{\updefault}{\color[rgb]{0,0,0}$t_1$}%
}}}}
\put(5017,-1320){\makebox(0,0)[lb]{\smash{{\SetFigFont{7}{8.4}{\rmdefault}{\mddefault}{\updefault}{\color[rgb]{0,0,0}$t_3$}%
}}}}
\put(4051,-1651){\makebox(0,0)[lb]{\smash{{\SetFigFont{7}{8.4}{\rmdefault}{\mddefault}{\updefault}{\color[rgb]{0,0,0}$\hat {\phi}_{t_1,t_2,t_3}(R_1,R_2,R_3,R_4,R_5,R_6)$}%
}}}}
\put(4796,-928){\makebox(0,0)[lb]{\smash{{\SetFigFont{7}{8.4}{\rmdefault}{\mddefault}{\updefault}{\color[rgb]{0,0,0}$t_2$}%
}}}}
\end{picture}%

%% file: racahfinal.bbl
\begin{thebibliography}{10}
\expandafter\ifx\csname url\endcsname\relax
  \def\url#1{\texttt{#1}}\fi
\expandafter\ifx\csname urlprefix\endcsname\relax\def\urlprefix{URL }\fi
\expandafter\ifx\csname href\endcsname\relax
  \def\href#1#2{#2} \def\path#1{#1}\fi

\bibitem{Witten:1988hf}
E.~Witten, {\it {Quantum Field Theory and the Jones Polynomial}},  {\em
  Commun.Math.Phys.} {\bf 121} (1989) 351.
  \newblock \href {http://dx.doi.org/10.1007/BF01217730}
  {\path{doi:10.1007/BF01217730}}.


\bibitem{RamaDevi:1992np}
P.~Rama~Devi, T.~Govindarajan, R.~Kaul, {Three-dimensional Chern-Simons theory
  as a theory of knots and links. 3. Compact semisimple group}, Nucl.Phys. B402
  (1993) 548--566.
\newblock \href {http://arxiv.org/abs/hep-th/9212110}
  {\path{arXiv:hep-th/9212110}}, \href
  {http://dx.doi.org/10.1016/0550-3213(93)90652-6}
  {\path{doi:10.1016/0550-3213(93)90652-6}}.


\bibitem{Ramadevi:1996:PHD}
P.~Ramadevi, {\em Chern-Simons theory as a theory of knots and links}.
\newblock PhD thesis, The Institute of Mathematical Sciences (IMSc), 1996.

\bibitem{Labastida:1990bt}
J.~Labastida, P.~Llatas, A.~Ramallo, {Knot operators in Chern-Simons gauge
  theory}, Nucl.Phys. B348 (1991) 651--692.
\newblock \href {http://dx.doi.org/10.1016/0550-3213(91)90209-G}
  {\path{doi:10.1016/0550-3213(91)90209-G}}.

\bibitem{Kirillov:1989}
A.~N. Kirillov and N.~Y. Reshetikhin, {\it {Representation algebra $U_q(sl_2)$,
  $q$-orthogonal polynomials and invariants of links}},  {\em New Developments
  in the Theory of Knots} (1989). ed. T. Kohno, World Scientific, Singapore.

\bibitem{Kaul:1993hb}
R.~Kaul, {Chern-Simons theory, colored oriented braids and link invariants},
  Commun.Math.Phys. 162 (1994) 289--320.
\newblock \href {http://arxiv.org/abs/hep-th/9305032}
  {\path{arXiv:hep-th/9305032}}, \href {http://dx.doi.org/10.1007/BF02102019}
  {\path{doi:10.1007/BF02102019}}.


\bibitem{Ooguri:1999bv}
H.~Ooguri, C.~Vafa, {Knot invariants and topological strings}, Nucl.Phys. B577
  (2000) 419--438.
\newblock \href {http://arxiv.org/abs/hep-th/9912123}
  {\path{arXiv:hep-th/9912123}}, \href
  {http://dx.doi.org/10.1016/S0550-3213(00)00118-8}
  {\path{doi:10.1016/S0550-3213(00)00118-8}}.


\bibitem{Labastida:2000yw}
J.~Labastida, M.~Marino, C.~Vafa, {Knots, links and branes at large N}, JHEP
  0011 (2000) 007.
\newblock \href {http://arxiv.org/abs/hep-th/0010102}
  {\path{arXiv:hep-th/0010102}}.

\bibitem{Ramadevi:2000gq}
P.~Ramadevi, T.~Sarkar, {On link invariants and topological string amplitudes},
  Nucl.Phys. B600 (2001) 487--511.
\newblock \href {http://arxiv.org/abs/hep-th/0009188}
  {\path{arXiv:hep-th/0009188}}, \href
  {http://dx.doi.org/10.1016/S0550-3213(00)00761-6}
  {\path{doi:10.1016/S0550-3213(00)00761-6}}.

\bibitem{Witten:2011zz}
E.~Witten, {Fivebranes and Knots.}
\newblock\href {http://arxiv.org/abs/1101.3216}
  {\path{arXiv:1101.3216}}.

\bibitem{Brini:2011wi}
A.~Brini, B.~Eynard, M.~Marino, {Torus knots and mirror symmetry.}
\newblock\href
  {http://arxiv.org/abs/1105.2012} {\path{arXiv:1105.2012}}.


\bibitem{Aganagic:2011sg}
M.~Aganagic, S.~Shakirov, {Knot Homology from Refined Chern-Simons
  Theory.}
  \newblock\href {http://arxiv.org/abs/1105.5117} {\path{arXiv:1105.5117}}.


\bibitem{Gaiotto:2011nm}
D.~Gaiotto, E.~Witten, {Knot Invariants from Four-Dimensional Gauge
  Theory.}
  \newblock\href {http://arxiv.org/abs/1106.4789} {\path{arXiv:1106.4789}}.


\bibitem{Marino:2001re}
M.~Marino, C.~Vafa, {Framed knots at large N.}
\newblock\href
  {http://arxiv.org/abs/hep-th/0108064} {\path{arXiv:hep-th/0108064}}.

\bibitem{Borhade:2003cu}
P.~Borhade, P.~Ramadevi, T.~Sarkar, {U(N) framed links, three manifold
  invariants, and topological strings}, Nucl.Phys. B678 (2004) 656--681.
\newblock \href {http://arxiv.org/abs/hep-th/0306283}
  {\path{arXiv:hep-th/0306283}}, \href
  {http://dx.doi.org/10.1016/j.nuclphysb.2003.11.023}
  {\path{doi:10.1016/j.nuclphysb.2003.11.023}}.

\bibitem{birman}
J.~S. Birman, Braids, Links and Mapping Class groups, Annals of Mathematics
  Studies, Princeton, NJ: Princeton Univ. Press, 1975.

\bibitem{Labastida:2001ts}
J.~M.~F. Labastida, M.~Marino, {A New point of view in the theory of knot and
  link invariants.}
  \newblock\href {http://arxiv.org/abs/math/0104180}
  {\path{arXiv:math/0104180}}.

\bibitem{Paul:2010wr}
C.~Paul, P.~Borhade, P.~Ramadevi, {Composite Representation Invariants and
  Unoriented Topological String Amplitudes}, Nucl.Phys. B841 (2010) 448--462.
\newblock \href {http://arxiv.org/abs/1008.3453} {\path{arXiv:1008.3453}},
  \href {http://dx.doi.org/10.1016/j.nuclphysb.2010.08.013}
  {\path{doi:10.1016/j.nuclphysb.2010.08.013}}.

\bibitem{Freyd:1985dx}
P.~Freyd, D.~Yetter, J.~Hoste, W.~B.~R. Lickorish, K.~Millett, et~al., {A new
  polynomial invariant of knots and links}, Bull.Am.Math.Soc. 12 (1985)
  239--246.
\newblock \href
  {http://dx.doi.org/http://dx.doi.org/10.1090/S0273-0979-1985-15361-3}
  {\path{doi:10.1090/S0273-0979-1985-15361-3}}.

\bibitem{PT}
J.~H. Przytycki, K.~P. Traczyk, {Conway Algebras and Skein Equivalence of
  Links}, Proc. Amer. Math. Soc. 100 (1987) 744--748.

\bibitem{Itoyama:2012fq}
H.~Itoyama, A.~Mironov, A.~Morozov, An.~Morozov, {HOMFLY and superpolynomials
  for figure eight knot in all symmetric and antisymmetric representations},
  JHEP 1207 (2012) 131.
\newblock \href {http://arxiv.org/abs/1203.5978} {\path{arXiv:1203.5978}},
  \href {http://dx.doi.org/10.1007/JHEP07(2012)131}
  {\path{doi:10.1007/JHEP07(2012)131}}.

\bibitem{Itoyama:2012qt}
H.~Itoyama, A.~Mironov, A.~Morozov, An.~Morozov,  {Character expansion for HOMFLY
  polynomials. III. All 3-Strand braids in the first symmetric representation},
  Int.J.Mod.Phys. A27 (2012) 1250099.
\newblock \href {http://arxiv.org/abs/1204.4785} {\path{arXiv:1204.4785}},
  \href {http://dx.doi.org/10.1142/S0217751X12500996}
  {\path{doi:10.1142/S0217751X12500996}}.

\end{thebibliography}
